\documentclass[
 reprint,
 superscriptaddress,
 preprintnumbers,
 amsmath
 amssymb,
 prx,
 floatfix,
]{revtex4-2}

\DeclareMathAlphabet{\mathsfit}{T1}{\sfdefault}{\mddefault}{\sldefault}
\SetMathAlphabet{\mathsfit}{bold}{T1}{\sfdefault}{\bfdefault}{\sldefault}

\usepackage{graphicx}
\graphicspath{{fig/}}   
\usepackage{dcolumn}
\usepackage{bm}
\usepackage{makecell}
\usepackage{braket} 
\usepackage{siunitx}
\usepackage{color}
\usepackage{array}
\usepackage{xr}
\usepackage{upgreek}
\usepackage[colorlinks]{hyperref}
\usepackage[capitalize]{cleveref}
\usepackage[caption=false]{subfig}
\usepackage{multirow} 

\newcommand{\supp}{Supplementary Material}

\newcommand{\tr}{\mathrm{Tr}}

\begin{document} 

\title{Probing entanglement across the energy spectrum of a hard-core Bose-Hubbard lattice}

\def\RLEaffil{Research Laboratory of Electronics, Massachusetts Institute of Technology, Cambridge, MA 02139, USA}
\def\LLaffil{MIT Lincoln Laboratory, Lexington, MA 02421, USA}
\def\Physaffil{Department of Physics, Massachusetts Institute of Technology, Cambridge, MA 02139, USA}
\def\EECSaffil{Department of Electrical Engineering and Computer Science, Massachusetts Institute of Technology, Cambridge, MA 02139, USA}
\def\Wellesleyaffil{Department of Physics, Wellesley College, Wellesley, MA 02481, USA}
\def\Maryaffil{Laboratory for Physical Sciences, College Park, MD 20740, USA}

\author{Amir~H.~Karamlou}
\email{karamlou@mit.edu}
\affiliation{\RLEaffil}
\affiliation{\EECSaffil}

\author{Ilan~T.~Rosen}
\affiliation{\RLEaffil}

\author{Sarah~E.~Muschinske}
\affiliation{\RLEaffil}
\affiliation{\EECSaffil}

\author{Cora~N.~Barrett}
\affiliation{\RLEaffil} 
\affiliation{\Wellesleyaffil}

\author{Agustin~Di~Paolo}
\affiliation{\RLEaffil}

\author{Leon~Ding}
\affiliation{\RLEaffil}
\affiliation{\Physaffil}

\author{Patrick~M.~Harrington}
\affiliation{\RLEaffil}

\author{Max~Hays}
\affiliation{\RLEaffil}

\author{Rabindra~Das}
\affiliation{\LLaffil}

\author{David~K.~Kim}
\affiliation{\LLaffil}

\author{Bethany~M.~Niedzielski}
\affiliation{\LLaffil}

\author{Meghan~Schuldt}
\affiliation{\LLaffil}

\author{Kyle~Serniak}
\affiliation{\RLEaffil}
\affiliation{\LLaffil}

\author{Mollie~E.~Schwartz}
\affiliation{\LLaffil}

\author{Jonilyn~L.~Yoder}
\affiliation{\LLaffil}

\author{Simon~Gustavsson}
\affiliation{\RLEaffil}

\author{Yariv~Yanay}
\affiliation{\Maryaffil}

\author{Jeffrey~A.~Grover}
\affiliation{\RLEaffil}

\author{William~D.~Oliver}
\email{william.oliver@mit.edu}
\affiliation{\RLEaffil}
\affiliation{\EECSaffil}
\affiliation{\Physaffil}
\affiliation{\LLaffil}

\date{\today}

\begin{abstract}

Entanglement and its propagation are central to understanding a multitude of physical properties of quantum systems. 
Notably, within closed quantum many-body systems, entanglement is believed to yield emergent thermodynamic behavior.
However, a universal understanding remains challenging due to the non-integrability and computational intractability of most large-scale quantum systems.
Quantum hardware platforms provide a means to study the formation and scaling of entanglement in interacting many-body systems.
Here, we use a controllable~$4 \times 4$ array of superconducting qubits to emulate a two-dimensional hard-core Bose-Hubbard lattice.
We generate superposition states by simultaneously driving all lattice sites and extract correlation lengths and entanglement entropy across its many-body energy spectrum.
We observe volume-law entanglement scaling for states at the center of the spectrum and a crossover to the onset of area-law scaling near its edges.

\end{abstract}

\maketitle

\section{Introduction}

\begin{figure*}[t]

\subfloat{\label{fig:subsystem_concept}}
\subfloat{\label{fig:hcbh_concept}}
\subfloat{\label{fig:hcbh_spectrum}}
\subfloat{\label{fig:volume_scaling_concept}}
\subfloat{\label{fig:sV_sA_concept}}
\subfloat{\label{fig:flip-chip_concept}}
\subfloat{\label{fig:qubit_tier}}
\subfloat{\label{fig:interposer_tier}}
\includegraphics{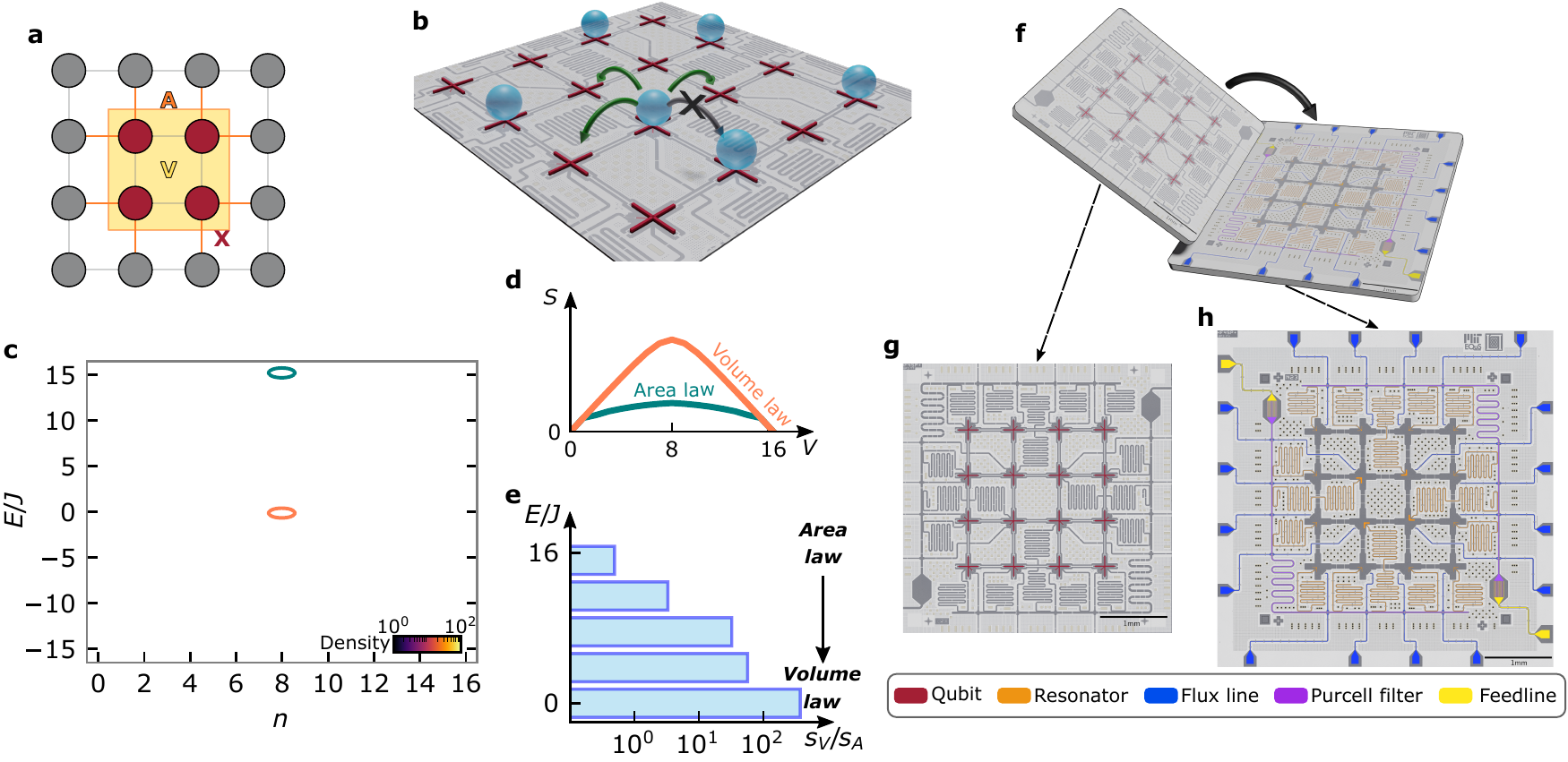}
\caption{
\textbf{Experimental concept} \textbf{(a)} Schematic for an example subsystem~$X$ of 4 qubits within a 16-qubit lattice. The subsystem has a volume of 4 (maroon sites) and an area of 8 (orange lines). \textbf{(b)} 2D hard-core Bose-Hubbard lattice emulated by the superconducting quantum circuit. Each site can be occupied by, at most, a single particle. \textbf{(c)} Energy~$E$ spectrum of the hard-core Bose-Hubbard lattice emulated by our device, shown in the rotating frame resonant with the lattice sites. The energy spectrum is partitioned into distinct subspaces defined by the total particle number~$n$. \textbf{(d)} Scaling of the entanglement entropy~$S$ with subsystem volume~$V$ for an eigenstate at the center of the energy spectrum (orange line corresponding to the energy eigenstate highlighted by an orange oval in \textbf{c}) and an eigenstate at the edge of the energy spectrum (teal line corresponding to the energy eigenstate highlighted by a teal oval in \textbf{c}). \textbf{(e)} Change in the entanglement behavior, quantified by the geometric entropy ratio~$s_V/s_A$, for states with~$n=8$. 
\textbf{(f)} Schematic for the flip-chip sample consisting of 16 superconducting qubits. Optical images of the qubit tier \textbf{(g)} and the interposer tier \textbf{(h)} are illustrated with the qubits and the different signal lines false-colored.}
\label{fig:concept}
\end{figure*}

Entanglement is a uniquely quantum property that underpins descriptions of interacting quantum systems as statistical ensembles~\cite{deutsch_1991, srednicki_1994, rigol_2008, dalessio_2016}.
Within closed many-body quantum systems, entanglement amongst constituent subsystems introduces uncertainty to their individual states, even when the full system is in a pure state~\cite{amico_2008,eisert_2010}.
For this reason, entropy measures are commonly used to quantify quantum entanglement in interacting many-body systems and have been directly probed in a number of platforms~\cite{islam_2015, kaufman_2016, neill_2016, linke_2018,lukin_2019, brydges_2019, nakagawa_2018}.
The study of entanglement in interacting many-body quantum systems is central to the understanding of a range of physical phenomena in condensed-matter systems~\cite{amico_2008}, quantum gravity~\cite{page_1993,nishioka_2009}, and quantum circuits~\cite{choi_2020}.
The scaling of the entanglement entropy with subsystem size provides insight into classifying phases of quantum matter~\cite{li_2008,bauer_2013,khemani_2017} and the feasibility of numerically simulating their time dynamics~\cite{eisert_2010}. 

In a closed system, the bipartite entanglement entropy of a subsystem quantifies the amount of entanglement between the subsystem and the remainder of the system.
For certain many-body states, such as the ground state of 1-dimensional local Hamiltonians~\cite{brandao_2013}, the entanglement entropy is proportional to the size of the boundary between a subsystem and the remaining system; this boundary is referred to as the area of a subsystem. 
Such states are said to have area-law entanglement scaling~\cite{eisert_2010}. 
For other states, entanglement entropy increases proportionally to the bulk size (volume) of a subsystem, a behavior referred to as volume-law scaling. 
To characterize the entanglement scaling in a lattice of quantum objects, such as natural or artificial atoms, we consider the area and volume entanglement entropy per lattice site, represented by~$s_A$ and~$s_V$, respectively. 
Disregarding logarithmic corrections, which are theoretically expected in certian contexts~\cite{barthel2006, miao2021}, for a given state, the entanglement entropy~$S(\rho_X)$ of a subsystem~$X$ can be expressed using the ansatz
\begin{equation}
    S(\rho_X) = s_A A_X + s_V V_X,
    \label{eq:sV_sA}
\end{equation}
where~$A_X$ is the subsystem area and~$V_X$ is the subsystem volume (see Fig.~\ref{fig:subsystem_concept} for an example). 
The ratio of volume to area entropy per site,~$s_V/s_A$, quantifies the extent to which that state obeys area- or volume-law entanglement scaling~\cite{yanay_2020}. 
Quantum states with area-law entanglement scaling have local correlations, whereas systems obeying volume-law entanglement scaling contain correlations that extend throughout the system and are therefore more challenging to study using classical numerical methods~\cite{schollwock_2005,perez-garcia_2007}.

In this work, we study the entanglement scaling of states residing in different energetic regions of the two-dimensional (2D) hard-core Bose-Hubbard (HCBH) lattice. 
The Bose-Hubbard model is particle-number conserving, allowing its energy spectrum to be partitioned into sectors with definite particle number~$n$. The ``hard-core" condition arises from strong on-site particle-particle interactions that mandate that each lattice site may be occupied by at most a single particle (Fig.~\ref{fig:hcbh_concept}). 
In 2D, the HCBH model is non-integrable and may exhibit eigenstate thermalization~\cite{srednicki_1994, santos2010, nandkishore_2015} and quantum information scrambling~\cite{braumuller_2022}.
Highly excited many-body states---states residing near the center of the energy spectrum (energy $E\approx 0$; orange oval in Fig.~\ref{fig:hcbh_spectrum})---are expected to exhibit volume-law scaling of the entanglement entropy, following the Page curve~\cite{page_1993} (orange line in Fig.~\ref{fig:volume_scaling_concept}). For subsystems smaller than half the size of the system, the entropy grows linearly in subsystem size. 
The entropy is maximized when the subsystem comprises half of the entire system and decreases as the subsystem size is further increased.
In contrast, the entanglement entropy of states residing near the edges of the HCBH energy spectrum does not follow the Page curve, but rather, grows less rapidly with subsystem volume (exemplified by the teal line in Fig.~\ref{fig:volume_scaling_concept} showing the entropy of the eigenstate highlighted by a teal oval in Fig.~\ref{fig:hcbh_spectrum}). 
For states with intermediate energies, a crossover is expected from area-law scaling at the edges of the energy spectrum to volume-law scaling at its center (Fig.~\ref{fig:sV_sA_concept})~\cite{miao2021, yanay_2020}.
Although Fig.~\ref{fig:sV_sA_concept} illustrates the crossover for the $n=8$ particle-number manifold, the crossover similarly occurs in other manifolds (see numerical simulations in Fig.~S12 of the \supp{}).

The crossover from area- to volume-law entanglement scaling can be observed by studying the exact eigenstates of the HCBH model.
However, preparing a specific eigenstate generally requires a deep quantum circuit~\cite{plesch_2011} or a slow adiabatic evolution~\cite{saxberg_2022}.
Alternatively, we can explore the behavior of the entanglement entropy by preparing superpositions of eigenstates across multiple particle-number manifolds~\cite{yanay_2020}.
Here, we prepare such superposition states by simultaneously and weakly driving 16 superconducting qubits arranged in a $4 \times 4$ lattice. 
By varying the detuning of the drive frequency from the lattice frequency, we generate superposition states occupying different regions of the HCBH energy spectrum. 
Measurements of correlation lengths and entanglement entropies indicate volume-law entanglement scaling for states prepared at the center of the spectrum and a crossover to the onset of area-law entanglement scaling for states prepared at the edges.

\section{Experimental system}

Superconducting quantum processors provide strong qubit-qubit interactions at rates exceeding individual qubit decoherence rates, making them a platform well-suited for emulating many-body quantum systems~\cite{roushan_2017, xu_2018, ma_2019, karamlou_2022, braumuller_2022, mi_2022, saxberg_2022, zhang_2023b}.
In this experiment, we use a 2D lattice of 16 capacitively-coupled superconducting transmon qubits~\cite{koch_2007} fabricated using a flip-chip process~\cite{rosenberg_2017}, as depicted in Fig.~\ref{fig:flip-chip_concept}. 
The qubits are located on a qubit tier (Fig.~\ref{fig:qubit_tier}), and the readout and control lines are located on a separate interposer tier (Fig.~\ref{fig:interposer_tier}, see \supp{} for device details). 
The circuit emulates the 2D Bose-Hubbard model, described by the Hamiltonian
\begin{equation}
\hat H _{\mathrm{BH}}/\hbar = \sum_i \epsilon_i\hat n _i + \sum_i \frac{U_i}{2} \hat n _i(\hat n_i-1) + \sum_{\langle i,j\rangle}J_{ij}\hat a _i^\dagger \hat a_j \,,
\label{eq:bosehubbard}
\end{equation}
where~$\hat a_i^\dagger$ ($\hat a_i$) is the creation (annihilation) operator for qubit excitations at site $i$, $\hat n_i=\hat a_i^\dagger \hat a_i$ is the respective excitation number operator, and qubit excitations correspond to particles in the Bose-Hubbard lattice.
The first term represents the site energies~$\epsilon_i=\omega_i - \omega_{r}$, which are given by the transmon transition frequencies~$\omega_i$ with respect to a frame rotating at frequency~$\omega_{r}$. 
The second term accounts for the on-site interaction with strength~$U_i$ arising from the anharmonicity of transmon qubit~$i$, representing the energy cost for two particles to occupy the same site with an average strength~$U/2\pi = \SI{-218 \pm 6}{MHz}$. 
The final term of the Hamiltonian describes the particle exchange interaction (within a rotating wave approximation) between neighboring lattice sites with strength~$J_{ij}$, realized by capacitively coupling adjacent qubits with an average strength of~$J/2\pi = \SI{5.9\pm0.4}{MHz}$ at qubit frequency~$\omega/2\pi=\SI{4.5}{GHz}$. 
While the coupling strengths between qubits are fixed, we switch particle exchange off for state preparation and readout by detuning the qubits to different frequencies (inset in Fig.~\ref{fig:coherent-like_state_population}). 
Our system features site-resolved, multiplexed single-shot dispersive qubit readout~\cite{blais_2004} with an average qubit state assignment fidelity of~$93\%$, which, together with site-selective control pulses, allows us to perform simultaneous tomographic measurements of the qubit states.

In our system, on-site interactions are much stronger than exchange interactions,~$J \ll |U|$.
By restricting each site to two levels, and mapping the bosonic operators to qubit operators, we transform the Hamiltonian in Eq.~\ref{eq:bosehubbard} to the hard-core limit:
\begin{equation}
\hat H_{\mathrm{HCBH}}/\hbar=\sum_{\langle i,j\rangle}J_{ij}\hat\sigma _i^+ \hat\sigma _j^- -\sum_i\frac{\epsilon_i}2\hat\sigma _i^z\,,
\label{eq:hardcoreBH}
\end{equation}
where~$\hat\sigma _i^+$ ($\hat\sigma _i^-$) is the raising (lowering) operator for a qubit at site~$i$, and~$\hat\sigma _i^z$ is the Pauli-$Z$ operator. 
The energy relaxation rate~$\Gamma_1$ and dephasing rate~$\Gamma_\phi$ in our lattice are small compared to the particle exchange rate, with~$\Gamma_1 \approx 10^{-3} J$ and~$\Gamma_\phi \approx 10^{-2} J$, which enables us to prepare quantum many-body states and probe their time dynamics faithfully.

\section{Coherent-like states}

\begin{figure*}[t]
\subfloat{\label{fig:coherent-like_state_population}}
\subfloat{\label{fig:excitation_disribution}}
\subfloat{\label{fig:sim_delta_0J}}
\subfloat{\label{fig:sim_delta_1J}}
\subfloat{\label{fig:sim_delta_2J}}
\subfloat{\label{fig:average correlations}}
\subfloat{\label{fig:correlation_length}}
\includegraphics{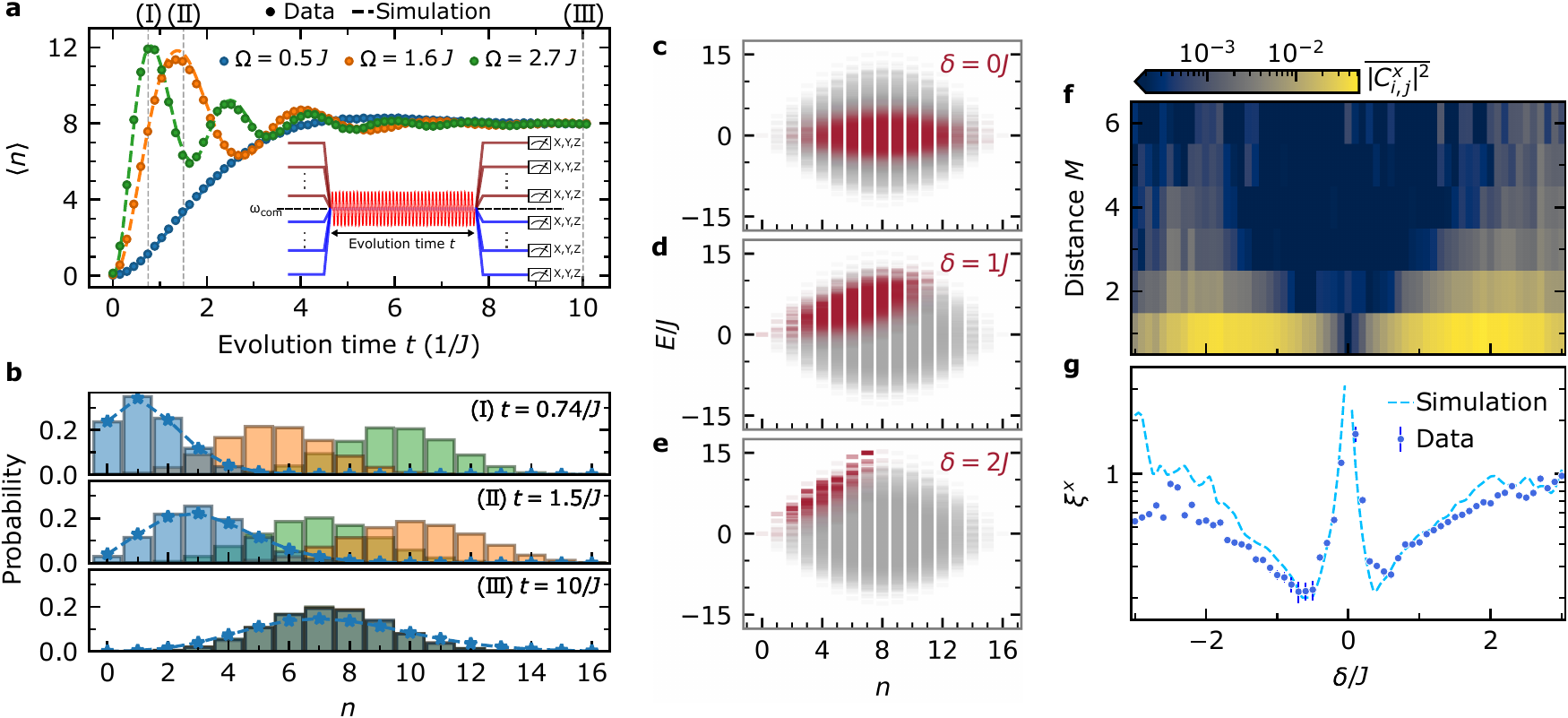}
\caption{
\textbf{Coherent-like state preparation.} \textbf{(a)} Total number of particles~$\langle n \rangle$ in the uniform lattice while driving the system on resonance for time~$t$. Simulations do not include decoherence. The lattice reaches half-filling at equilibrium. The experiments are executed using the pulse sequence shown in the inset. \textbf{(b)} Probability of measuring different numbers of excitations in the lattice at three different times. The blue stars are from a Poisson fit to the excitation number distribution for~$\Omega=J/2$, with the dashed lines as guides to the eye.
Simulated overlap of the prepared coherent-like state in steady state ($t=10/J$) with drive strength~$\Omega=J/2$ and drive detuning~$\delta=0$ \textbf{(c)},~$\delta=1J$ \textbf{(d)}, and~$\delta=2J$ \textbf{(e)} with the hard-core Bose-Hubbard energy eigenstates. The different shades of red indicate the magnitude of the overlap between the prepared superposition states and energy eigenstates. Note the spectra are shown in the rotating frame of the lattice sites, not of the drive.
\textbf{(f)} Average 2-point correlator squared along the~$x$-basis,~$\overline{|C^{x}_{i,j}|^2}$, between qubit pairs at distance~$M$ for drive duration~$t=10/J$, strength~$\Omega=J/2$ and detuning~$\delta$ from the lattice frequency. \textbf{(g)} Correlation length~$\xi^{x}$ extracted using the 2-point correlators at different values of~$\delta$.}
\label{fig:coherent-like_state_preperation}
\end{figure*}

We generate a superposition of many eigenstates of the lattice, which we refer to as a coherent-like state, by simultaneously driving the qubits via a common control line. 
Applied to the lattice initialized with no excitations, the drive acts as a displacement operation of the Fock basis defined by the number of excitations in the lattice~\cite{yanay_2020}, hence, it will result in a state analogous to coherent states of light. 
The common drive is applied to the system via the feedlines coupled to each qubit through their respective readout resonators.
Selecting the rotating frame frequency $\omega_r$ to be the frequency of the drive, the Hamiltonian of the driven lattice is
\begin{equation}
    \hat{H}/\hbar = \sum_{\langle i,j\rangle}J_{ij}\hat\sigma _i^+ \hat\sigma _j^- + \frac{\delta}{2} \sum_i\hat\sigma _i^z + \Omega  \sum_j (\alpha_j \hat\sigma_j^- + \mathrm{h.c.})\,,
    \label{eq:driven_hamiltonian}
\end{equation}
where~$\delta = \omega_{r}-\omega_{\mathrm{com}}$ is the detuning between the drive and the qubit frequencies (all qubits are biased on-resonance at $\omega_{\mathrm{com}}$).
The drive strength~$\Omega$ can be tuned by varying the amplitude of the applied drive pulse. 
The common drive couples independently to each qubit with a complex coefficient~$\alpha_j$ that depends on the geometric circuit parameters of the lattice.

We first study the time dynamics of the average number of excitations~$\langle n \rangle$ in the lattice under a resonant drive ($\delta = 0$) in Fig.~\ref{fig:coherent-like_state_population}. 
The driven lattice reaches a steady state of half-filling, with an average particle number~$\langle n \rangle=8$, after driving the lattice for time~$t\approx10/J$. 
Once in steady-state, the drive adds and removes excitations coherently from the system at the same rate. 
Hamiltonian parameters $J_{ij}$ and $\alpha_j$ were characterized through a procedure described in Section~S4 of the \supp{}, and the excellent agreement of numerical simulations of time evolution under Eq.~\ref{eq:driven_hamiltonian} with experimental data confirms their accuracy.
In Fig.~\ref{fig:excitation_disribution}, we report the discrete probability distribution of measuring a different number of excitations in the lattice at three different times. 
The probability of a particular excitation number approximately follows a Poisson distribution for a weak drive $\Omega = J/2$ (blue stars with dashed lines as guides to the eye), signifying that the quantum state is in a coherent-like superposition of excitation-number states.

The coherent-like state comprises a swath of the HCBH energy spectrum that depends on the drive detuning.
Driving the lattice on resonance prepares a coherent-like state at the center of the HCBH energy spectrum (Fig.~\ref{fig:sim_delta_0J}). 
By varying the drive detuning, we can generate states that are a superposition of the eigenstates closer to the edge of the energy band (Fig.~\ref{fig:sim_delta_1J},\ref{fig:sim_delta_2J}). 
The standard deviation in the energy of the coherent-like state depends on the strength of the drive: a stronger drive increases the bandwidth of populated energy eigenstates within each particle number subspace. 
Therefore, to probe the entanglement properties across the lattice energy spectrum, we choose a relatively weak drive with strength~$\Omega=J/2$ for the remainder of the main text. 
We choose a drive duration~$t=10/J$, which is short compared to the timescale of decoherence yet long enough to allow the coherent-like state to reach its steady-state distribution (which occurs at roughly~$t=6/J$ according to simulations shown in Section~S15 of the \supp{}).

\section{Correlation lengths}

A system of interacting particles exhibits correlations between its constituent subsystems.
The hard-core Bose-Hubbard Hamiltonian is equivalent to an XY~Hamiltonian, where quantum order is reflected by the transverse correlations~\cite{chen_2023}.
In order to quantify transverse correlations for coherent-like states generated with detuning~$\delta$, we make single-shot measurements of identically prepared systems and then calculate the 2-point correlators~$C^{x}_{i,j} \equiv \langle \hat{\sigma}^x_i \hat{\sigma}^x_j \rangle - \langle \hat{\sigma}^x_i \rangle \langle \hat{\sigma}^x_j \rangle$ between different qubit pairs. 
In Fig.~\ref{fig:average correlations}, we show the magnitude-squared 2-point correlator values~$\overline{|C^{x}_{i,j}|^2}$, averaged over qubits at the same Manhattan (1-norm) distance~$M$.
When~$\delta/J$ is small, generating a superposition state occupying the center of the energy band,~$\overline{|C^{x}_{i,j}|^2}$ becomes small for all~$M$, matching the expectation for states with volume-law entanglement scaling.
The 2-point correlator has an upper-bound set by the mutual information between the sites $I(i:j)=S(\rho_i)+S(\rho_j)-S(\rho_{ij})$.
In a volume-law state, the entropy of small subsystems is equal to their volume~\cite{page_1993}, hence we expect the mutual information and, in turn, the correlator between any qubit pair will vanish.

Using the 2-point correlators, we extract the correlation length~$\xi^{x}$ by fitting~$|C^{x}_{i,j}|^2 \propto \exp(-M/\xi^{x})$ (Fig.~\ref{fig:correlation_length}). 
The correlation length quantifies the dependence of correlations on the distance between the subsystems within our system. 
As we increase the magnitude of the drive detuning~$\delta$, skewing the superposition state to the edge of the energy spectrum, we generally find that the correlation length increases.
The states prepared with $-J/2 \lesssim  \delta \lesssim J/2$, however, follow the opposite trend, and the extracted correlation length diverges around $\delta=0$.
In this regime where the coherent-like state occupies the center of the energy spectrum, the 2-point correlators asymptotically reach zero, so the extracted 2-point correlation length loses meaning.
We attribute the slight asymmetry of the observed correlation lengths about~$\delta=0$ to a slight offset of the energy spectrum of our lattice towards positive energies, which, as shown in Section~S5 of the \supp{}, is a consequence of next-nearest neighbor exchange interactions.

\begin{figure*}[t]

\subfloat{\label{fig:subsystem_examples}}
\subfloat{\label{fig:entanglement_scaling_2d}}
\subfloat{\label{fig:entanglement_scaling_line_cut}}
\subfloat{\label{fig:sV_sA}}
\subfloat{\label{fig:sV_sA_ratio}}
\includegraphics{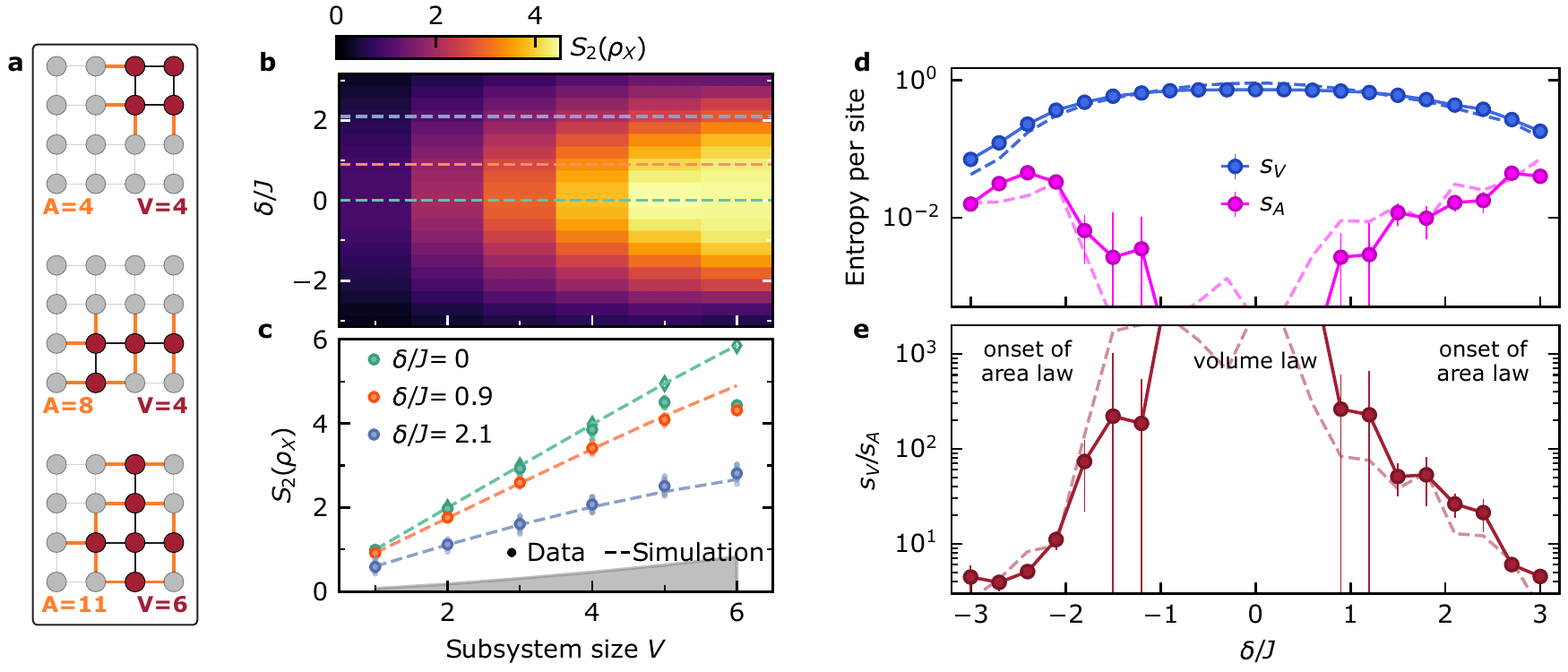}
\caption{
\textbf{Entanglement scaling behavior} 
\textbf{(a)} Examples of subsystems measured in our lattice. 
\textbf{(b)} The average subsystem entanglement entropy~$S_2(\rho_X)$ for subsystem with volume~$V$ for a state prepared with drive time~$t=10/J$, strength~$\Omega=J/2$ and detuning~$\delta$ from the lattice frequency. 
\textbf{(c)} Subsystem entanglement entropy~$S_2(\rho_X)$ along dashed lines in panel \textbf{(b)} for detunings $\delta/J=$0 (green), 0.9 (orange), and 2.1 (blue). Solid circles are experimental data averaged across all subsystems of each volume, with data for individual subsystems indicated by smaller, transparent circles. For $\delta/J=$0 (green), open diamonds are Monte Carlo simulations of 20,000 samples indicating that deviations at subsystem sizes~5 and~6 arise partially from insufficient sampling. The gray region indicates the estimated classical entropy from dephasing (see Section~S14 of the \supp{} for details). 
\textbf{(d)} The volume entanglement entropy,~$s_V$, and area entanglement entropy,~$s_A$, per site extracted using 163 different subsystems of various volumes and areas for states prepared with drive detuning~$\delta$. The error bars indicate~$\pm 1$ standard error of the fit parameter. \textbf{(e)} The geometric entropy ratio~$s_V/s_A$ is used for quantifying the behavior of entanglement. States prepared with~$\delta=0$ exhibit a strong volume-law scaling, and the states prepared with larger drive detuning values show a weaker volume-law scaling with an increasing area-law scaling. Dashed lines in \textbf{(c-e)} indicate results from numerical simulation.}
\label{fig:entanglement}
\end{figure*}

\section{Entanglement scaling behavior}

In order to study the entanglement scaling of the coherent-like states, we reconstruct the density matrices of~$163$ unique subsystems, listed in Section~S9 of the \supp{}, via a complete set of tomography measurements. 
The measured subsystems contain up to 6~qubits (see Fig.~\ref{fig:subsystem_examples} for examples). 
We quantify the entanglement entropy between subsystem~$X$ and the remaining system through the second R\'enyi entropy~\cite{renyi_1961}
\begin{equation}
    S_2(\rho_X) = -\log_2  \tr \left(\rho_X^2 \right),
    \label{eq:renyi_entropy}
\end{equation}
where~$\rho_X$ is the reduced density matrix of a subsystem~$X$ of the entire quantum system~$\rho$. 

We first study the scaling of the entanglement entropy with the subsystem volume~$V_X$, defined as the number of qubits within the subsystem, for coherent-like states prepared with different drive detunings~$\delta$ (Fig.~\ref{fig:entanglement_scaling_2d}). 
For coherent-like states prepared with~$\delta=0$, we observe nearly maximal scaling of the entropy with subsystem size~$S_2(\rho_X)\approx V_X$, whereas with increasing~$\delta$ the entropy grows more slowly with subsystem size (Fig.~\ref{fig:entanglement_scaling_line_cut}).

\begin{figure*}[t!]
\subfloat{\label{fig:Schmidt_coefficients}}
\subfloat{\label{fig:Schmidt_coefficient_ratios}}
\subfloat{\label{fig:number_of_coefficients}}
\includegraphics{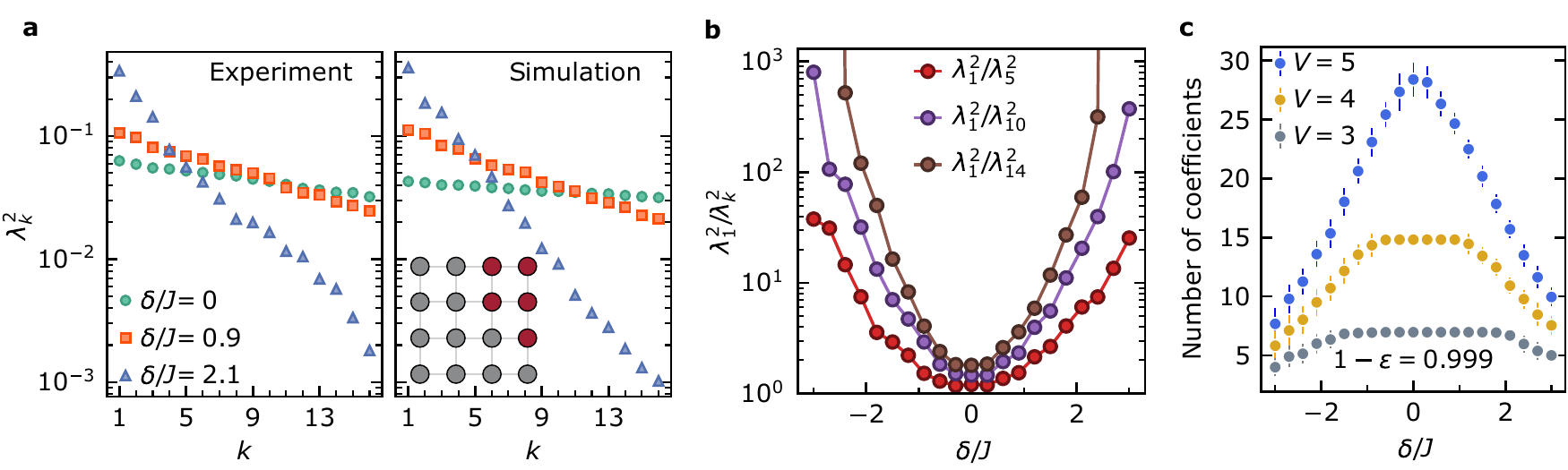}
\caption{
\textbf{Schmidt coefficients scaling across the spectrum.} \textbf{(a)} The Schmidt coefficients for the decomposition of a five-qubit subsystem (highlighted with maroon color in the inset) and the remaining lattice.
Coherent-like states are prepared with different drive detunings $\delta/J = 0, 0.9, 2.1$.
The coefficients in the left panel are calculated from experimental data and agree well with the simulated coefficients in the right panel.
\textbf{(b)} The ratio of the largest and the $k^{\mathrm{th}}$-largest Schmidt coefficient $\lambda^2_1/\lambda^2_{k}$, for $k=5,10,14$, of coherent-like states prepared with different drive detunings $\delta$ in experiment. \textbf{(c)} The number of Schmidt coefficients required to represent the bipartition of the lattice with subsystem volume $V=3,4,5$ with accuracy $1-\epsilon=0.999$. Each data point is averaged over all the measured subsystems of the same size, with the error bars indicating $\pm 1$ standard deviation.}
\label{fig:correlations}
\end{figure*}

There is excellent agreement between the expected and the measured entropy for subsystems of volume~$\leq 4$, yet there is a discrepancy for the largest subsystems where the coherent-like states are prepared near the center of the spectrum (see Fig.~\ref{fig:entanglement_scaling_line_cut}).
The discrepancy arises, in large part, from having a finite number of measurement samples when reconstructing subsystem density matrices.
Tomographic reconstruction of the state of a subsystem requires determining the distribution of measurement outcomes for each Pauli string.
When the system has volume-law entanglement, the measurement outcome distribution becomes more uniform.
Density matrix reconstruction therefore becomes more sensitive to sampling error.
This becomes especially relevant as the subsystem size increases.
Therefore, for the largest subsystems, the entropy of the reconstructed density matrix will be smaller than the actual entropy of the subsystem, as seen for $V=5,6$ in Fig.~\ref{fig:entanglement_scaling_line_cut}. 
We simultaneously measured all subsystems by taking 2,000 measurement samples for $3^6$ Pauli strings, as visualized in Fig.~S15 of the \supp{}.
For subsystems of volume $V$, we can therefore extract 2,000$\times 3^{6-V}$ measurement samples---sufficient for $V=1$-$4$, but less so for $V=5,6$. 
In hindsight, after the experiment had concluded, we realized that we could have obtained better agreement for $V=5,6$ if we had used 20,000 samples (open diamonds in Fig.~\ref{fig:entanglement_scaling_line_cut}), which would have been straightforward to implement experimentally. 
In Section~S11 of the \supp, we discuss this sampling dependence in more detail and show, using Monte Carlo simulations, that a more accurate reconstruction of highly entangled states follows from larger numbers of measurement samples of each Pauli string. 

We next determine the scaling of entanglement entropy with subsystem volume $s_V$ and area $s_A$ per~Eq.~\ref{eq:sV_sA}. 
Using the R\'enyi entropies of the density matrices reconstructed from experimental data, we extract~$s_V$ and~$s_A$ by measuring the rate of change of~$S_2(\rho_X)$ with~$V_X$ and the subsystem area~$A_X$, respectively, with the other held fixed. Here, $A_X$ is defined as the number of nearest-neighbor bonds intersecting the boundary of the subsystem $X$. The linear fitting procedure used to determine $s_{V,A}$ is detailed in Section~S10 of the \supp{}. 
In Fig.~\ref{fig:sV_sA} we observe that as the magnitude of~$\delta$ becomes larger,~$s_V$ decreases and~$s_A$ increases. 
While extraction of~$s_V$ is reliable at all drive detunings, at small drive detuning values ($-J<\delta<J$) the entanglement entropy does not exhibit a notable dependence on the area of the subsystem in our finite lattice, hence we are not able to reliably fit~$s_A$. 
By considering the geometric entropy ratio~$s_V/s_A$ we observe a change in the behavior of entanglement entropy within our system (Fig.~\ref{fig:sV_sA_ratio}). 
The states prepared at the center of the energy spectrum with~$\delta=0$ exhibit a strong volume-law scaling, and the states prepared closer to the edge of the energy spectrum show weaker volume-law scaling and increasing area-law scaling.

The entanglement spectrum, obtained via Schmidt decomposition, further quantifies the structure of entanglement across a bipartition of a closed quantum system~\cite{li_2008}.
The quantum state $\ket{\psi}$ can be represented as a sum of product states of the orthonormal Schmidt bases $\ket{k_X}$ for a given subsystem $X$ and $\ket{k_{\bar{X}}}$ for the remaining lattice~\cite{nielsen_2000}: 
\begin{equation}
    \ket{\psi} = \sum_k \lambda_k \ket{k_X} \ket{k_{\bar{X}}}\,,
\end{equation}
where positive scalars $\lambda_k$ are the Schmidt coefficients with $\sum_k \lambda_k^2=1$.
The Schmidt coefficients form the entanglement spectrum and provide a proxy for the degree of entanglement between the two subsystems. 
For a subsystem $X$ maximally entangled with the remaining lattice, all the Schmidt coefficients will have an equal value $\lambda_k = 1/\sqrt{2^{V_X}}$, where $V_X$ is the volume of $X$ (assuming that $V_X < V_{\bar{X}}$).

We obtain the entanglement spectrum for a bipartition of our lattice by diagonalizing the measured partial density matrix of a subsystem.
In Fig.~\ref{fig:Schmidt_coefficients}, we study the entanglement formed within states prepared across the energy spectrum. 
We report the first 16~Schmidt coefficients squared $\lambda^2_k$, in decreasing order, for a subsystem highlighted in maroon and the remaining lattice. 
We observe that for states obeying volume-law entanglement scaling at the center of the spectrum, the variation in coefficient magnitudes is small compared to states closer to the edge of the spectrum, in close agreement with numerical simulation. 
To quantify this difference, in Fig.~\ref{fig:Schmidt_coefficient_ratios} we show the ratio of the largest and the $k^{\mathrm{th}}$ largest Schmidt coefficient, $\lambda^2_1/\lambda^2_{k}$, for $k=5,10,14$, of coherent-like states prepared with different drive detunings $\delta$. 
We observe that a small number of Schmidt states contain nearly all the weight of the decomposition for the area-law-like states, whereas, for the volume-law states, the Schmidt coefficients are roughly equal.
This variation signals a change in the extent of the entanglement distribution across the system.

In Fig.~\ref{fig:number_of_coefficients}, we report the number of coefficients required to approximate the state of the lattice bipartition with accuracy $1-\epsilon=0.999$ (see Section~S13 in \supp) for subsystems with volume $V=3,4,5$. 
We find that the states at the edge of the energy spectrum can be accurately represented with fewer coefficients, while the number of coefficients needed to approximate states at the center of the band approaches the dimension of the Hilbert space of the subsystem, $2^V$.

\section{Conclusion}

In this work, we study the entanglement scaling properties of the 2D hard-core Bose-Hubbard model, emulated using a 16-qubit superconducting quantum processor. 
By simultaneously driving all qubits, we generate coherent superposition states that preferentially incorporate eigenstates from regions of the many-body energy spectrum that we tune between the center and edges.
We probe the transverse quantum correlation lengths and the entanglement scaling behavior of the superposition states. 
We observe a crossover from volume-law scaling of entanglement entropy near the center of the band, marked by vanishing two-point correlators, to the onset of area-law entropy scaling at the edges of the energy band, accompanied by finite-range correlations.

A coherent-like state comprises eigenstates across a swath of the HCBH energy spectrum.
Decreasing the drive amplitude will decrease the width of the coherent-like state.
Because excitations (particles) are added more slowly to the lattice, the coherent-like state will need to evolve for more time before reaching steady-state.
Therefore, coherent-like states can be prepared with narrower spectral width as the coherence of superconducting processors improves, providing finer energy resolution of the area-to-volume-law crossover.

A central outstanding question is when and how statistical ensembles arise from unitary time evolution~\cite{deutsch_1991, srednicki_1994, rigol_2008, abanin_2019a, lami_2023}. 
While an analytical relation between entanglement and thermodynamic entropy exists for integrable systems~\cite{alba_2017}, an analogous insight for non-integrable systems could lead to a deeper understanding of the emergence of thermodynamic behavior in quantum systems.
In recent years, random circuit protocols have emerged as a potential tool for addressing this question in a digital context~\cite{fisher_2023}. 
The protocol introduced in the present work may be viewed as an analog (i.e. continuous time evolution) counterpart to random circuit experiments, which can be utilized to generate highly entangled states rapidly.
While tomography itself has exponential time complexity, our techniques are extensible to larger system sizes.
In Section~S12 of the \supp{}, we present numerical simulations indicating that the area-to-volume-law transition should remain clear in larger systems without increasing the volume of subsystems used for tomography.
Because the maximum subsystem volume is fixed and the tomographic readout of all subsystems is simultaneous, our protocol has constant runtime as a function of system size.
Therefore, as quantum technology moves to larger systems, but before full error correction is available, our protocol will elucidate emergent thermalization in a regime where results are classically intractable.

Finally, understanding the structure of entanglement within a quantum system determines the effective degrees of freedom that need to be considered to simulate the quantum states.
Area-law states can generally be numerically simulated efficiently using tensor network methods~\cite{eisert_2010,schollwock_2005,perez-garcia_2007}, whereas the computational complexity of classically simulating volume-law states scales exponentially with system size.
It is the latter complexity that underpins the promise of quantum computational advantage.
In this work, we have introduced an extensible, hardware-efficient analysis of the scaling structure of entanglement that could serve as the basis for an algorithm-agnostic means to verify the computational complexity of programs executed on large-scale quantum devices.

\textit{Note:} During the preparation of this manuscript, we became aware of related studies in a 1D trapped-ion simulator~\cite{joshi_2023}.

\section*{Acknowledgements}

The authors are grateful to Soonwon Choi, Aram Harrow, Steve Disseler, Kushal Seetharam, Christopher McNally, Lauren Li, Lamia Ateshian, David Rower, and Katherine Van Kirk for fruitful discussions, and to Terry Orlando and Adam Kaufman for a careful reading of the manuscript.
This material is based upon work supported in part by the U.S. Department of Energy, Office of Science, National Quantum Information Science Research Centers, Quantum System Accelerator (QSA); in part by the Defense Advanced Research Projects Agency under the Quantum Benchmarking contract; in part by U.S. Army Research Office Grant W911NF-18-1-0411; and by the Department of Energy and Under Secretary of Defense for Research and Engineering under Air Force Contract No. FA8702-15-D-0001.
AHK acknowledges support from the NSF Graduate Research Fellowship Program.
CNB acknowledges support from the STC Center for Integrated Quantum Materials, NSF Grant No. DMR-1231319 and the Wellesley College Samuel and Hilda Levitt Fellowship.
SEM is supported by a NASA Space Technology Research Fellowship.
ITR, PMH, and MH are supported by an appointment to the Intelligence Community Postdoctoral Research Fellowship Program at the Massachusetts Institute of Technology administered by Oak Ridge Institute for Science and Education (ORISE) through an interagency agreement between the U.S. Department of Energy and the Office of the Director of National Intelligence (ODNI).
Any opinions, findings, conclusions, or recommendations expressed in this material are those of the author(s) and do not necessarily reflect the views of the Department of Energy, the Department of Defense, or the Under Secretary of Defense for Research and Engineering.

\section*{Author contribution}

AHK performed the experiments and analyzed the data with ITR.
AHK and ITR performed the numerical simulations and analysis used for characterizing the system and validating the experimental results with assistance from ADP and YY.
AHK and CNB developed the experiment control tools used in this work.
AHK, CNB, ADP, LD, PMH, MH, and JAG contributed to the experimental setup.
SEM, PMH, and MS designed the~$4 \times 4$ qubit array.
RD, DKK, and BMN fabricated the~$4 \times 4$ flip-chip sample with coordination from KS, MES, and JLY.
SG, JAG, and WDO provided experimental oversight and support.  
All authors contributed to the discussions of the results and to the development of the manuscript.

\section*{Competing interests}

The authors declare no competing interests.

\section*{Data availability}

The data that support the findings of this study are available from the corresponding author upon reasonable request and with the cognizance of our US Government sponsors who funded the work.

\section*{Code availability}

The code used for numerical simulations and data analyses is available from the corresponding author upon reasonable request and with the cognizance of our US Government sponsors who funded the work.

\bibliography{refs}

\end{document}


\title{Supplementary material for ``Probing entanglement across the energy spectrum of a hard-core Bose-Hubbard lattice''}

\def\RLEaffil{Research Laboratory of Electronics, Massachusetts Institute of Technology, Cambridge, MA 02139, USA}
\def\LLaffil{MIT Lincoln Laboratory, Lexington, MA 02421, USA}
\def\Physaffil{Department of Physics, Massachusetts Institute of Technology, Cambridge, MA 02139, USA}
\def\EECSaffil{Department of Electrical Engineering and Computer Science, Massachusetts Institute of Technology, Cambridge, MA 02139, USA}
\def\Wellesleyaffil{Department of Physics, Wellesley College, Wellesley, MA 02481, USA}
\def\Maryaffil{Laboratory for Physical Sciences, College Park, MD 20740, USA}

\author{Amir~H.~Karamlou}
\email{karamlou@mit.edu}
\affiliation{\RLEaffil}
\affiliation{\EECSaffil}

\author{Ilan~T.~Rosen}
\affiliation{\RLEaffil}

\author{Sarah~E.~Muschinske}
\affiliation{\RLEaffil}
\affiliation{\EECSaffil}

\author{Cora~N.~Barrett}
\affiliation{\RLEaffil} 
\affiliation{\Wellesleyaffil}

\author{Agustin~Di~Paolo}
\affiliation{\RLEaffil}

\author{Leon~Ding}
\affiliation{\RLEaffil}
\affiliation{\Physaffil}

\author{Patrick~M.~Harrington}
\affiliation{\RLEaffil}

\author{Max~Hays}
\affiliation{\RLEaffil}

\author{Rabindra~Das}
\affiliation{\LLaffil}

\author{David~K.~Kim}
\affiliation{\LLaffil}

\author{Bethany~M.~Niedzielski}
\affiliation{\LLaffil}

\author{Meghan~Schuldt}
\affiliation{\LLaffil}

\author{Kyle~Serniak}
\affiliation{\RLEaffil}
\affiliation{\LLaffil}

\author{Mollie~E.~Schwartz}
\affiliation{\LLaffil}

\author{Jonilyn~L.~Yoder}
\affiliation{\LLaffil}

\author{Simon~Gustavsson}
\affiliation{\RLEaffil}

\author{Yariv~Yanay}
\affiliation{\Maryaffil}

\author{Jeffrey~A.~Grover}
\affiliation{\RLEaffil}

\author{William~D.~Oliver}
\email{william.oliver@mit.edu}
\affiliation{\RLEaffil}
\affiliation{\EECSaffil}
\affiliation{\Physaffil}
\affiliation{\LLaffil}

\date{\today}

\maketitle


\tableofcontents

\section{Experimental setup}\label{sec:exp}

We house our superconducting qubit array within a BlueFors XLD-600 dilution refrigerator. 
In our setup, the mixing chamber (MXC) base temperature reaches as low as 14mK, however, we use a heater with a PID controller to stabilize the base temperature at 20mK. 
This mitigates temperature fluctuations during experiments. 

We use a 24-port microwave package~\cite{huang_2021} to host the sample, consisting of a copper casing, a multilayer interposer, a shielding cavity in the package center, and a set of microwave SMP connectors. 
The signal and ground connections between the package and the sample are provided using superconducting aluminum wirebonds. 

Semi-rigid coaxial cables with $0.086"$ outer conductor diameter and SMA connectorization transmit qubit control and readout signals between the mixing chamber stage and room temperature electronics. 
Cables connecting stages at different temperatures have low thermal conductivity to minimize the heat flow from warmer stages. 
We use KEYCOM ULT-05 cables between room temperature and the 4~K stage. 
All input lines (readout input, qubit control, and TWPA pump lines) use stainless-steel (SS-SS) cables between the 4~K and the mixing chamber stages. 
Readout output lines use low-loss superconducting niobium-titanium (NbTi) cables between the 4~K and mixing chamber stages to maximize the signal-to-noise ratio of the readout signal. 
We use hand-formable, non-magnetic, copper coaxial cables (EZ Form Cable, EZ-Flex.86-CU) between the mixing chamber stage and the sample package.

We attenuate and filter the signal at different stages of the dilution refrigerator to reduce the noise incident on the qubits. 
For the qubit control lines, we use a 20~dB attenuator (XMA Corporation) at the 4~K, a 1~dB attenuator at the still, and a $\SI{1}{GHz}$ low-pass filter (Mini-Circuits VLFG-1000+) at the mixing chamber stage. 
The 1~dB attenuator at the still was selected to thermalize the center conductor of the coaxial line while maintaining the ability to apply DC current through the line without causing excessive heating. 
The VLFG-1000+ low-pass filter reduces the noise at higher frequencies while allowing an RF drive to reach the qubit.
The filter response in the range between $4-\SI{5}{GHz}$ varies by only 3~dB, approximately providing $40$~dB of attenuation. 
Our readout input lines contain a 20~dB attenuator at 4~K, a 10~dB attenuator at the still, and 40~dB of attenuation at the mixing chamber stage, followed by a $\SI{12}{GHz}$ low-pass filter(RLC F-30-12.4-2).
This attenuation and filtering scheme mitigates the thermal-photon population in the resonators~\cite{yan_2016}, which limits the qubit coherence times due to photon shot noise. 

The readout circuit in the samples used in our experiment is set up for measuring in reflection mode.  
The resonator probe tone is coupled to the sample via a circulator (LNF-CIC4\_12A).
The reflected readout signal gets routed to the readout output chain using the third port of the circulator. 
The signal is amplified at the mixing chamber stage using a near–quantum-limited Josephson traveling-wave parametric amplifier (TWPA)~\cite{macklin_2015} fabricated at MIT Lincoln Laboratory and mounted inside the $\mu$-metal shield. 
The TWPA pump tone is combined with the readout signal using a directional coupler (Marki C20-0116). 
The signal then travels through two isolators (Quinstar 0XE89) followed by a $\SI{12}{GHz}$ low-pass filter (RLC F-30-12.4-2) and a $\SI{3}{GHz}$ high-pass filter (RLC F-19704) to prevent noise from coupling back to the sample through the readout chain. 
Afterward, the signal is once again amplified using a low-noise high-electron-mobility transistor (HEMT) amplifier at the 4~K stage before being transmitted outside the cryostat to room temperature measurement electronics.

\begin{figure*}[ht!]
\includegraphics{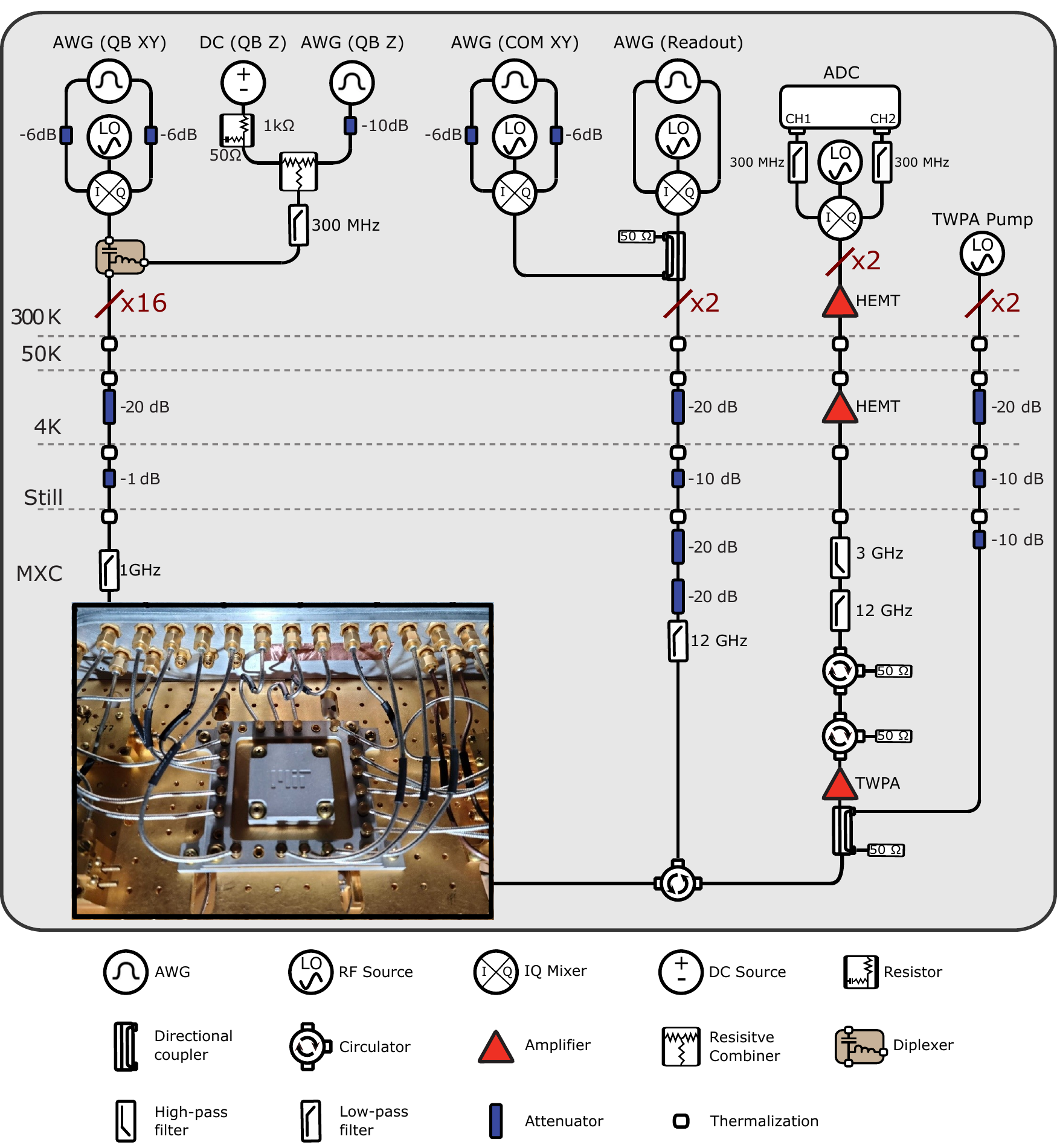}
\caption{\textbf{Measurement setup}. }
\label{fig:measurement_setup}
\end{figure*}

We use a microwave source (Rohde \& Schwarz SGS100A) with an internal IQ mixer to drive the readout resonators. 
We use a single-sideband mixing scheme to address multiple resonators coupled to the same feedline using a single microwave local oscillator (LO).
The intermediate frequency (IF) signals are generated using two arbitrary waveform generator (AWG) channels (Keysight M3202) and are interfered with the LO using the IQ mixer to obtain an RF signal with multiple frequency components, each of which can address a different resonator. 
In order to avoid saturating the IQ mixer, we attenuate the output signal of the AWG channels using 10~dB attenuators. 

At room temperature, the resonator response signal is amplified by a HEMT amplifier (MITEQ AMF-5D-00101200-23-10P) and sent through a band-pass filter (K\&L 033F8) to filter out the TWPA pump tone. 
The signal is then down-converted using a heterodyne demodulation scheme~\cite{krantz_2019} using an IQ mixer (Marki IQ-4509LXP) and the readout probe LO. 
The demodulated signals resulting from the interference between the readout output signal and the LO are then filtered using low-pass filters (Mini-circuits VLFX-300+), passing only the terms at different frequencies corresponding to the signal from each resonator.
The two output signals are then digitized on two separate digitizer channels (Keysight M3102), and the signal for each frequency component is processed using a built-in field-programmable gate array (FPGA) to obtain the static in-phase, $I$, and quadrature, $Q$, signals.

We use the secondary LO output of the microwave source for the demodulation LO. 
In order to reduce the signal noise, we use two high-pass filters (Mini-Circuits VHF-6010+) in series. 
We use a combination of attenuators and an amplifier (Mini-Circuits ZVE-8G+) to ensure that the LO power input to the IQ mixer is in the 10--13~dBm range required for optimal mixer performance. 

A 24-channel DC voltage source (QDevil QDAC-I) with in-line resistors generates the current required to apply a static flux to each qubit in our setup. 
In addition to this static biasing of the qubit, we can apply a baseband, fast-flux pulse to each qubit individually with an AWG channel (Keysight M3202), which allows us to tune the qubit frequency of the qubits on the timescale of nanoseconds. 
We attenuate the flux pulse produced by the AWG using a 10~dB attenuator in order to reduce the flux noise experienced by the qubit. 

For individual $X,Y$ control of each qubit, we use a microwave source (Rohde \& Schwarz SGS100A) with an internal IQ mixer. 
We calibrate the $I$ and $Q$ quadrature offsets, the gain imbalance, and the skew of our IQ mixer by minimizing both the LO leakage and the signal level of the unwanted sideband.
We use single-sideband mixing to generate our drive pulses and use an additional AWG channel to gate the RF output in order to mitigate the adverse effects of mixer LO leakage. 
In our setup, we use four different AWG channels to fully control each qubit: 1 channel for fast $Z$ control, and 3 channels for $X,Y$ control.

We combine the DC bias (used for tuning the static frequency of the qubit), baseband pulse (used for fast tuning of the qubit frequency), and RF pulse (used for charge driving the qubit) at RT. 
The DC and baseband flux signals are combined using a resistive combiner (Mini-circuits \textit{ZFRSC-42B-S+}). 
In order to match the impedance of the resistive combiner to $\SI{50}{\ohm}$ to prevent distortions in the baseband pulse, we use a homemade resistor box on the DC side, consisting of a $\SI{1}{k \ohm}$ resistor in parallel with a $\SI{50}{\ohm}$ resistor in series with a $\SI{10}{\micro  F}$ capacitor connected to ground. 
The output flux signal is sent through a $\SI{300}{MHz}$ low-pass filter (Mini-circuit VLFX-300+) and is combined with the RF signal using a diplexer (QMC-CRYODPLX-0218). 
We use a $\SI{1}{GHz}$ low-pass filter (VLFG-1000+) with a flat response of approximately 40~dB of rejection throughout the qubit operation frequency range.
Using this filter, we are able to filter out high-frequency sources of noise while maintaining the ability to apply high-fidelity single-qubit gates.

The advantage of this approach is that we only need a single microwave coaxial line for full control of each qubit.
In addition, by combining DC and baseband flux signals at RT we are able to more efficiently characterize and compensate for the flux transients arising from signal combination with a fast oscilloscope or a digitizer. 
However, the impedance mismatch caused by the reflective low-pass filter at the mixing chamber gives rise to standing modes between the filter and the chip ground at the end of the flux line. 
As a result, the effective qubit-drive coupling depends on the drive frequency. 
We characterize this dependence by biasing the qubit at different frequencies and Rabi driving on resonance. 

The active components used in our experimental setup are summarized in Table~\ref{tab:active_components}. 

\begin{table}[h!]
 \centering
\begin{tabular}{ ||c|c|c|| } 
\hline
\textbf{Quantity} & \textbf{Electrical component} & \textbf{Part number}\\ 
\hline
\hline
2$\times$ & PXI chassis & Keysight M9019\\ 
 \hline
 18$\times$ & AWG & Keysight M3202 \\
\hline
 2$\times$ & Digitizer & Keysight M3102 \\ 
\hline
 18$\times$ & Microwave source & Rohde \& Schwarz SGS100A \\ 
 \hline
1$\times$ & Microwave source & Holzworth HS9004A (4 channels) \\ 
 \hline
 1$\times$ & DC source & QDevil QDAC-I (24 channels) \\ 
 \hline
1$\times$ & Reference clock & SRS Model FS725\\ 
 \hline
1$\times$ & Digitizer pre-amplifier & SR445A 350MHz pre-amplifier \\ 
 \hline
2$\times$ & Amplifier & MITEQ AMF-5D-00101200-23-10P \\ 
 \hline
1$\times$ & Demodulation LO amplifier & Mini-Circuits ZVE-8G+\\ 
 \hline
2$\times$ & TWPA & Made at MIT LL \\ 
 \hline
\end{tabular}
\caption{\textbf{Active electrical components used in the experimental setup.}}
\label{tab:active_components}
\end{table}

\section{Sample}

The $4 \times 4$ superconducting transmon array is fabricated using a 3D integration flip-chip process~\cite{rosenberg_2017} (Fig.~1f in the main text).
The qubits are located on a $4 \times 4~\mathrm{mm}$ qubit tier as shown in Fig.~1g, and the readout and control lines are located on a separate $5 \times 5~\mathrm{mm}$ interposer tier as shown in Fig.~1h.
The two layers are separated using $\SI{3}{\micro  m}$ superconducting indium bump bonds and silicon hard stops. 

The interposer tier contains the flux bias lines and readout resonators that are respectively inductively and capacitively coupled to the qubits across the gap separating the two chips.
The ground plane of the interposer layer is etched away in the regions corresponding to the location of each qubit, with an additional $\SI{10}{\micro  m}$ gap to avoid unwanted capacitance between the qubits and the interposer ground plane. 

The qubit tier comprises a 2D lattice of capacitively-coupled single-ended transmon qubits~\cite{koch_2007} arranged in $4 \times 4$ array.
The capacitive coupling between each transmon is facilitated using an intermediary piece of superconductor, resulting in a nearest-neighbor qubit-qubit exchange interaction with average strength $J/2\pi=(5.89\pm 0.4)\,\mathrm{MHz}$, measured at qubit frequencies of $\SI{4.5}{GHz}$.
The ground plane of the qubit tier is etched away in the regions corresponding to the location of the signal line on the interposer tier, with an additional gap equal to the width of each line.

Multiplexed, dispersive qubit-state readout is performed through individual capacitively coupled coplanar resonators.
The sample contains two readout feedlines, each containing a single-port Purcell filter coupled to eight readout resonators.
We use $\lambda/4$ readout resonators for all qubits along the edge of the lattice to minimize their footprint on the sample.
The central qubits are read out with a $\lambda/2$ resonator.
This is done to minimize the parasitic couplings; the central qubits' resonator crossing is located in the middle of the $\lambda/2$ resonator at its voltage node.
The Purcell filter is designed to have a large bandwidth of $\SI{0.5}{GHz}$ at a resonance frequency of $\SI{6.3}{GHz}$, such that the readout resonators can be distributed over a $\sim\SI{400}{MHz}$ band while limiting Purcell decay to a rate $\lesssim 1/(\SI{300}{\micro s})$.
The resonators have a measured average linewidth of $\kappa/2\pi = (1.17 \pm 0.3)\mathrm{MHz}$, with an average dispersive shift of $\chi/2\pi = (0.87 \pm 0.11)\mathrm{MHz}$ characterized with each qubit biased at $\SI{4.5}{GHz}$.

The 3D integration process allows us to route each flux bias line directly above the corresponding superconducting quantum interference device (SQUID).
This positioning allows us to attain an average mutual inductance between the qubit SQUID loops and their respective flux bias lines of $\SI{1.15}{pH}$ while reducing the SQUID area to $4 \times \SI{4}{\micro m}$ compared to planar transmon arrays~\cite{braumuller_2022,karamlou_2022}.
Decreasing the area of the SQUID reduces the susceptibility of the qubit to flux noise, the offset due to uncontrollable magnetic fields in the environment, and crosstalk.
We measure more than an order of magnitude reduction in the crosstalk levels of the flip-chip sample compared to a planar $3 \times 3$ array of transmons~\cite{braumuller_2022, karamlou_2022}.

The sample is fabricated on a silicon substrate by dry etching an MBE-grown, 250-nm-thick aluminum film in an optical lithography process, forming all larger circuit elements such as the qubit capacitor pads, resonators, and the signal lines for qubit readout and control.
The qubit SQUID loops are fabricated with an electron beam lithography process and a double-angle shadow evaporation technique~\cite{dolan_1977} to form the Josephson junctions.
We use a wire width of $\SI{5}{\micro m}$ for the SQUIDs to minimize flux noise from local magnetic spin defects on the SQUID surface and interfaces~\cite{braumueller_2020}.

\begin{table}[ht!]
\centering
\label{tab:sample_parameters}
{\renewcommand{\arraystretch}{1.7}   
\begin{tabular}{ p{4.5cm}  p{1.5cm} p{1.5cm}  p{1.5cm}  p{1.5cm} p{1.5cm} p{1.5cm}  p{1.5cm}  p{1.5cm}  }
\toprule

Parameters & QB1 & QB2 & QB3 & QB4 & QB5 & QB6 & QB9 & QB10 \\
\hline
$\omega_{\mathrm{q}}^{\mathrm{max}}/2\pi$ (GHz) & 4.867 & 4.89 & 4.801 & 4.845 & 4.706 & 4.833 & 5.061 & 4.972 \\
$U/2\pi$ (MHz) & -217.3 & -217.6 & -217.0 & -241.9 & -217.9 & -214.5 & -217.9 & -210.1 \\
$\omega_{\mathrm{res}}/2\pi$ (GHz) & 6.207 & 6.358 & 6.237 & 6.496 & 6.367 & 6.337 & 6.285 & 6.431 \\
$\mathcal{F}_{\mathrm{gg}}$ & 0.96 & 0.96 & 0.96 & 0.95 & 0.96 & 0.95 & 0.95 & 0.96 \\
$\mathcal{F}_{\mathrm{ee}}$ & 0.92 & 0.88 & 0.92 & 0.93 & 0.89 & 0.88 & 0.86 & 0.81 \\
Readout assignment fidelity & 0.94 & 0.93 & 0.94 & 0.94 & 0.94 & 0.91 & 0.91 & 0.9 \\
$T_1$ ($\SI{}{\micro s}$) at operating point  & 23.8 & 24.1 & 16.7 & 34.9 & 17.6 & 15.2 & 14.7 & 17.1 \\
$T^*_2$ ($\SI{}{\micro s}$) at operating point & 3.6 & 2.7 & 1.8 & 2.6 & 2.9 & 3.8 & 2.7 & 1.1 \\
Single-qubit RB (individual) & 99.78\% & 99.88\% & 99.84\% & 99.89\% & NA & 99.84\% & 99.88\% & NA \\
Single-qubit RB (simultaneous) & 99.42\% & 99.88\% & 99.72\% & 99.71\% & NA & 99.81\% & 99.87\% & NA \\
\hline
\hline
Parameters & QB7 & QB8 & QB11 & QB12 & QB13 & QB14 & QB15 & QB16 \\
\hline
$\omega_{\mathrm{q}}^{\mathrm{max}}/2\pi$ (GHz) & 4.699 & 4.944 & 4.892 & 4.865 & 5.077 & 5.009 & 4.785 & 4.948 \\
$U/2\pi$ (MHz) & -217.8 & -217.6 & -213.3 & -219.2 & -218.6 & -216.6 & -219.3 & -218.4 \\
$\omega_{\mathrm{res}}/2\pi$ (GHz) & 6.425 & 6.28 & 6.336 & 6.356 & 6.503 & 6.25 & 6.359 & 6.196 \\
$\mathcal{F}_{\mathrm{gg}}$ & 0.95 & 0.95 & 0.96 & 0.95 & 0.97 & 0.96 & 0.96 & 0.96 \\
$\mathcal{F}_{\mathrm{ee}}$ & 0.92 & 0.91 & 0.9 & 0.91 & 0.83 & 0.9 & 0.9 & 0.87 \\
Readout assignment fidelity & 0.93 & 0.93 & 0.93 & 0.94 & 0.9 & 0.94 & 0.93 & 0.93 \\
$T_1$ ($\SI{}{\micro s}$) at operating point & 22.1 & 25.6 & 26 & 23.5 & 22.1 & 25.9 & 28.2 & 26.7 \\
$T^*_2$ ($\SI{}{\micro s}$) at operating point & 2.8 & 2.7 & 2.3 & 2.1 & 2.4 & 1.3 & 2.4 & 1.1 \\
Single-qubit RB (individual) & 99.78\% & 99.82\% & 99.88\% & 99.86\% & 99.77\% & 99.84\% & 99.9\% & 99.86\% \\
Single-qubit RB (simultaneous) & 99.63\% & 99.73\% & 99.78\% & 99.14\% & 99.83\% & 99.57\% & 99.39\% & 99.34\% \\
\hline
\hline
\end{tabular}
}
\caption{\textbf{Sample parameters}. We show the maximum transmon transition frequencies $\omega_{\mathrm{q}}^{\rm max}$ at the upper flux-insensitive point, the qubit anharmonicities $U$ (measured at $\omega_{\mathrm{q}}^{\rm max}$), the readout resonator frequencies $\omega_{\mathrm{res}}$, the probabilities $\mathcal{F}_{ij}$ of measuring the qubit in state $i$ after preparing it in state $j$, the readout assignment fidelity $(\mathcal{F}_{\mathrm{gg}}+\mathcal{F}_{\mathrm{ee}})/2$, the measured $T_1$s, $T^*_2$s. The reported readout assignment fidelities are limited by the state preparation error due to the initial thermal population of the qubits. In addition, we report the individual and simultaneous single-qubit randomized benchmarking (RB) fidelities. Due to the significant RF crosstalk between the drive targeting QB5 and QB10, we do not apply single-qubit gates to these qubits in our experiments.}
\end{table}

Detailed sample parameters for individually isolated qubits are summarized in Table~\ref{tab:sample_parameters}.  We show the maximum transmon transition frequencies $\omega_{\mathrm{q}}^{\rm max}$ at the upper flux insensitive point, the qubit anharmonicities $U$ (measured at $\omega_{\mathrm{q}}^{\rm max}$), the readout resonator frequencies $\omega_{\mathrm{res}}$, the probabilities $\mathcal{F}_{ij}$ of measuring the qubit in state $i$ after preparing it in state $j$, the readout assignment fidelity $(\mathcal{F}_{\mathrm{gg}}+\mathcal{F}_{\mathrm{ee}})/2$, the measured $T_1$s and $T^*_2$s at qubit frequencies around the bias point used in the experiment. We anticipate that the reported readout assignment fidelities are limited by the state preparation error due to the initial thermal population of the qubits. The relatively low $\mathcal{F}_{\mathrm{gg}}$ is a result of the high thermal population caused by noticeable heating at the MXC stage when DC biasing the qubits. This heating is caused by sending a DC current through the microwave cable between the still and MXC stages, dissipating approximately $\sim \SI{500}{\micro W}$ of power at the MXC when all the qubits are biased at $\SI{4.5}{GHz}$. 
In addition, we report the individual and simultaneous single-qubit randomized benchmarking (RB) fidelity for the gates used for tomographic readout. The microwave charge pulses for the gates are applied via the flux lines. Due to the significant RF crosstalk between the drives targeting QB5 and QB10, we do not calibrate and apply simultaneous gates for tomographic readout on these qubits.

\section{Control and readout calibration}

\subsection{Flux crosstalk calibration}

Flux-tunable transmon qubits comprise two Josephson junctions in a SQUID loop in parallel with a shunting capacitor. We can control the effective Josephson energy $E_J$ of the SQUID, and therefore the transition frequency between the ground and the first-excited state of the transmon, by threading magnetic flux through the SQUID. 
The frequency of the transmon in response to an applied external flux $\Phi_\mathrm{ext}$ is approximately given by~\cite{koch_2007}:
\begin{equation}
    f(\Phi_\mathrm{ext}) = \left(f^{\text{max}} + \frac{E_{C}}{h}\right)\sqrt[4]{d^2 + (1-d^2)\:\text{cos}^2\left(\pi \frac{\Phi_\mathrm{ext}}{\Phi_0}\right) } - \frac{E_{C}}{h}\,,
    \label{eq:transmon_spectrum}
\end{equation}
where $E_{C}$ is the transmon charging energy and $f^{\text{max}}=(\sqrt{8 E_J E_C}-E_C)/h$ is the maximum qubit frequency.
The asymmetry parameter $d$ of the SQUID junctions is given by $d = |(E_{J,2}-E_{J,1})/(E_{J,2}+E_{J,1})|$, where $E_{J,1}$ and  $E_{J,2}$ are the Josephson energies of the two SQUID junctions.
The above equation is an approximation because it holds only in the limit $E_J\gg E_C$.

We thread flux through the SQUID by applying a current through a local flux line that terminates near the SQUID.
We use a voltage source applied over a series resistor of resistance $1~\text{k}\Omega$ at room temperature to generate a stiff current source for each flux line.
There is a linear relation between the voltage applied to a flux line and the magnetic flux experienced by the SQUID:
\begin{equation}
    \Phi_\mathrm{ext} = V/V^{\Phi_0}+ \Phi_{\text{offset}}\,,
    \label{eq:volts_to_flux}
\end{equation}
where $V^{\Phi_0}$ is the voltage required to tune the qubit by one magnetic flux quantum $\Phi_0$, and $\Phi_{\text{offset}}$ is a flux offset due to non-controllable sources of static magnetic field.

For an array of transmons, the applied current in a flux line targeting one SQUID may thread unwanted magnetic flux through other SQUIDs.
We can model this flux crosstalk as a linear process, where the fluxes $\vec{\Phi}_\mathrm{ext}$ experienced by the SQUIDs are related to the voltages $\vec{V}$ applied to the flux lines:
\begin{equation}
    \vec{\Phi}_\mathrm{ext} = {(\boldsymbol{V}^{\Phi_0})^{-1}\boldsymbol{S}}\vec{V} + \vec{\Phi}_{\text{offset}}\,,
    \label{eq:crosstalk_eq}
\end{equation}
where $\boldsymbol{V}^{\Phi_0}$ is a diagonal matrix with $\boldsymbol{V}^{\Phi_0}_{i,i}$ corresponding to the $V^{\Phi_0}$ of qubit $i$.
${\boldsymbol{S}}$ is the flux crosstalk matrix, with $\boldsymbol{S}_{i,j}=\partial V_i /\partial V_j$ representing the voltage response of qubit $i$ to a voltage signal applied to qubit $j$.
In this representation, the diagonal elements of ${\boldsymbol{S}}$ are 1. 
By characterizing $\boldsymbol{S}$, we can compensate for the crosstalk and set the qubit frequencies precisely. 

\begin{figure}[ht!]
    \centering
    \subfloat{\label{fig:dc_xtalk}}
    \subfloat{\label{fig:ff_xtalk}}
    \includegraphics{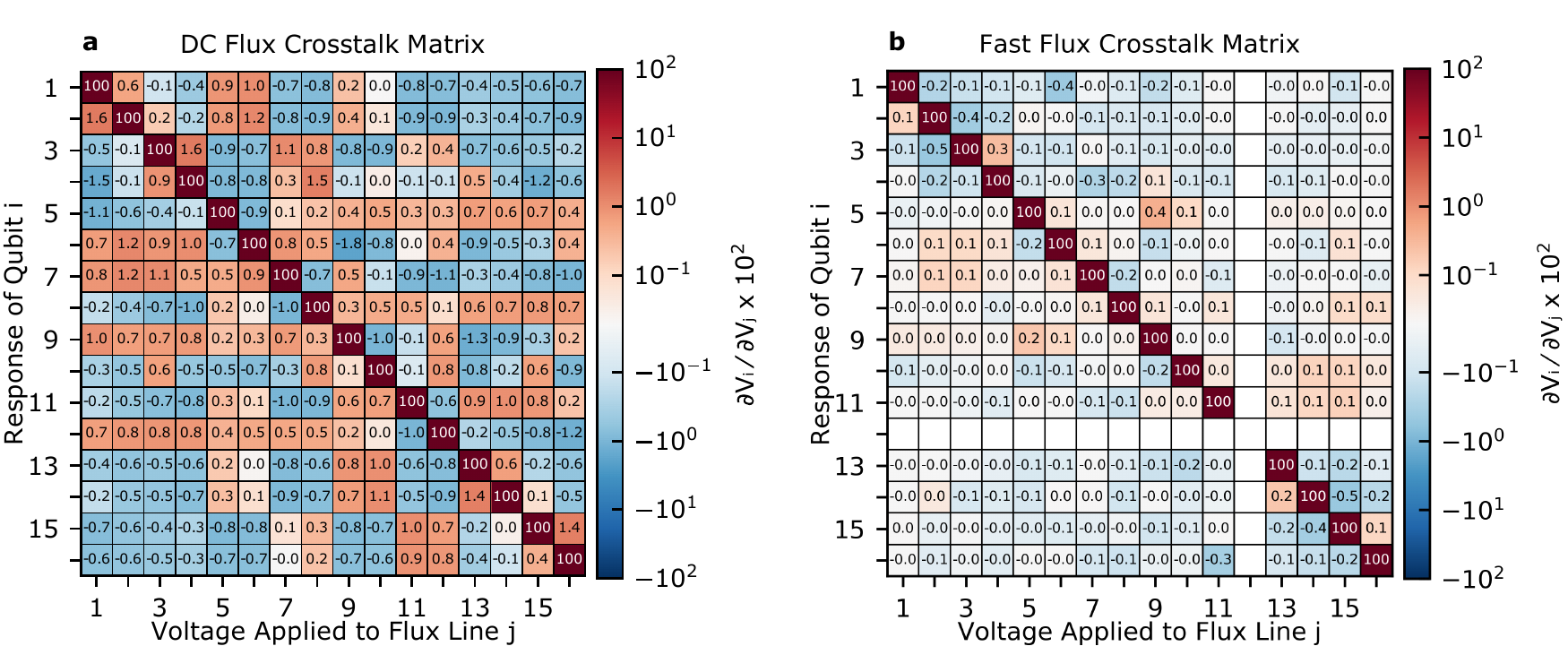}
    \caption{\textbf{Learned Flux Crosstalk Matrices}. The learned \textbf{(a)} DC flux crosstalk matrix and \textbf{(b)} fast flux crosstalk matrix for the 16-qubit device, each rescaled by a factor of 100 such that each element is a percentage. The fast flux microwave line for qubit 12 was broken, so we have no crosstalk information for qubit 12.}
    \label{fig:xtalk_matrices}
\end{figure}

The first step of calibrating flux crosstalk is characterizing the transmon spectrum of each qubit in the array.
We do this by sweeping the voltage applied to the flux line and measuring the qubit frequency.
By fitting this data with Eq.~\ref{eq:transmon_spectrum} and Eq.~\ref{eq:volts_to_flux}, we extract the transmon spectrum parameters $f^{\text{max}}$, $E_C$, and $d$, and the voltage to flux conversion parameters $V/V^{\Phi_0}$ and $\Phi_{\text{offset}}$.

Next, we calibrate flux crosstalk with a learning-based protocol described in~\cite{barrett_2023}. 
We start with an initial guess crosstalk matrix, which can either be an estimate from a previous measurement or the identity matrix.

We select a quasi-random vector of target frequencies $\vec{f}$, subject to a few constraints.
First, we require that the frequency for each qubit falls in the range of $100$~MHz below its sweet spot to $1$~GHz below its sweet spot.
Second, to reduce frequency shifts due to resonant exchange interaction between qubits, we require that the frequency detuning between neighboring qubits is at least $\SI{200}{MHz}$ and that the frequency detuning between any two qubits in the array is at least $50$~MHz.
We use our initial $\boldsymbol{S}$ and equations Eq.~\ref{eq:transmon_spectrum} and Eq.~\ref{eq:crosstalk_eq} to solve for the voltages $\vec{V}$ needed to target $\vec{f}$.
We apply $\vec{V}$, then measure the qubit frequencies via spectroscopy.
Even though we detune our qubits, they still experience frequency shifts.
We use previously characterized qubit-qubit couplings to calculate the uncoupled qubit frequencies.
Finally, we convert the uncoupled frequencies to fluxes experienced. 

By repeating this process $M$ times, we obtain a size-$M$ training set of applied voltages and measured fluxes: $\{\vec{V}_i, \vec{\Phi}_{\text{meas},i} \}_{i=1:M}$. We learn our crosstalk matrix using the crosstalk relation (Eq.~\ref{eq:crosstalk_eq}), by leveraging the fact that the estimated flux ${(\boldsymbol{V}^{\Phi_0})^{-1}\boldsymbol{S}}\vec{V}_i + \vec{\Phi}_{\text{offset}}$ ought to equal the measured flux $\vec{\Phi}_{\text{meas},i}$. 
We learn $\boldsymbol{S}$ row-by-row.
We learn the elements of the $k^{\text{th}}$ row of $\boldsymbol{S}$ by minimizing the mean-squared-error cost function
\begin{align}
    & C(\boldsymbol{S}_k) = 
    \frac{1}{M} \sum\limits_{i=1}^M \Bigg|\Bigg| (\vec{\Phi}_{\text{meas},i})_k
    - \Big[ {(\boldsymbol{V}^{\Phi_0}_{k,k})^{-1}\boldsymbol{S}_k} \vec{V}_i+
    (\vec{\Phi}_{\text{offset}})_k \Big]\Bigg|\Bigg|^2\,.
\end{align}
in a gradient descent optimizer (we use the L-BFGS optimizer in \textsf{PyTorch}).
The optimization will converge as the estimated fluxes approach the measured fluxes.
Once we have repeated this minimization for all rows, we have our optimized $\boldsymbol{S}$.
In Fig.~\ref{fig:dc_xtalk}, we report the DC flux crosstalk matrix for the 16-qubit device.

This same calibration procedure is utilized to calibrate crosstalk for fast flux pulse control, as well.
In order to learn the fast flux crosstalk for each qubit, we tune that qubit to a target frequency using a $\SI{100}{ns}$ fast flux pulse and measure the qubit frequency via spectroscopy.
Then, using the same pulse amplitude for the target qubit, we apply flux pulses with random amplitudes through the other flux lines and measure the change in the target qubit's frequency.
We use the training set consisting of the voltages applied and the changes in the target qubit frequency for learning the fast flux crosstalk matrix.
In Fig.~\ref{fig:ff_xtalk}, we report the fast flux crosstalk matrix for the 16-qubit device.

\subsection{Time-domain flux pulse shaping and transient calibration}

Baseband-frequency flux pulses experience distortions as they travel to the sample.
The distortion caused by room-temperature components can be characterized and compensated for using a fast oscilloscope with a sampling rate better than $\SI{500}{MHz}$.
The distortion introduced by the components in the cryostat is generally temperature-dependent and needs to be characterized while the fridge is operating.
By using the qubit as a sensor, we can characterize the distortion of a square flux pulse ($Z$~pulse) using Ramsey-type measurements~\cite{rol_2020}.
The step response for the signal generated by the AWG, $s_{\rm{AWG}}(t)$, the signal reaching the qubit, $s_{\rm{qubit}}(t)$, and the time-dependent distortion, $n(t)$, are related by
\begin{equation}
    s_{\rm{qubit}}(t) = s_{\rm{AWG}}(t) \: \big(1+n(t) \big),
\end{equation}

To characterize $n(t)$, the target qubit is biased to its sweet spot using static flux and excited with a $\frac{\pi}{2}$-pulse around the $x$-axis followed by a $Z$ pulse for time $t$.
The dynamic frequency change of the qubit as a response to the $Z$ pulse can be extracted by measuring both $\langle X \rangle $ and $\langle Y \rangle $, denoting the expectation values of the qubit state projected along the $x$-axis and $y$-axis of the Bloch sphere.
The phase accumulated by the qubit during the $Z$ pulse can be described by $\phi (t)=\int_0^t \Delta(t^\prime) \rm{d}t^\prime$, where $\Delta(t)=\Delta_0 + \delta(t)$ is the frequency detuning, consisting of the target detuning $\Delta_0$ and the added frequency distortion $\delta(t)$.
The $\langle X \rangle$ and $\langle Y \rangle$ measurements are used to obtain $\cos\phi(t)$ and $\sin\phi(t)$, respectively, which we use for constructing the phasor $e^{i\phi(t)} = \cos\phi(t) + i \sin\phi(t)$.
We can then extract the frequency distortion as $\delta(t)=\frac{\mathrm{d}}{\mathrm{d}t} \arg(e^{i\phi(t)})$.
By using the qubit spectrum, we map $\delta(t)$ to the distortion in the signal reaching the qubit $n(t)$. 

In order to obtain an analytical expression, we fit the extracted $n(t)$ with a sum of typically three or more exponential damping terms with time constants $\tau_i$ and settling amplitudes $A_i$:
\begin{equation}
    n(t)=\prod_i A_i e^{-t/\tau_i}
\end{equation}
In our experiments, we use the analytical expression of $n(t)$ to pre-distort the output signal of the AWG in order to compensate for the system transients.

One method of analytically pre-distorting the signal is using Fourier transformations.
Using the signal step-response $h(t)=\Theta(t)(1+n(t))$, we can obtain the impulse-response of our system:
\begin{equation*}
    I(t)=\dot{h}(t)=\delta(t)+\delta(t)\prod_i A_i e^{-t/\tau_i} - \Theta(t)\sum_j \left( \frac{1}{\tau_j} \prod_i A_i e^{-t/\tau_i} \right).
\end{equation*}
We can solve for the pre-distorted signal needed to make the qubit feel a constant flux using the convolution relation $s_{\text{qubit}}(t)=s_{\text{AWG}}(t) * I(t)$, therefore,
\begin{equation*}
    s_{\text{AWG}}(t) = \mathcal{F}^{-1}\left\{\frac{\mathcal{F}\{s_{\text{qubit}}(t)\}}{\mathcal{F}\{I(t)\}}\right\}
\end{equation*}
where $\mathcal{F}\{I(t)\}=\mathcal{F}\{\delta(t)\}+ \mathcal{F}\{ \delta(t) \prod_i A_i e^{-t/\tau_i}\} - \sum_j  \frac{1}{\tau_j} \mathcal{F}\{ \Theta(t) \prod_i A_i e^{-t/\tau_i} \}$. We can calculate an expression for $\mathcal{F}\{I(t)\}$:
\begin{equation*}
    \mathcal{F}\{I(t)\}=\prod_i \left(1 + \frac{i \omega A_i \tau_i}{i \omega \tau_i+1} \right).
\end{equation*}

In practice, in order to accurately pre-distort the pulse shape using Fourier transformations we need to zero-pad the waveform such that the total length of the waveform is at least $5 \times \max_i(\tau_i)$.
As a result, when correcting for long-timescale transients with $\tau \propto \mathcal{O}(\SI{100}{\micro s})$ this procedure becomes inefficient.

Instead, we use an infinite-impulse-response (IIR) filter for pulse pre-distortion~\cite{rol_2020} with the difference equation:
\begin{equation}
    a_0 y[n] = \sum^N_{i=0} b_i x[n-i] - \sum^M_{j=1} a_j y[n-j]
\end{equation}
where $y[n]$ is the output signal, $x[n]$ is the input signal, $N$ is the feed-forward filter order, and $M$ is the feed-back filter order.
A first-order IIR filter ($N=M=1$) can be used to correct the exponential transients.
We find the IIR coefficients $a_0$, $a_1$, $b_0$, and $b_1$ for a single exponential pole with amplitude $A$ and time-constant $\tau$:

\begin{equation}
    a_0 = 1 + A, \quad
    a_1 = -\alpha - A, \quad
    b_0 = 1, \quad
    b_1 = -\alpha\,,
\end{equation}
where $\alpha=e^{-1/f_s\tau}$ depends on the AWG sampling rate $f_s$. 

We use the SciPy library in Python to apply the IIR filter with the coefficient calculated above to the waveforms in our pulse generation software.
For systems with more than one exponential transient term, we concatenate multiple IIR filters, each corresponding to a term.

\subsection{Pulse alignment}

Differences in the pulse path lengths and the delays in control electronics cause discrepancies in the time that flux ($Z$) and charge ($XY$) pulses reach the target qubit.
To accurately execute the desired pulse sequence, we characterize the relative delay of each signal line and adjust pulse timings in software to compensate. 

\begin{figure}[t!]
    \centering
    \subfloat{\label{fig:z_delay}}
    \subfloat{\label{fig:xy_delay}}
    \includegraphics{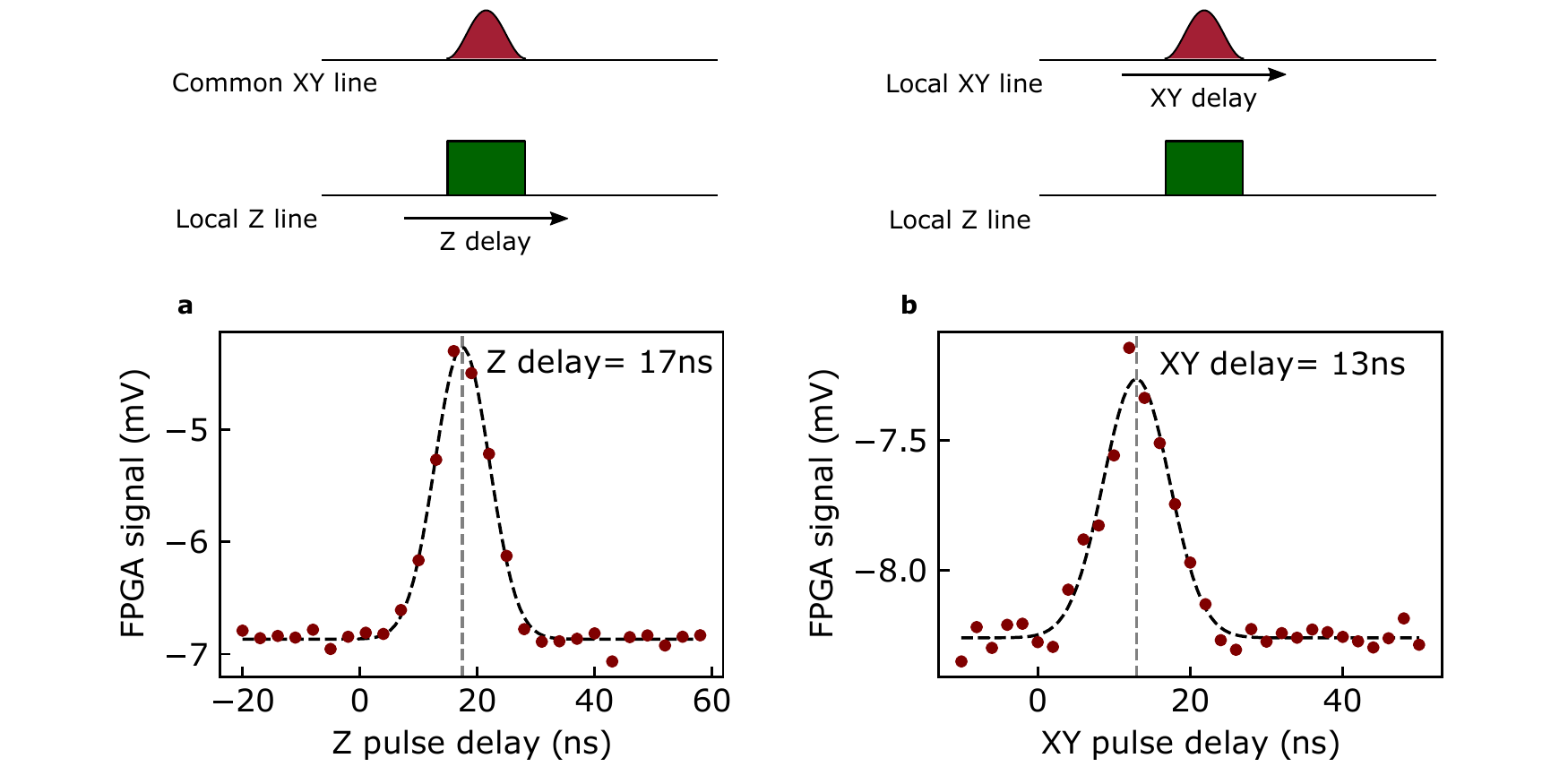}
    \caption{\textbf{Pulse alignment calibrations}. \textbf{(a)} The timing for the $Z$ pulse applied through each line is calibrated by sweeping the delay time of a $Z$ pulse by tuning the qubit to the sweet spot while applying an XY pulse through the common resonator feedline at that frequency. \textbf{(b)} The timing for the XY pulse applied through each local line is similarly calibrated by applying a $Z$ pulse with a constant delay characterized in \textbf{(a)} and sweeping the delay time of the XY pulse.}
    \label{fig:pulse_alignment}
\end{figure}

We first characterize the delay of a $Z$ pulse applied to each qubit with respect to a common $XY$ drive through the resonator feedline.
An $XY$ pulse through the resonator feedline reaches each qubit roughly simultaneously; it, therefore, serves as a common reference for pulse delay timing.
To calibrate the $Z$ pulse delay of a particular qubit, we first isolate the qubit from the rest of the lattice and away from the flux sweet spot.
We then apply a short $\SI{13}{ns}$ $Z$ pulse with the amplitude required to tune the qubit to the sweet spot, and an $XY$ pulse of the same length at the sweet-spot frequency.
The amplitude of the $XY$ pulse is less than that required for a $\pi$-pulse.
We measure the qubit resonator response as a function of the $Z$ pulse delay (Fig.~\ref{fig:z_delay}); a peak in the measured response indicates the delay at which the qubit is maximally excited, which occurs when the two pulses are aligned.
(When the two pulses are misaligned, part, or all, of the $XY$ drive is off-resonant with the qubit frequency, resulting in a lower qubit excited state population.)
By fitting the data in Fig.~\ref{fig:z_delay} to a Gaussian curve, we find the optimal $Z$ delay of $\SI{17}{ns}$ with respect to the $XY$ pulse applied through the common line.

Next, we calibrate the delay for $XY$ pulses sent to each qubit through its respective local line, using the calibrated $Z$ pulse delays for reference.
Similar to the procedure described above, we tune the qubit to the sweet spot using a $Z$-pulse and apply an $XY$ pulse, now through the local line, at the qubit sweet spot frequency.
Now with the $Z$ pulse timing fixed at the calibrated value, we sweep the delay of the local $XY$ pulse and extract the optimal local $XY$ pulse delay with respect to the common $XY$ pulse timing, as shown in Fig.~\ref{fig:xy_delay}.

\subsection{Readout optimization}

We use frequency multiplexed heterodyne mixing to read out multiple resonators coupled to the same feedline with a common local oscillator (LO) frequency~\cite{krantz_2019}.
To maximize the readout fidelity, we maximize the separation in in-phase/quadrature ($I$/$Q$) space between the readout signal conditioned on the qubit ground and excited state by optimizing the frequency and amplitude of each sideband tone.

\begin{figure}
    \centering
    \subfloat{\label{fig:assignment_fidelity}}
    \subfloat{\label{fig:separation_fidelty}}
    \includegraphics{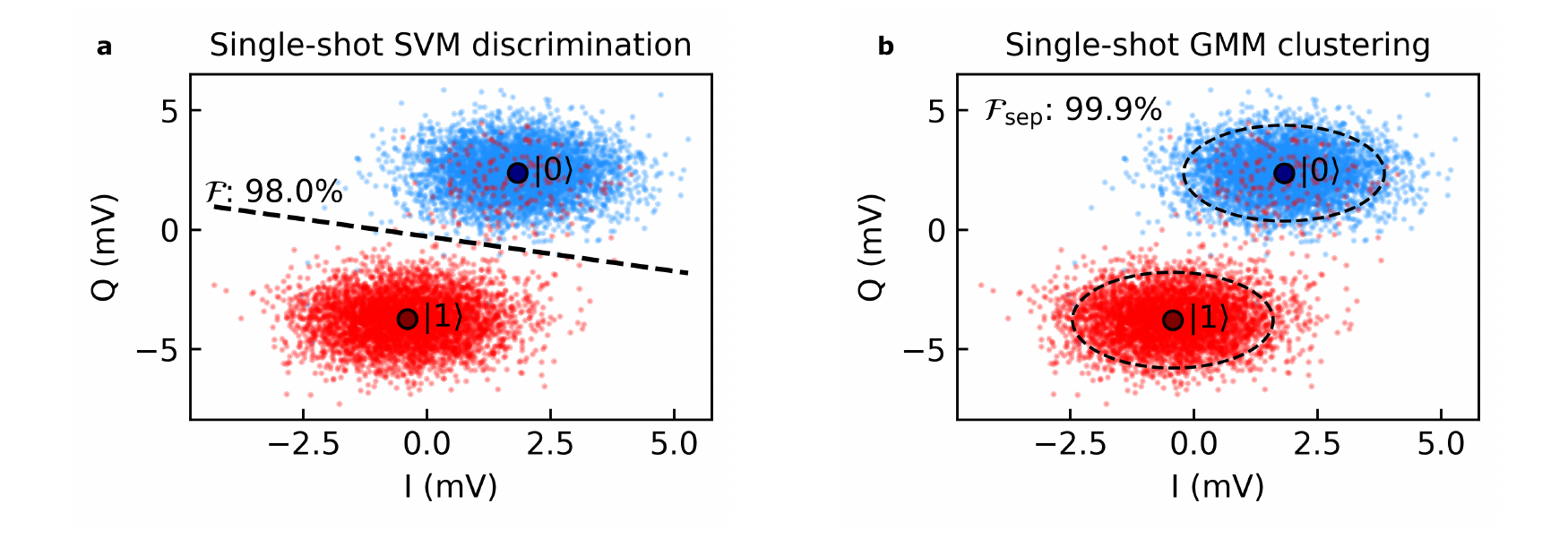}
    \caption{\textbf{Readout single-shot data}. \textbf{(a)} We use an SVM to draw a discrimination line (dashed line) between the single-shot points prepared in the $|0\rangle$ state (blue points) and $|1\rangle$ state (red points) prior to readout. From this data, we obtain the readout fidelity $\mathcal{F}_{\rm RO} =98.0\%$. \textbf{(b)} The same single-shot dataset is grouped into two clusters using a GMM with a cluster separation fidelity of $\mathcal{F}_{\rm sep}=99.9\%$}
    \label{fig:readout_optimization_discrimination}
\end{figure}

In order to distinguish between the single-shot clusters resulting from different qubit states we first take single-shot measurements with the qubit prepared in the $|0\rangle$ state and the $|1\rangle$ state by applying a $\pi-$pulse to the qubit.
Using this data, we train a support vector machine (SVM)~\cite{pedregosa_2011} to find the hyperplane that classifies each single-shot point as 0 or 1.
The hyperplane parameters are calculated in a manner that maximizes the probability of correctly classifying each data point.
We quantify the readout quality using the readout fidelity:
\begin{equation}
    \mathcal{F}_{\rm RO} = \frac{f_{00} + f_{11}}{2}\,,
\end{equation}
where $f_{00}=P(0|\:|0\rangle)$ is the probability of correctly classifying the qubit $|0\rangle$ state as $0$, and $f_{11}=P(1|\:|1\rangle)$ is the probability of classifying the $|1\rangle$ state as $1$ after readout.

Fig.~\ref{fig:assignment_fidelity} illustrates an example set of single-shot data acquired by preparing the qubit in the $|0\rangle$ state (blue) and in the $|1\rangle$ state (red) and the corresponding line used to discriminate between the two states.
For this dataset, the readout fidelity is $\mathcal{F}_{\rm RO} =98.0\%$, with $f_{00}=99.6\%$ and $f_{11}=96.5\%$.
The discrepancy between $f_{11}$ and $f_{00}$ is a result of qubit relaxation during the readout process.

In addition to using $\mathcal{F}_{\rm RO}$ to quantify the readout quality, we also consider cluster separation.
We use a Gaussian Mixture Model (GMM) probability distribution to group the single-shot data into two clusters, independent of the intended state.
We fit the mean ($\boldsymbol{\mu}$) and the covariance matrix $\boldsymbol{\Sigma}$ for each Gaussian distribution representing a cluster using the expectation-maximization (EM) algorithm~\cite{pedregosa_2011}.
A multivariate Gaussian distribution in the $I/Q$ plane with mean $\boldsymbol{\mu}$ and covariance matrix $\boldsymbol{\Sigma}$ has density:
\begin{equation}
    P_{\boldsymbol{\mu},\boldsymbol{\Sigma}}(\boldsymbol{x}) = \frac{1}{2\pi \sqrt{|\boldsymbol{\Sigma}|}} \rm exp \big( -\frac{1}{2} (\boldsymbol{x}-\boldsymbol{\mu})^T \boldsymbol{\Sigma}^{-1} (\boldsymbol{x}-\boldsymbol{\mu}) \big).
\end{equation}

In our model, we use the same covariance matrix $\boldsymbol{\Sigma}$ for both clusters.
We use the spatial overlap between the Gaussian distributions to quantify the cluster separation by defining the separation fidelity $\mathcal{F}_{\rm sep}$ as
\begin{equation} \label{eq:F_sep}
    \mathcal{F}_{\rm sep} = 1 - \int \rm min \big[ P_{\boldsymbol{\mu_0},\boldsymbol{\Sigma}}(\boldsymbol{x}) , P_{\boldsymbol{\mu_1},\boldsymbol{\Sigma}}(\boldsymbol{x}) \big] d\boldsymbol{x}\,. 
\end{equation}

In Fig.~\ref{fig:separation_fidelty}, we group the readout single-shot readout into two clusters.
The ellipse around each cluster encloses the area corresponding to two standard deviations away from the center of the cluster.
Using Eq.~\ref{eq:F_sep} we obtain a separation fidelity of $\mathcal{F}_{\rm sep}=99.9\%$ between the two readout clusters.

\begin{figure}
    \centering
    \subfloat{\label{fig:ass_fid_opt}}
    \subfloat{\label{fig:sep_fid_opt}}
    \subfloat{\label{fig:ro_fid_comparison}}
    \includegraphics{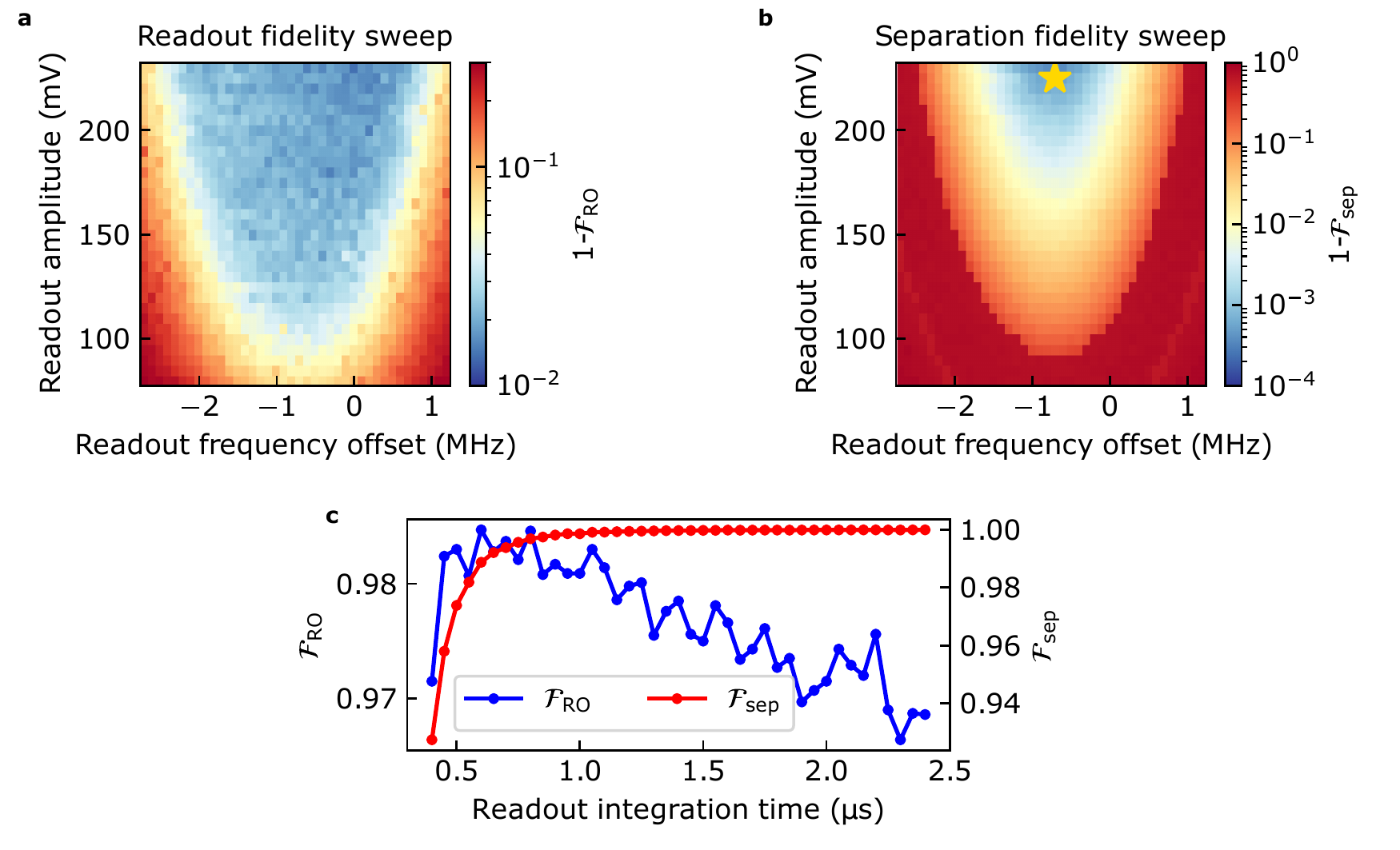}
    \caption{\textbf{Readout optimization}. \textbf{(a)} Experimentally measured readout optimization landscape using the readout fidelity. The plateau in the quantity is caused by qubit relaxation during readout. \textbf{(b)} Experimentally measured readout optimization landscape using the cluster separation fidelity. The star depicts the optimal parameters. \textbf{(c)} $\mathcal{F}_{\rm RO}$ and $\mathcal{F}_{\rm sep}$ at different readout integration times.}
    \label{fig:readout_optimization_sweep}
\end{figure}

To optimize the readout in our calibration procedure, we vary the sideband modulation pulse frequency and amplitude used for multiplexed readout.
Both the readout fidelity $\mathcal{F}_{\rm RO}$ and the separation fidelity $\mathcal{F}_{\rm sep}$ can be used as the optimization metric.
We measure the readout optimization landscape by sweeping the readout frequency offset and readout amplitude for a given qubit and calculating $1-\mathcal{F}_{\rm RO}$ (Fig.~\ref{fig:ass_fid_opt}) and $1-\mathcal{F}_{\rm sep}$ (Fig.~\ref{fig:sep_fid_opt}).
The optimization landscape using both metrics is convex, which allows for the use of a conventional optimizer, such as Nelder-Mead, to find the optimal readout sideband frequency and amplitude efficiently.
The landscape of the cluster separation fidelity is more sensitive to changes in the readout parameters compared to the readout fidelity, which is a result of the limit imposed on the readout fidelity caused by qubit relaxation during readout.
Therefore, separation fidelity is a better metric to use for the optimal readout sideband frequency and amplitude while keeping the readout time fixed. It is important to limit the maximum readout amplitude to ensure that the readout remains in the dispersive regime
In the dispersive readout scheme, with the readout resonator frequency above the transmon frequency, we find the magnitude of the optimal readout frequency offset to be roughly equal to the measured qubit-resonator dispersive shift $\chi$ as predicted for $\chi \lesssim \kappa/2$~\cite{krantz_2019}.

Using these optimal parameters, we vary the readout integration time to maximize the readout fidelity (Fig.~\ref{fig:ro_fid_comparison}).
Increasing the integration time results in an increase in the single-shot cluster separation, however, the readout fidelity suffers more from qubit relaxation.
Therefore, the optimal integration time is selected in a way to balance these two effects by maximizing $\mathcal{F}_{\rm RO}$.

\subsection{State preparation and measurement (SPAM) error mitigation}

In order to mitigate SPAM errors from our measured observable, we use a stochastic $\beta$-matrix (often also referred to as a $T$ matrix in the literature) for each qubit, where 
\begin{equation}
    \beta = \begin{pmatrix} f_{00} & 1 - f_{11}\\ 1-f_{00} & f_{11} \end{pmatrix}
\end{equation}
where $f_{00}$ and $f_{11}$ are the probabilities of correctly preparing and measuring the qubit in the $|0\rangle$ and $|1\rangle$ states respectively. There are two main sources of our experimental error captured by this matrix: incoherent thermal population of the qubit prior to the experiment sequence, and readout errors. 

In the absence of readout crosstalk, the $\beta$-matrix for an $n$-qubit system is the tensor product of the $\beta$-matrix for each individual qubit: $\overline{\overline{\beta}}=\beta_1 \otimes ... \otimes \beta_n$. By inverting this matrix, and applying it to the measured bitstring probabilities we can correct for SPAM errors in the measured observable~\cite{geller_2021}. 

\subsection{Idling frequency layout}

\begin{figure}[h!]
    \centering
    \includegraphics{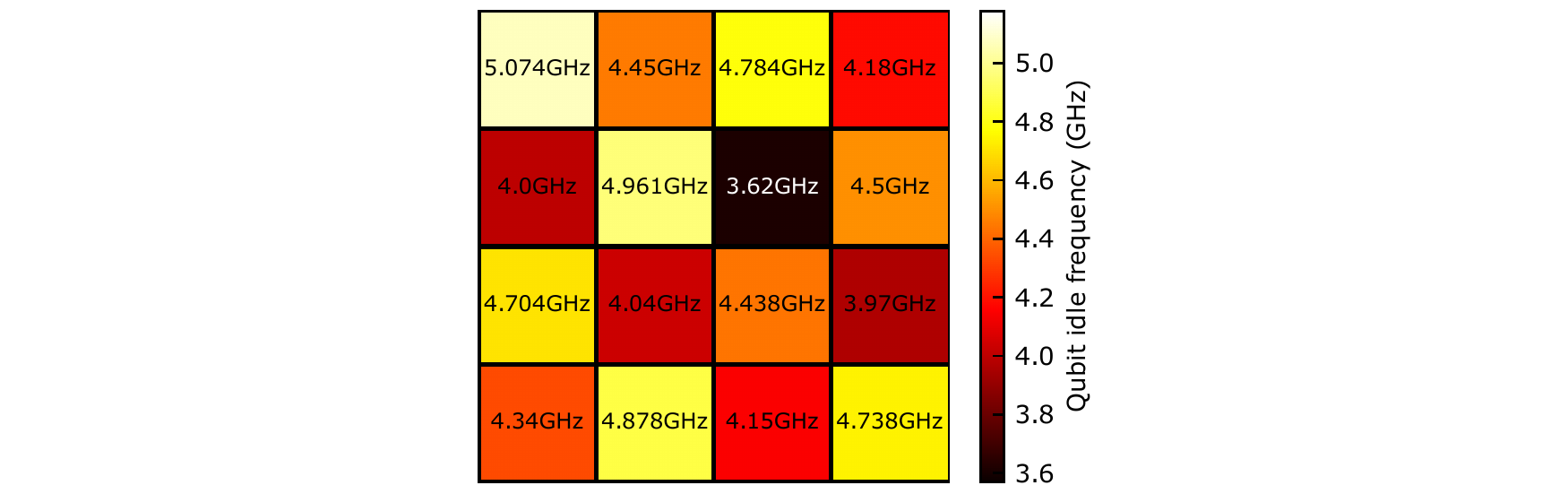}
    \caption{\textbf{Lattice qubit idle frequencies}.}
    \label{fig:qubit_idle_frequency}
\end{figure}

During the emulation of the Bose-Hubbard Hamiltonian, all qubits are brought on resonance to allow particle exchange. During state preparation and readout steps of the experiment sequence, the qubits are detuned from one another to turn off particle exchange.
The capability to apply simultaneous high-fidelity single-qubit gates to qubits at their respective idling frequencies is necessary for performing tomographic readout, which is required to measure correlations and state purities.
The idling frequencies of the qubits during the state preparation and readout steps of the experimental sequence satisfy the following criteria:
\begin{itemize}
    \item Each qubit is detuned from its nearest neighbors by at least $\SI{300}{MHz}$ to prevent particle exchange during readout.
    \item Each qubit is detuned from its next-nearest neighbors by at least $\SI{40}{MHz}$.
    \item Qubit frequencies are selected to minimize charge-driving crosstalk effects to enable high-fidelity simultaneous single-qubit gates.
    \item Qubit 12 is biased at $\SI{4.5}{GHz}$ (qubit 12 suffered from elevated pulse distortion due to a partial break in its flux line).
    \item Each qubit frequency is detuned from any two-level system (TLS) defects.
\end{itemize}
The idling qubit frequency layout used in our experiment is shown in Fig.~\ref{fig:qubit_idle_frequency}.
At these frequencies, we apply single-qubit gates on all qubits, excluding qubits 5 and 10, for which the qubit was too weakly coupled to the line used for charge driving. 
The randomized benchmarking (RB) fidelities for the calibrated gates are reported in Tab.~\ref{tab:sample_parameters}, with an average individual RB fidelity of $99.8\%$ and an average simultaneous RB fidelity of $99.6\%$.

\section{Hamiltonian Characterization} 

The Hamiltonian of the driven system is described by 
\begin{equation}
    \hat{H}/\hbar = \sum_{\langle i,j\rangle}J_{ij}\hat\sigma _i^+ \hat\sigma _j^- + \frac{\delta}{2} \sum_i\hat\sigma _i^z + \Omega  \sum_j (\alpha_j \hat\sigma_j^- + \mathrm{h.c.})\,,
    \label{eq:driven_hamiltonian}
\end{equation}
as shown in Eq.~5 of the main text.
The particle exchange strengths $J_{ij}$ and the amplitudes and phases of the relative drive couplings $\alpha_i$ are determined by the circuit layout, and remain constant in our experiments.
To accurately simulate the behavior of our lattice, we carefully characterize each of these parameters.

\subsection{Particle exchange strengths}

We measure the particle exchange strengths $J_{ij}$ for nearest and next-nearest neighboring qubits in the lattice (that is, for qubits $i$ and $j$ of relative Manhattan distance 1 and 2). 

Each value $J_{ij}$ is measured via the resonant iSWAP rate between qubits $i$ and $j$ at $\omega_{\mathrm{com}}/2\pi=\SI{4.5}{GHz}$.
With all other qubits detuned, we first add an excitation to qubit $i$ by applying a $\pi$-pulse.
Qubits $i$ and $j$ are then brought on resonance at frequency $\omega_{\mathrm{com}}$.
After allowing the qubits to interact for time $t$, the qubits are once again detuned and measured in the $z$ basis.
The coupling strength $J_{ij}$ is determined by the time $T$ required for a full excitation swap, $J=\pi/2T$~\cite{blais_2021}.

\begin{figure*}[ht!]
\includegraphics[width=15cm]{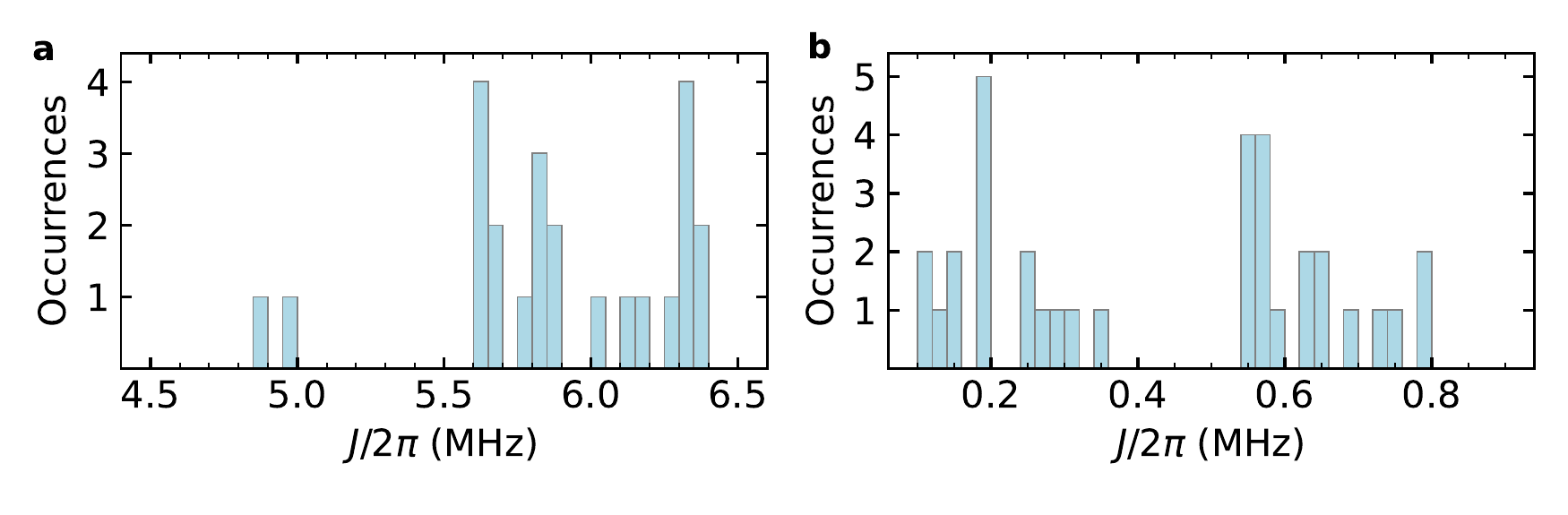}
\caption{Histogram of coupling strengths among (a) nearest-neighboring sites and (b) next-nearest-neighboring sites. In a 4-by-4 square lattice, there are a total of 24 and 34 such couplings, respectively.}
\label{sfig:J_hist}
\end{figure*}

A histogram of the measured values $J_{ij}$ for (next-)nearest neighbors are shown in Fig.~\ref{sfig:J_hist}(a(b)).
The average nearest coupling in our device is $J/2\pi = \SI{5.9\pm0.4}{MHz}$ at $\omega_{\rm com}/2\pi=\SI{4.5}{GHz}$. The next-nearest neighbor couplings are typically in the range of $J/10$ to $J/40$.
The qubits at the center of the lattice exhibit a larger coupling to their next-nearest neighbor, which can be attributed to larger stray capacitances.

\subsection{Drive coupling amplitude}

\begin{figure*}[ht!]
\subfloat{\label{fig:rabi_sweep}}
\subfloat{\label{fig:drive_rate_v_amplitude}}
\subfloat{\label{fig:relative_couplings}}
\includegraphics{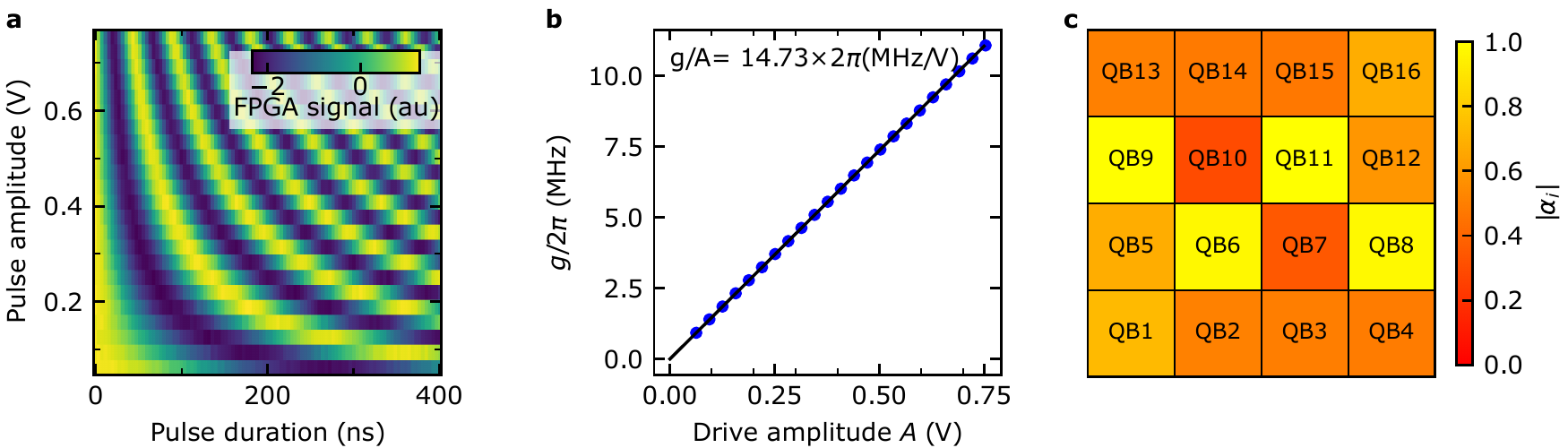}
\caption{\textbf{Characterizing the drive coupling amplitudes.} \textbf{(a)} Rabi oscillations observed by sweeping the drive pulse duration and amplitude. \textbf{(b)} Rabi rates extracted for different values of the pulse amplitude. \textbf{(c)} The amplitude of the relative drive coupling $|\alpha_i|$ for different qubits in the lattice.}
\label{fig:drive_coupling_characterization_1}
\end{figure*}

To determine the coupling amplitude $g_i = \Omega \times |\alpha_i|$ between qubit $i$ and the common drive, we use a Rabi measurement.
We drive each qubit on resonance at $\omega_{\rm com}/2\pi=\SI{4.5}{GHz}$, with all other qubits far-detuned to avoid particle exchange.
We sweep the drive amplitude $A$ and the drive pulse duration, and measure the Rabi rate at each drive amplitude (Fig.~\ref{fig:rabi_sweep}).
Next, we find the Rabi rate for different drive amplitudes and use a linear fit to extract $g_i/A$ (Fig.~\ref{fig:drive_rate_v_amplitude}).
We repeat this procedure for all qubits in our lattice to find the Rabi rate dependence on the common drive amplitude.
By asserting that $|\alpha|=1$ for the qubit with the strongest coupling to the common drive (qubit 11), we report the relative drive coupling magnitudes for all the lattice qubits in Fig.~\ref{fig:relative_couplings} (see Tab.~\ref{tab:drive_parameters} for the values).
The common drive coupling strength $\Omega$ is therefore related to the drive amplitude $A$ through the linear relationship
\begin{equation}
    \Omega(A) = A \times 2\pi \times 49.82 (\mathrm{MHz}).
\end{equation}
We notice that when driving all 16 qubits simultaneously while on resonance, we need to apply a $+7.5\%$ correction to the relation above in order for all of our numerical simulations to match the experimental data.

\subsection{Drive coupling phase}

\begin{figure*}[t!]
\subfloat{\label{fig:interferometry_pulse_sequence}}
\subfloat{\label{fig:interferometry_sweep}}
\subfloat{\label{fig:plaquet_optimization}}
\subfloat{\label{fig:plaquet_phase_selection}}
\includegraphics{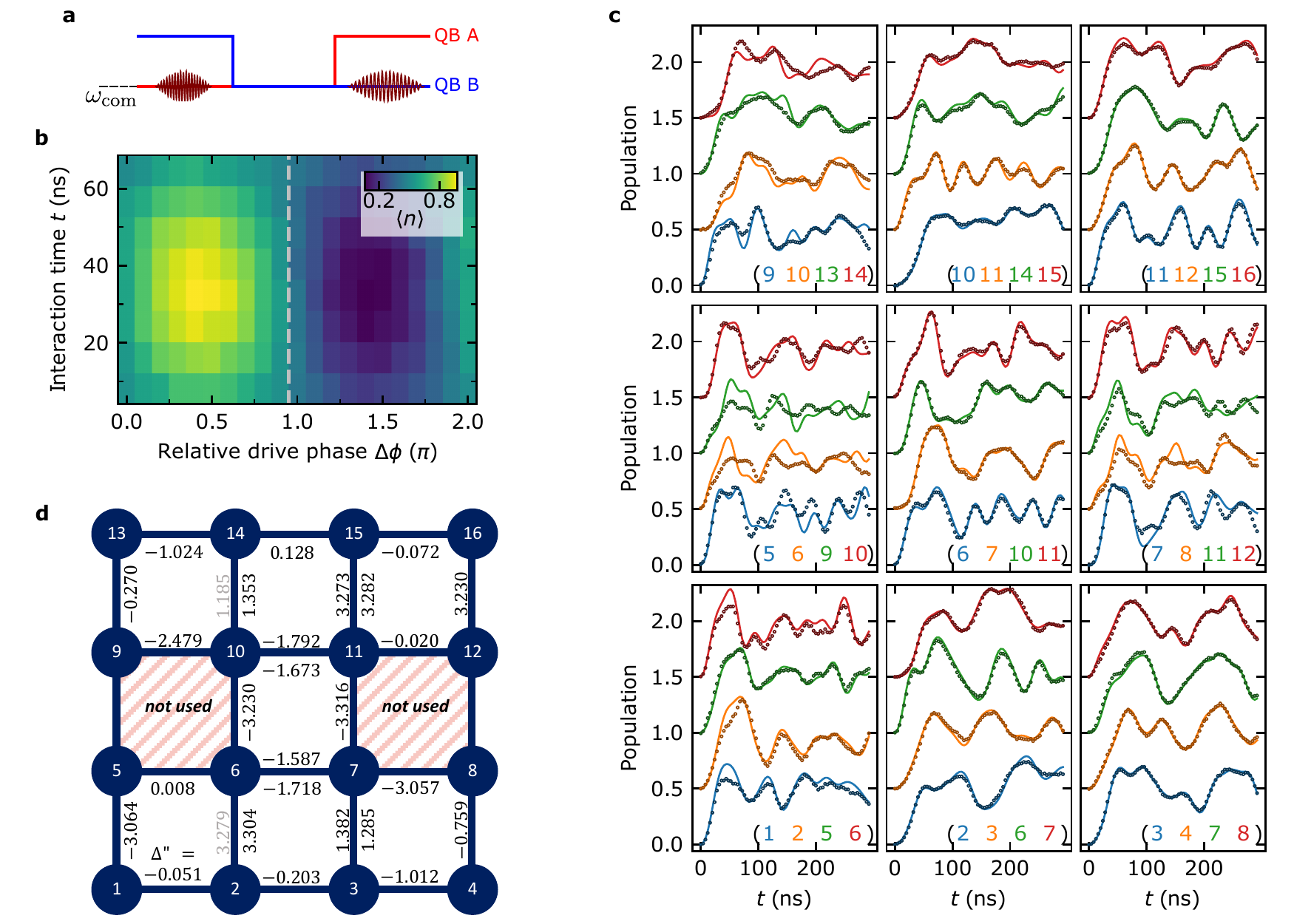}
\caption{\textbf{Characterizing the drive coupling phase.} \textbf{(a)} Pulse sequence used for interferometric pairwise relative drive coupling phase characterization. \textbf{(b)} Measured population on qubit ``B" as a function of sweeping the qubit-qubit exchange interaction time $t$ and the relative drive phase $\Delta \phi$ of the $\frac{\pi}{2}$-pulse applied to qubit ``B" before the readout. The gray dashed line indicates the relative common drive phase between the two qubits. \textbf{(c)} Population versus time for the $9$ different $2 \times 2$ plaquettes; experimental data are shown as circles and the numerical simulations with the optimized drive coupling phases are shown as lines with colors corresponding to the qubit indices shown in parentheses. \textbf{(d)} The phase difference between the adjacent qubits extracted from the respective $2 \times 2$ plaquette fits. The two numbers in grey were unused because those bonds were captured through a higher-quality fit in the neighboring plaquette. The two red-shaded plaquettes were unused because of poor fit quality.}
\label{fig:drive_coupling_characterization_2}
\end{figure*}

We use a two-step approach to characterize the coupling phase $\phi_i = \arg{\alpha_i}$ between the common drive and each qubit. 
We need to calculate 15 coupling phases $\phi_i$ relative to $\phi_1$ (the global phase is irrelevant).
Single-qubit experiments cannot provide information on coupling phases, so to determine them, we need to conduct drive coupling characterization experiments on multi-qubit subsystems of the entire lattice.

We initially characterize the relative drive coupling phase between different pairs of neighboring qubits using an interferometric approach.
For such experiments, we isolate a qubit pair by far-detuning the rest of the lattice.
We begin by biasing qubit ``A" at $\omega_{\mathrm{com}}/2\pi=\SI{4.5}{GHz}$ and applying a $\frac{\pi}{2}$-pulse through the common line while detuning qubit ``B" to prevent them particle exchange.
After this step, the two-qubit system is in the state
\begin{equation*}
    (\ket{0} - i e^{i \phi_A} \ket{1}) \otimes \ket{0}\,,
\end{equation*}
where $\phi_A$ is the phase of the common drive coupled to qubit ``A".
Afterward, we bring qubit ``B" on resonance with qubit ``A", allowing the two qubits to undergo an iSWAP interaction for time $t$. After a full iSWAP period, the state of the system is 
\begin{equation*}
    \ket{0} \otimes (\ket{0} - e^{i\phi_A} \ket{1})\,.
\end{equation*}
Next, we detune qubit ``A", while keeping qubit ``B" at $\omega_{\mathrm{com}}$ and apply a $\frac{\pi}{2}$-pulse through the common line to qubit ``B" resulting in the state
\begin{equation*}
    \ket{0} \otimes \left( \frac{1-ie^{i(\phi_A-\phi_B)}}{2}\ket{0} - \frac{i e^{i \phi_B}+e^{i\phi_A}}{2}\ket{1} \right)\,,
\end{equation*}
where $\phi_B$ is the phase of the common drive coupled to qubit ``B".
By sweeping the phase $\Delta \phi$ of the drive and measuring the population on qubit ``B" we can find $\phi_A-\phi_B$.
The excited state population of the qubit will depend on $\Delta \phi$
\begin{equation*}
    p(\ket{1}) = \frac{1+\sin\left[(\phi_A-\phi_B)-\Delta \phi\right]}{2}.
\end{equation*}
Hence, when $\Delta \phi=(\phi_A-\phi_B)$ the measured population is $p(\ket{1})=1/2$ with a negative slope.
We show the pulse sequence for this measurement in Fig.~\ref{fig:interferometry_pulse_sequence}, and representative measurement results using qubits 9 and 10 in Fig.~\ref{fig:interferometry_sweep}.
The gray dashed line indicates $\Delta \phi=(\phi_{9}-\phi_{10})$.

Using these relative phases measured between the different qubit pairs, we calculate an absolute phase for each qubit with reference to the drive phase on qubit 1, with $\phi_1=0$.

We then take a second step to improve the accuracy of characterizing the drive coupling phases.
We isolate different $2 \times 2$ plaquettes within our lattice and set the corresponding qubit frequencies to $\omega_\mathrm{com}$.
We apply a weak common drive with strength $\Omega=J/2$ and measure the population on each qubit within the plaquettes as a function of time.
By comparing the measured qubit populations to numerical simulations, we fine-tune the drive coupling phases. 
To accomplish this, we define the cost function
\begin{equation}
    \mathcal{C}(\vec{\phi}) = \sum_{i=1}^4\sum_{t=0}^{t_f} \left(P_i^\text{data}(t) - P_i^\text{sim}(t, \vec{\phi}) \right)^2,
\end{equation}
where $P_i^\text{data}(t)$ is the measured population of qubit $i$ after time $t$ and $P_i^\text{sim}(t,\vec{\phi})$ is the corresponding simulated value, as parameterized by the three relative drive coupling phases $\vec{\phi}$ in the plaquette.
Starting with the characterized drive coupling phases using the pairwise interferometry experiments, we use a gradient-descent-based optimizer to find phase values that minimize the cost function.

We utilize the JAX~\cite{jax2018github} and OPTAX~\cite{deepmind2020jax} packages in Python to perform simulations and optimization.
We leverage automatic differentiation to compute gradients and determine parameter updates with an Adam optimizer.
The initial values provided by the first drive coupling phase calibration step are crucial for gradient-descent-based minimization of the cost function since $\mathcal{C}$ is non-convex in $\vec{\phi}$. 
We report the experimental data from the different plaquettes and the numerical simulations with the optimized drive coupling phases in Fig.~\ref{fig:plaquet_optimization}. Generally, we find an excellent agreement between the data and numerical simulation.
However, for two $2\times 2$ plaquettes, formed by qubits (5, 6, 9, 10) and qubits (7, 8, 11, 12), we observed poor convergence of the optimizer, hence, we exclude the results from those two plaquettes from estimating the phases.

The second step of characterization overdetermines the relative phases: we performed the second step for seven $2\times 2$ plaquettes.
Each such measurement yields three independent phase parameters, producing a total of 21 extracted relative phases.
Yet there are only 15 physically independent phase values.
In order to estimate the most likely drive coupling phases, we take the quality of each fit for each plaquette into consideration (see Fig.~\ref{fig:plaquet_phase_selection}).
In Tab.~\ref{tab:drive_parameters} we report the final characterized values for $\phi_i = \arg(\alpha_i)$ that we use for the numerical simulations of our $4 \times 4$ lattice.

\begin{figure*}[t]
\includegraphics{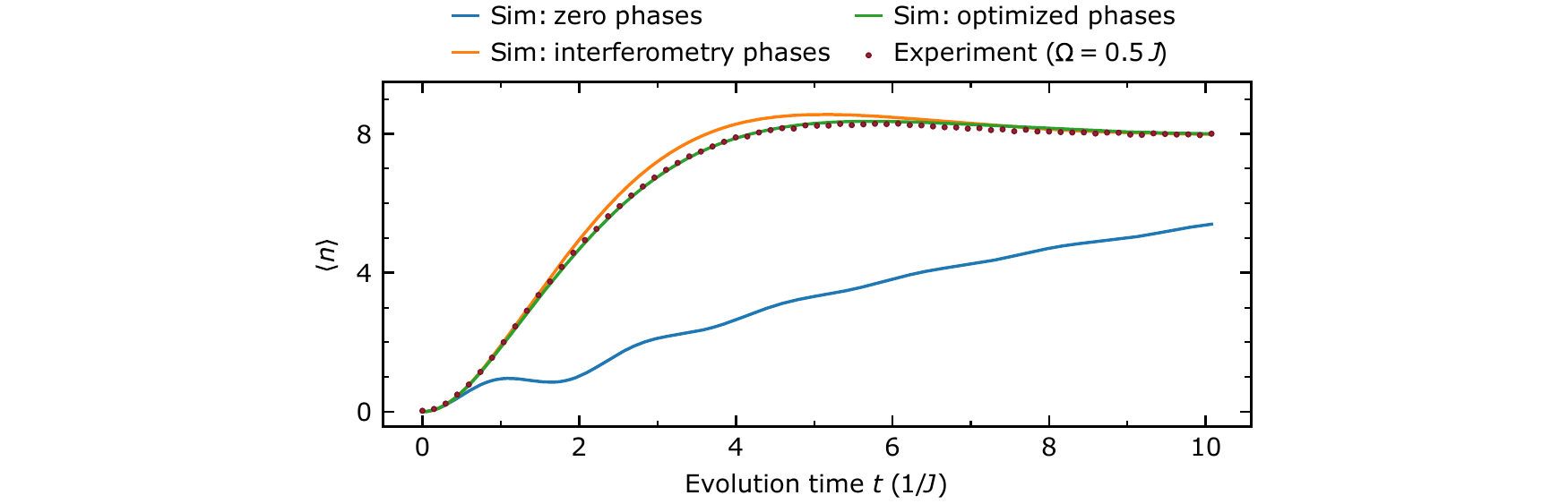}
\caption{\textbf{Experimental data comparison to simulation.} with (blue line) all phases are set to zero, (orange line) using phases characterized with the pairwise interferometry experiments, and (green line) with fine-tuned phase values.}
\label{fig:simulation_phase_comparison}
\end{figure*}

To emphasize the importance of accurately characterizing the drive coupling phase to numerically simulate the behavior of our driven lattice, we consider the time dynamics of the average number of excitations $\langle n \rangle$ in the lattice under a resonant drive.
In Fig.~\ref{fig:simulation_phase_comparison} we compare experimental data for drive strength $\Omega=J/2$ with numerical simulations with (i) all phases set to zero, (ii) using phases characterized with the pairwise interferometry experiments and (iii) with fine-tuned phase values.
We observe that with the fine-tuned phases the numerical simulations of our $4 \times 4$ lattice are in excellent agreement with our experimental results.

\begin{table}[ht]
\centering
\label{tab:drive_parameters}
{\renewcommand{\arraystretch}{1.5}   
\begin{tabular}{ p{2cm}  p{0.75cm} p{0.75cm}  p{0.75cm}  p{0.75cm} p{0.75cm} p{0.75cm}  p{0.75cm}  p{0.75cm}   p{0.75cm} p{0.75cm}  p{0.75cm}  p{0.75cm} p{0.75cm} p{0.75cm}  p{0.75cm}  p{0.75cm}}
\toprule
Qubit index & 1 & 2 & 3 & 4 & 5 & 6 & 7 & 8 & 9 & 10 & 11 & 12 & 13 & 14 & 15 & 16 \\
\hline
$|\alpha|$ & 0.73 & 0.51 & 0.49 & 0.49 & 0.68 & 0.97 & 0.35 & 0.97 & 1.0 & 0.3 & 1.0 & 0.59 & 0.5 & 0.48 & 0.47 & 0.68 \\
arg$(\alpha)$ & 0.0 & 6.22 & 0.1 & 5.38 & 3.23 & 3.26 & 1.41 & 4.63 & 2.88 & 0.45 & 3.96 & 4.0 & 2.54 & 1.44 & 1.14 & 1.01 \\
\hline
\hline
\end{tabular}
}
\caption{\textbf{Characterized common drive coupling.} amplitudes ($|\alpha|$) and phases ($\mathrm{arg}(\alpha)$).}
\end{table}

\section{Hard-core Bose-Hubbard model energy spectrum}

\begin{figure*}[ht!]
\subfloat{\label{fig:hcbh_spectrum}}
\subfloat{\label{fig:hcbh_spectrum_no_NNN}}
\subfloat{\label{fig:hcbh_spectrum_skew}}
\includegraphics{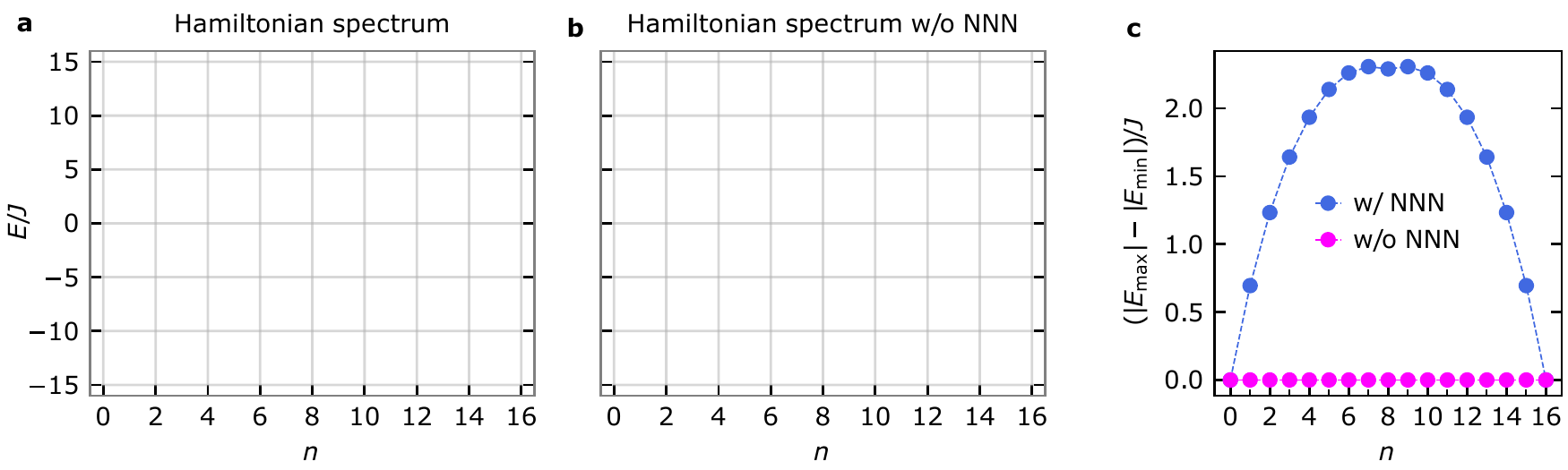}
\caption{\textbf{Energy spectrum comparison}. The energy spectrum of the characterized system Hamiltonian \textbf{(a)}, and the Hamiltonian without the next-nearest coupling (NNN) terms  \textbf{(b)}. The NNN couplings cause the spectrum to skew towards higher positive energies. \textbf{(c)} The difference between the magnitude of the highest eigenenergy $E_{\mathrm{max}}$ and the lowest eigenenergy $E_{\mathrm{min}}$ for the Hamiltonian with and without NNN terms.}
\label{fig:hcbh_spectrum_comparison}
\end{figure*}

We study the energy spectrum of the hard-core Bose-Hubbard Hamiltonian $\hat{H}_{\mathrm{HCBH}}$ with uniform site energies ($\epsilon_i=0$) described in Eq.~4 of the main text.
We use the measured qubit-qubit coupling strengths, including nearest neighbor and next-nearest neighbor exchange interactions, characterized at $\SI{4.5}{GHz}$.
In order to efficiently diagonalize the Hamiltonian of our 16-qubit lattice, we first project the $2^{16} \times 2^{16}$ matrix into a subspace spanned by a fixed particle number $n$.
This projection can be accomplished using a $2^{16} \times {16 \choose n}$ matrix $U$, where the columns of $U$ are all the permutations of the vectors corresponding to the product states with exactly $n$ particles.
The Hamiltonian corresponding to the $n$ particle subspace is therefore calculated through
\begin{equation}
    \hat{H}^\prime_n = U^T \hat{H}_{\mathrm{HCBH}} U,
\end{equation}
where $\hat{H}^\prime_n$ is an ${16 \choose n} \times {16 \choose n}$ matrix.
By diagonalizing $\hat{H}^\prime_n$ we obtain the eigenenergy distribution for different particle numbers $n$ as shown in Fig.~\ref{fig:hcbh_spectrum}.
The largest reduced Hamiltonian is $\hat{H}^\prime_{n=8}$ (describing the subspace of $n=8$) and its width is $2^{16}/{16 \choose 8} \approx 5$ times less than the full Hamiltonian.
Therefore, we expect over $100 \times$ speed-up when diagonalizing the Hamiltonian of each particle number subspace independently compared to diagonalizing the full system Hamiltonian. 

We notice a mild skew towards the higher positive energy eigenenergies in the energy spectrum of our lattice obtained numerically (see Fig.~\ref{fig:hcbh_spectrum}).
In contrast, the eigenenergies of the system Hamiltonian excluding the NNN coupling terms are symmetric around zero energy (Fig.~\ref{fig:hcbh_spectrum_no_NNN}).
In Fig.~\ref{fig:hcbh_spectrum_skew} we show the difference between the magnitude of the highest eigenenergy $E_{\mathrm{max}}$ and the lowest eigenenergy $E_{\mathrm{min}}$ for the different particle-number subspaces.
We notice that for all particle-number subspaces, the eigenenergy distribution is skewed in the positive direction for our system Hamiltonian, which includes NNN couplings, whereas in the absence of NNN coupling, the eigenenergy distribution is symmetric around $E=0$.
Therefore, we conclude that the NNN coupling present in our lattice causes a skew in the eigenenergies of the system.

\section{Coherent-like states across the spectrum}

\begin{figure*}[ht!]
\subfloat{\label{fig:particle-number_entangelment_scaling}}
\subfloat{\label{fig:particle-number_sV_sA_ratio}}
\includegraphics{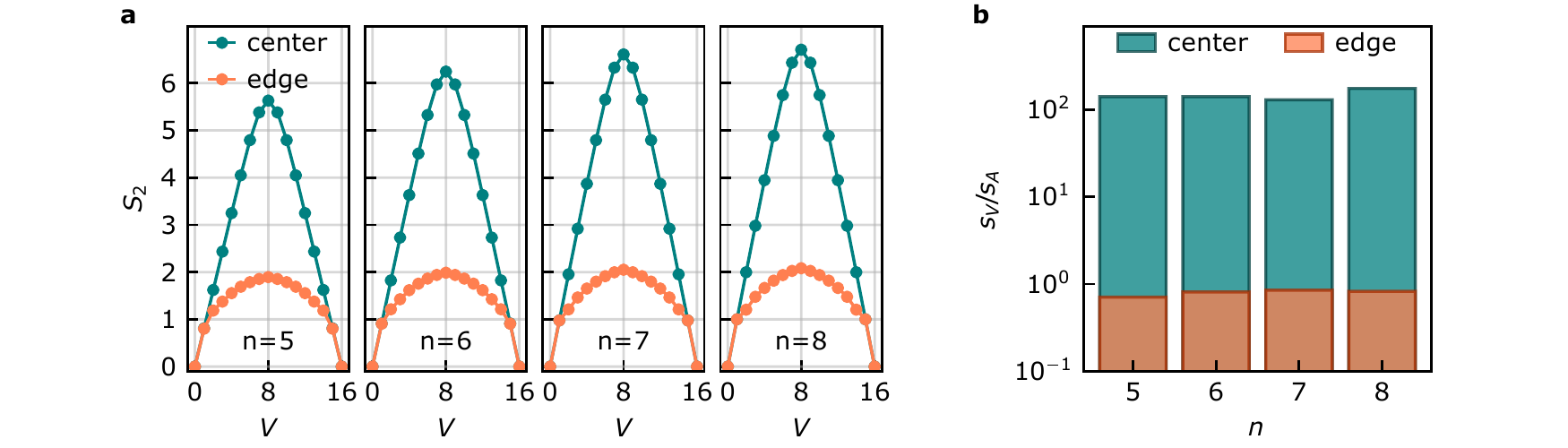}
\caption{\textbf{Entanglement scaling at different particle-number subspaces}. \textbf{(a)} Subsystem entropy versus volume of states at the center and edge of the energy band for subspaces with $n=$5, 6, 7, and 8 particles. \textbf{(b)} Geometric entropy ratio $s_V/s_A$ comparison for different particle-number subspaces.} 
\label{fig:particle-number_entangelment}
\end{figure*}

For each constant-particle-number subspace of the HCBH Hamiltonian, we observe a variation in the geometric entanglement from the edge to the center of the spectrum~\cite{yanay_2020}. To illustrate this variation, we report the average subsystem entropy as a function of volume for states at the edge and the center of the energy band of subspaces with $n=5,6,7,8$ particles in Fig.~\ref{fig:particle-number_entangelment_scaling}. The states at the center of the energy band exhibit a distinct Page curve, while the entropy of the states at the edge of the energy band shows a weak dependence on volume. Furthermore, in Fig.~\ref{fig:particle-number_sV_sA_ratio} we show the geometric entanglement ratio $s_V/s_A$ and notice the same trend between the states at the center and edge of the energy band for the subspaces designated by the different number of particles. The geometric entanglement behavior is consistent across different particle-number subspaces, allowing us to probe the entanglement scaling across the many-body spectrum using a superposition of different eigenstates.

Instead of preparing specific eigenstates, we can use a superposition of eigenstates in different regions of the energy spectrum to probe the entanglement properties in a many-body system. We accomplish this by applying a weakly driving a uniform lattice with some detuning $\delta$ from the frequency of the qubits. 

The hard-core Bose-Hubbard Hamiltonian for a lattice with $N$ sites can be represented in the eigenstate basis as 
\begin{equation}
    \hat H_{\mathrm{HCBH}}/\hbar=\sum_{n}^N \int d \epsilon \rho(n, \epsilon) \left( \omega_q + \epsilon \right) \ket{n, \epsilon} \bra{n, \epsilon}\,,
\end{equation}
where $\ket{n, \epsilon}$ is the eigenstate and $\rho(n, \epsilon)$ is the density of states with $n$ particles and energy $\epsilon$. In this basis, the driving operator $\hat \Sigma$ can be represented as
\begin{equation}
    \hat \Sigma = \sum_n \int d \epsilon d \epsilon^\prime \left( \rho(n+1, \epsilon) \rho(n, \epsilon^\prime) \bra{n+1, \epsilon} \hat \Sigma \ket{n,\epsilon} \right) \ket{n+1, \epsilon} \bra{n, \epsilon^\prime}\,.
\end{equation}
For a weak drive, the driving operator will couple eigenstates that are separated in energy by detuning $\epsilon^\prime - \epsilon = \delta$. Therefore, we can approximate the operators as a combination of energy-raising operators 
\begin{equation}
    e^{-i \omega_{\mathrm{d}} t} \hat \Sigma \approx \int d \epsilon e^{-i  H_{\mathrm{HCBH}} t} \hat A_\epsilon e^{i  H_{\mathrm{HCBH}} t}\,,
\end{equation}
with 
\begin{equation}
    \hat A_\epsilon = \sum_n \left( \sqrt{\rho(n+1, \epsilon_{n+1}) \rho(n, \epsilon_n)} \bra{n+1, \epsilon_{n+1}} \hat \Sigma \ket{n, \epsilon_n} \right) \ket{n+1, \epsilon_{n+1}} \bra{n, \epsilon_n}
\end{equation}
where $\epsilon_n = \epsilon \times \delta$. Each $\hat A_\epsilon$ couples states on the Hamiltonian spectrum shown in Fig.~\ref{fig:hcbh_spectrum} that are on a line with slope $\delta$ and intersection $n=0$ at energy $\epsilon$. For a lattice initialized with no particles, the only relevant operators are $\hat A_0$ and $\hat A_0^\dagger$. Therefore, for times $t$ much shorter than $1/\Omega$, the state of the system is approximately
\begin{equation}
    \ket{\psi(t)} \approx e^{-i \hat H_{\mathrm{HCBH}} t} e^{-i \Omega (\hat A_0 + \hat A_0^\dagger)t}\ket{\mathrm{vacuum}}
\end{equation}
which is similar to a multi-mode coherent state. Due to exchange and on-site interactions, the analogy to coherent states becomes weaker at longer times as the number of photons becomes larger.

\section{Impact of the drive coupling phase on experiments}

\begin{figure*}[h!]
\subfloat{\label{fig:entropy_phase_comparison}}
\subfloat{\label{fig:correlation_length_phase_comparison}}
\subfloat{\label{fig:correlation_matrix_phase_comparison}}
\includegraphics{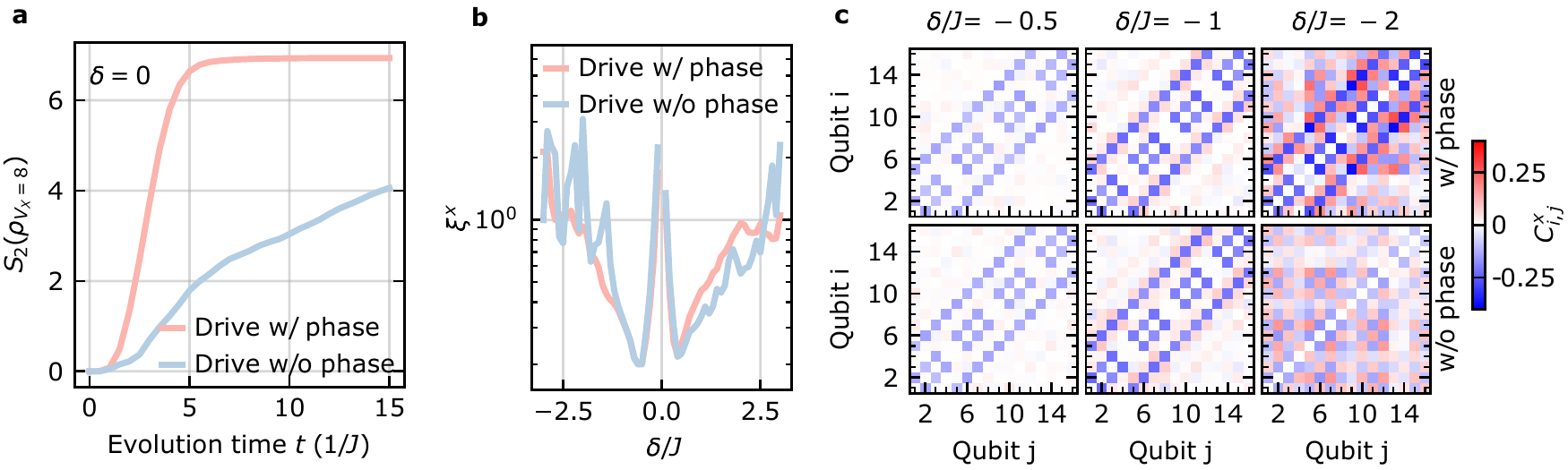}
\caption{\textbf{Impact of the drive coupling phase on correlations and entanglement.} Simulations of \textbf{(a)} the average entanglement entropy of the 8-qubit subsystems, \textbf{(b)} the correlation lengths, and \textbf{(c)} the correlation matrices within the $4 \times 4$ lattice when driven on-resonance with and without the characterized drive phase.}
\label{fig:drive_phase_impact}
\end{figure*}

In this section, we discuss the impact of the characterized drive coupling phases on our experiments using numerical simulations. First, we consider the impact of the drive coupling phase on the formation of entanglement entropy within a driven lattice. In Fig.~\ref{fig:entropy_phase_comparison} we show the simulated time evolution of the average entanglement entropy of the 8-qubit subsystems within the $4 \times 4$ lattice when driven on-resonance with drive strength $\Omega = J/2$. We see that the entropy within the lattice driven with the characterized phases reaches equilibrium much faster than a driven lattice with uniform phases. 

Next, we study the correlation lengths within the states prepared with and without the characterized drive phases (Fig.~\ref{fig:correlation_length_phase_comparison}). To ensure that the prepared states are in equilibrium, we simulate evolution of the lattice for $t=10/J$ under the drive with the characterized coupling phase values, and $t=50/J$ for the drive with uniform phases. We observe that coherence lengths follow the same trend in both preparation scenarios, with some minor deviations. Finally, we simulate the effect of the drive phase on the correlation matrix in Fig.~\ref{fig:correlation_matrix_phase_comparison}. The correlation matrices exhibit the same general pattern, however, the correlations between neighboring sites are stronger for states that are prepared using the drive with the characterized coupling phases.

\section{Measurement of global entanglement}

\begin{figure*}[ht!]
\subfloat{\label{fig:E_gl_2d}}
\subfloat{\label{fig:E_gl_line_cut}}
\includegraphics{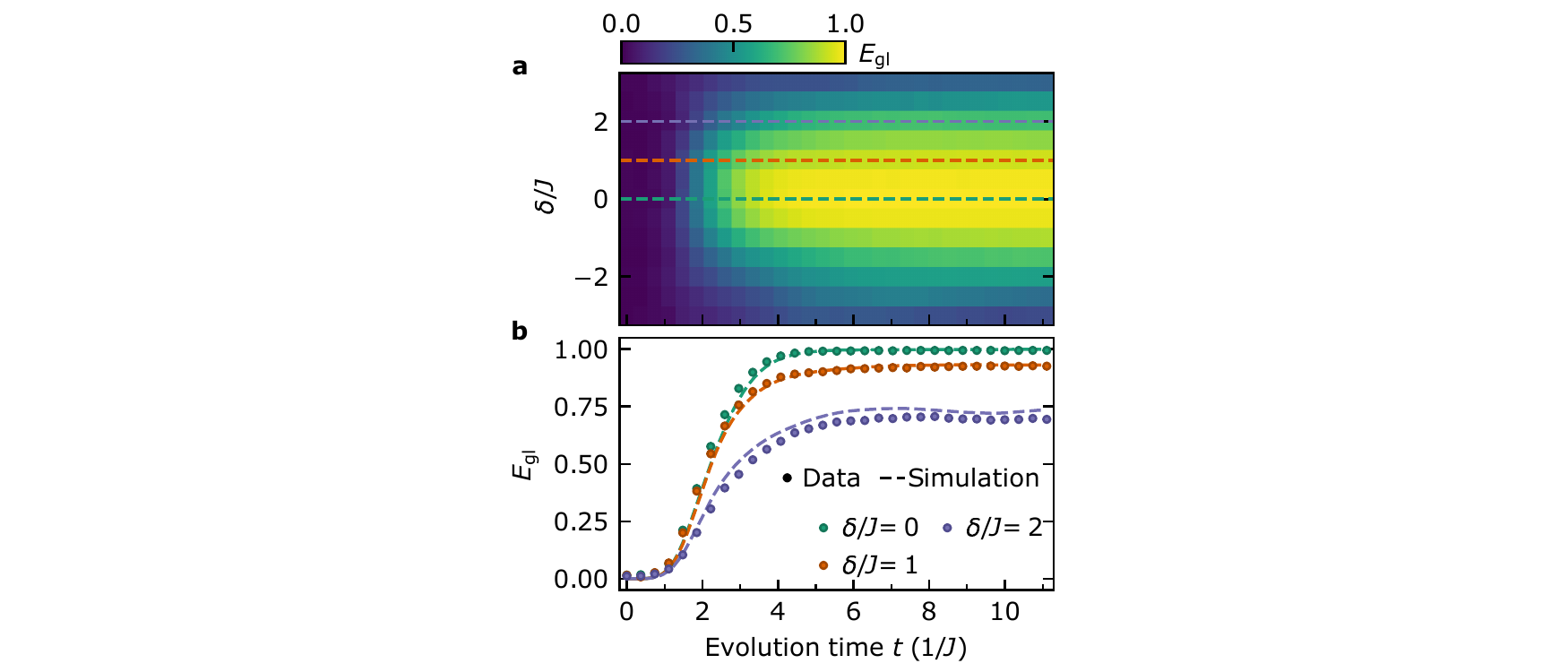}
\caption{\textbf{Entanglement formation time dynamics}. \textbf{(a)},\textbf{(b)}  The global entanglement build-up in the lattice after driving for time~$t$ with strength~$\Omega=J/2$ and detuning~$\delta$ from the lattice frequency. Simulations do not include decoherence.}
\label{fig:global_entanglement}
\end{figure*}

In order to quantify the total entanglement formed, we use the normalized average purity of all the qubits in the lattice~$E_{\mathrm{gl}}=2-\frac{2}{N}\sum_i \tr (\rho^2_i)$, which serves as a global entanglement metric~\cite{meyer_2002}.
This quantity can be measured with low experimental overhead. 
We measure~$E_{\mathrm{gl}}$ by reconstructing the density matrix for each qubit in the lattice via single-qubit tomography. 
In Fig.~\ref{fig:E_gl_2d} we report the measured~$E_{\mathrm{gl}}$ after driving the lattice for time~$t$ with a drive detuning~$\delta$. 
In Fig.~\ref{fig:E_gl_line_cut}, the close agreement between experimental data and numerical simulations, which do not take into account decoherence, suggests that the entanglement formed is between different sites of the lattice, rather than with the uncontrolled environmental degrees of freedom related to decoherence. 

While we use~$E_{\mathrm{gl}}$ to probe the formation of entanglement within our lattice, we cannot distinguish the nature of the entanglement (i.e. whether it is short-ranged or long-ranged).
The dynamics of the entanglement formation in the driven lattice reach a steady state after~$t\approx10/J$. 
We note that as the drive detuning~$\delta$ gets larger, the steady state value for~$E_{\mathrm{gl}}$ decreases. 
This decrease can be attributed to a reduction in the average number of excitations in the lattice with an increase in~$\delta$ (see Fig.~\ref{fig:detuned_drive_population}).

\section{Tomography subsystems for studying the entanglement behavior}

\begin{figure*}[ht!]
\subfloat{\label{fig:tomography_subsystems_sequence}}
\subfloat{\label{fig:tomography_subsystems_dist}}
\includegraphics{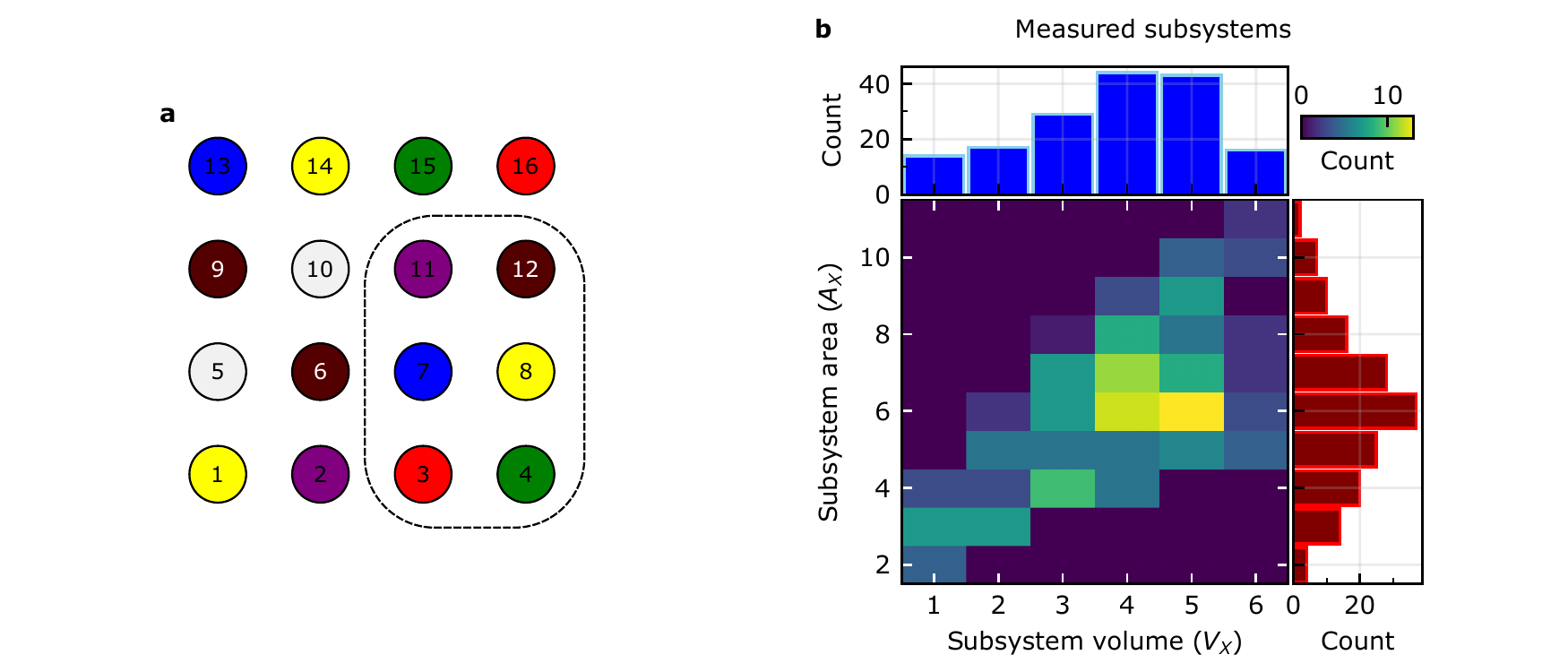}
\caption{\textbf{Tomography subsystems}. \textbf{(a)} Matching colors indicate the qubits in our lattice that have identical Pauli operators in each measured Pauli-string for tomography. \textbf{(b)} Distribution of the area ($A_X$) and the volume ($V_X$) of the measured subsystems. }
\label{fig:tomography_subsystems}
\end{figure*}

In order to tomographically reconstruct the density matrix for a 6-qubit subsystem, we need to measure $3^6=729$ Pauli-strings (e.g. $ZXXYZZ$).
We use a maximum-likelihood estimation (MLE) routine implemented in Qiskit~\cite{Qiskit} to reconstruct the density matrix using the measured Pauli-strings.
With the aid of our high-fidelity simultaneous single-qubit gates and readout, we can efficiently measure the Pauli-strings needed to reconstruct the density matrix for multiple 6-qubit subsystems using a single round of measurements.
Our approach involves initially listing the complete collection of Pauli-strings needed to reconstruct the 6-qubit subsystem $(3,4,7,8,11,12)$.
We then pair up each of the qubits in the rest of the lattice with a qubit in the subsystem so that their measurement basis is consistent for every measurement.
To illustrate how we match the measurement basis of our qubits, we utilize Fig.~\ref{fig:tomography_subsystems_sequence}, in which we use matching colors to indicate the qubits that have identical Pauli operators in each Pauli-string.
Qubits 5 and 10 are excluded from tomographic measurements as applying single-qubit gates to these two qubits causes substantial drive crosstalk affecting the other qubits in the system.

\begin{figure*}[ht!]
\includegraphics[width=0.9\textwidth]{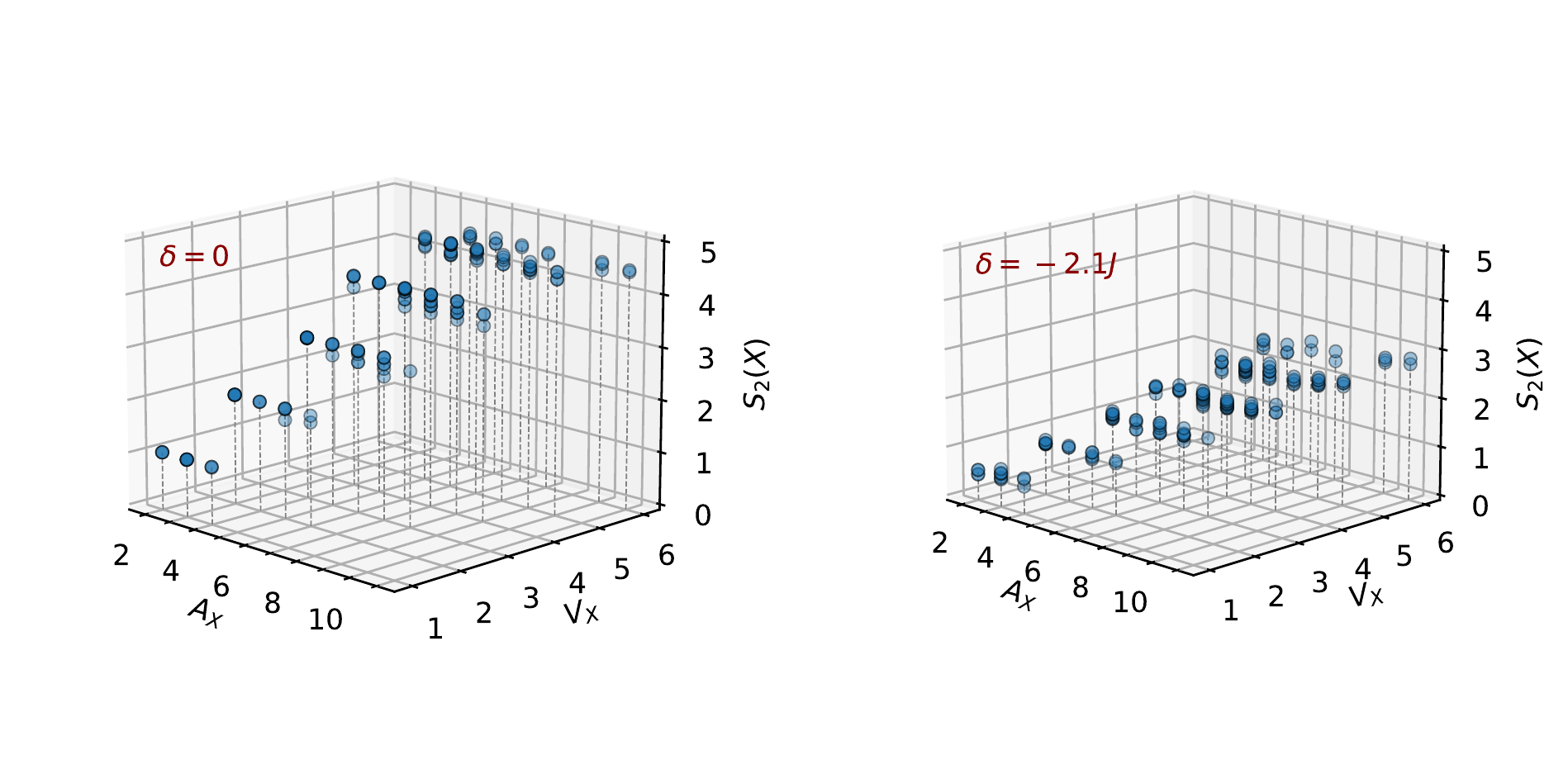}
\caption{\textbf{Visualizing entanglement in different subsystems}. Scatter plots of the second R\'enyi entropy extracted of all 163~subsystems, shown as a function of the area ($A_X$) and the volume ($V_X$) of the each subsystem. Subsystem entropies are shown for the coherent-like states prepared at (left) $\delta=0$ and (right) $\delta/J=-2.1$.}
\label{fig:entropy_scatterplot}
\end{figure*}

Next, we identify all the subsystems that can be reconstructed using the measured Pauli-string.
We include all subsystems of size one through six that are nearest-neighbor connected and in which each qubit possesses a distinct color (according to Fig.~\ref{fig:tomography_subsystems_sequence}), yielding $163$ unique subsystems.
We report a list of all these subsystems, sorted by size $V$ in Table~\ref{tab:tomography_subsystems}.
In Fig.~\ref{fig:tomography_subsystems_dist} we report the distribution of the area ($A_X$) and the volume ($V_X$) of the subsystems used for studying the entanglement scaling behavior in the main text.
In Fig.~\ref{fig:entropy_scatterplot}, we provide a visualization of the entropy extracted for all subsystems as a function of both area and volume.

In our experiments, we measure the expected value of each Pauli string with $2000$ shots. 
We note that the number of samples we can extract for each Pauli string depends on the subsystem size. 
To see this, notice that measurements of the 6-qubit Pauli strings $XYZXYX$, $XYZXYY$, and $XYZXYZ$ all include measurements of the 5-qubit Pauli string $XYZXY$; therefore, while measuring the former we get three times more measurements of the latter.
In general, our measurements yield $2000\times 3^{6-V}$ samples of each volume~$V$ subsystem included in the coloring shown in Fig.~\ref{fig:tomography_subsystems_sequence}.

\begin{table}[ht]
\centering
\vspace{8pt}
\label{tab:tomography_subsystems}
{\renewcommand{\arraystretch}{1.7}   
\begin{tabular}{ p{1.5cm}  p{15cm} }
\toprule

$V$ = 1 & (1), (2), (3), (4), (6), (7), (8), (9), (11), (12), (13), (14), (15), (16) \\
$V$ = 2 & (1,~2), (2,~3), (2,~6), (3,~4), (3,~7), (4,~8), (6,~7), (7,~8), (7,~11), (8,~12), (9,~13), (11,~12), (11,~15), (12,~16), (13,~14), (14,~15), (15,~16) \\
$V$ = 3 & (1,~2,~3), (1,~2,~6), (2,~3,~4), (2,~3,~6), (2,~3,~7), (2,~6,~7), (3,~4,~7), (3,~4,~8), (3,~6,~7), (3,~7,~8), (3,~7,~11), (4,~7,~8), (4,~8,~12), (6,~7,~8), (6,~7,~11), (7,~8,~11), (7,~8,~12), (7,~11,~12), (7,~11,~15), (8,~11,~12), (8,~12,~16), (9,~13,~14), (11,~12,~15), (11,~12,~16), (11,~14,~15), (11,~15,~16), (12,~15,~16), (13,~14,~15), (14,~15,~16) \\
$V$ = 4 & (1,~2,~3,~4), (1,~2,~3,~6), (1,~2,~3,~7), (1,~2,~6,~7), (2,~3,~4,~6), (2,~3,~4,~7), (2,~3,~4,~8), (2,~3,~6,~7), (2,~3,~7,~8), (2,~6,~7,~8), (3,~4,~6,~7), (3,~4,~7,~8), (3,~4,~7,~11), (3,~4,~8,~12), (3,~6,~7,~8), (3,~6,~7,~11), (3,~7,~8,~11), (3,~7,~8,~12), (3,~7,~11,~12), (3,~7,~11,~15), (4,~6,~7,~8), (4,~7,~8,~11), (4,~7,~8,~12), (4,~8,~11,~12), (4,~8,~12,~16), (6,~7,~8,~11), (6,~7,~11,~15), (7,~8,~11,~12), (7,~8,~11,~15), (7,~8,~12,~16), (7,~11,~12,~15), (7,~11,~12,~16), (7,~11,~14,~15), (7,~11,~15,~16), (8,~11,~12,~15), (8,~11,~12,~16), (8,~12,~15,~16), (9,~13,~14,~15), (11,~12,~14,~15), (11,~12,~15,~16), (11,~13,~14,~15), (11,~14,~15,~16), (12,~14,~15,~16), (13,~14,~15,~16) \\
$V$ = 5 & (1,~2,~3,~4,~6), (1,~2,~3,~4,~7), (1,~2,~3,~6,~7), (2,~3,~4,~6,~7), (2,~3,~4,~6,~8), (2,~3,~4,~7,~8), (2,~3,~4,~8,~12), (2,~3,~6,~7,~8), (2,~3,~7,~8,~12), (2,~4,~6,~7,~8), (3,~4,~6,~7,~8), (3,~4,~6,~7,~11), (3,~4,~7,~8,~11), (3,~4,~7,~8,~12), (3,~4,~7,~11,~12), (3,~4,~8,~11,~12), (3,~6,~7,~8,~11), (3,~6,~7,~11,~15), (3,~7,~8,~11,~12), (3,~7,~8,~11,~15), (3,~7,~11,~12,~15), (3,~7,~11,~14,~15), (4,~6,~7,~8,~11), (4,~7,~8,~11,~12), (4,~7,~8,~12,~16), (4,~8,~11,~12,~16), (6,~7,~8,~11,~15), (6,~7,~11,~14,~15), (6,~7,~11,~15,~16), (7,~8,~11,~12,~15), (7,~8,~11,~12,~16), (7,~8,~11,~15,~16), (7,~8,~12,~15,~16), (7,~11,~12,~14,~15), (7,~11,~12,~15,~16), (7,~11,~14,~15,~16), (8,~11,~12,~15,~16), (9,~11,~13,~14,~15), (9,~13,~14,~15,~16), (11,~12,~13,~14,~15), (11,~12,~14,~15,~16), (11,~13,~14,~15,~16), (12,~13,~14,~15,~16) \\
$V$ = 6 & (1,~2,~3,~4,~6,~7), (2,~3,~4,~6,~7,~8), (2,~3,~4,~7,~8,~12), (3,~4,~6,~7,~8,~11), (3,~4,~7,~8,~11,~12), (3,~6,~7,~8,~11,~15), (3,~6,~7,~11,~14,~15), (3,~7,~8,~11,~12,~15), (3,~7,~11,~12,~14,~15), (4,~7,~8,~11,~12,~16), (6,~7,~8,~11,~15,~16), (6,~7,~11,~14,~15,~16), (7,~8,~11,~12,~15,~16), (7,~11,~12,~14,~15,~16), (9,~11,~13,~14,~15,~16), (11,~12,~13,~14,~15,~16) \\

\hline
\end{tabular}
}
\caption{\textbf{List of reconstructed subsystems.}}
\end{table}

\clearpage

\section{Extracting \texorpdfstring{$s_V$}{Lg} and \texorpdfstring{$s_A$}{Lg}}
\label{sec:sV_sA_extraction}

\begin{figure*}[ht!]
\subfloat{\label{fig:S_2_v_V}}
\subfloat{\label{fig:s_V_extracted}}
\subfloat{\label{fig:S_2_v_A}}
\subfloat{\label{fig:s_A_extracted}}
\includegraphics{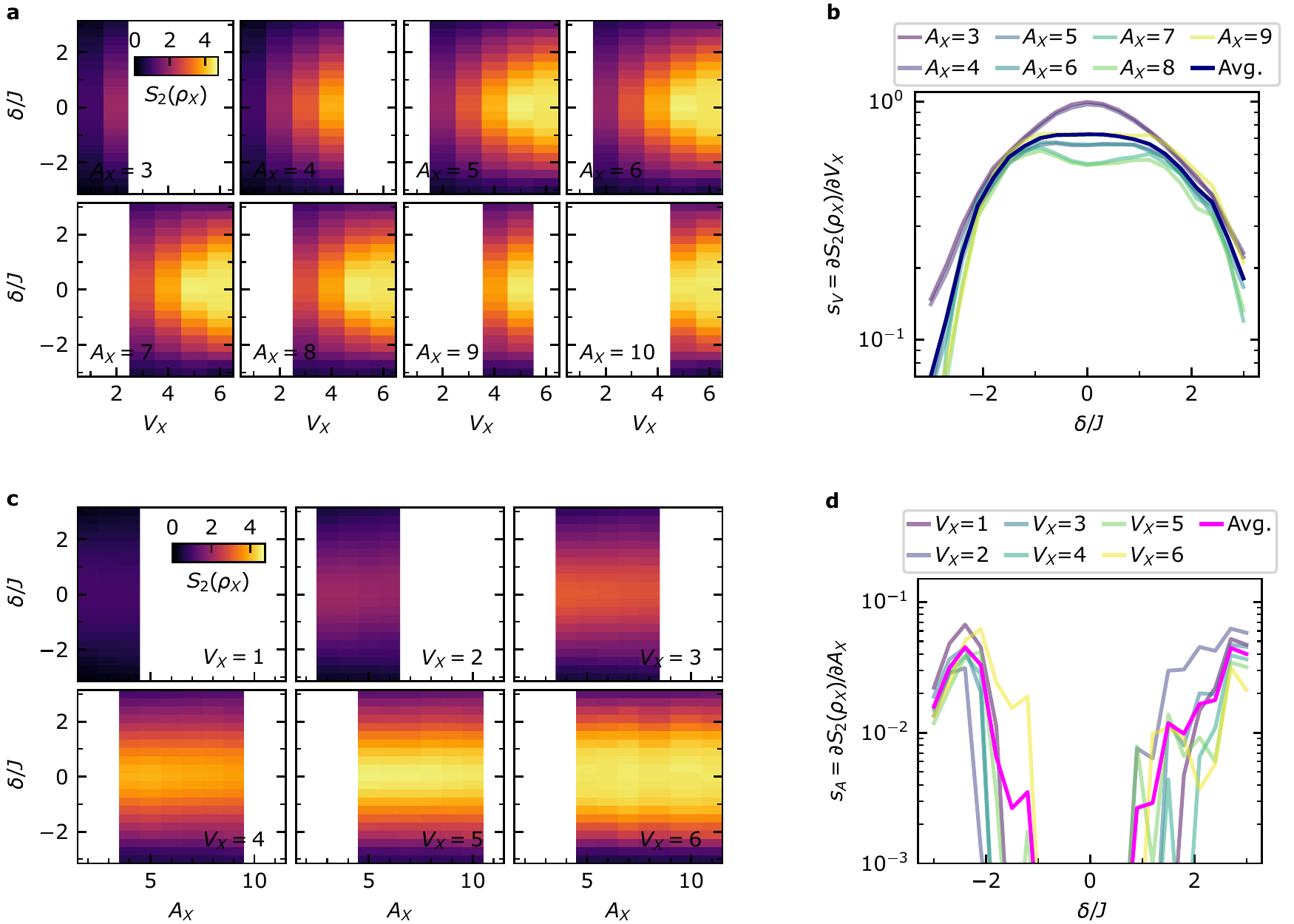}
\caption{\textbf{Extracting $s_V$ and $s_A$ from tomography data.} \textbf{(a)} Dependence of the average subsystem entropy on $V_X$ for subsystems with a fixed $A_X$. \textbf{(b)} The volume entropy per site $s_V$ at different drive detunings $\delta$ obtained via fitting. \textbf{(c)} Dependence of the average subsystem entropy on $A_X$ for subsystems with a fixed $V_X$. \textbf{(d)} The area entropy per site $s_A$ at different drive detunings $\delta$ obtained via fitting. }
\label{fig:sV_sA_extraction}
\end{figure*}

After preparing a specific coherent-like state, we extract the volume entropy per site $s_V$ and the area entropy per site $s_A$ using the measured entanglement entropy ($S_2$) of the different subsystems of our lattice through
\begin{align}
    s_V &= \frac{\partial S_2(\rho_X)}{\partial V_X} \vert_{A_X} \\
    s_A &= \frac{\partial S_2(\rho_X)}{\partial A_X} \vert_{V_X}.
\end{align}
To calculate $s_V$, we consider the rate of change of $S_2(\rho_X)$ as a function of the subsystem volume $V_X$ for subsystems of constant area ($A_X$).
In Fig.~\ref{fig:S_2_v_V} we show the dependence of the average subsystem entropy on $V_X$ for subsystems with a fixed $A_X$. We fit the slope of the increase in entanglement entropy for each $A_X$ and then calculate $s_V$ as an average of the fitted slopes for each drive detuning value $\delta$ (Fig.~\ref{fig:s_V_extracted}). 

We extract the value of $s_A$ using a similar approach.
In Fig.~\ref{fig:S_2_v_A} we show the dependence of the average subsystem entropy on $A_X$ for subsystems with a fixed $V_X$.
We fit the slope of the entanglement entropy as a function of the area for each $V_X$ and then calculate $s_A$ as an average of the fitted slopes for each drive detuning value $\delta$ (Fig.~\ref{fig:s_A_extracted}).
Due to the slope being too close to zero, we cannot reliably fit $\partial S_2(\rho_X)/\partial A_X$ in the detuning range $-1 < \delta/J < 1$. In order to avoid divergences in the $s_A$ and $s_V$ fits, we bound the fitted values in the range spanned by $[10^{-4}, 1]$.

\section{Measurement sampling statistics}\label{section:sampling}

In the main text, we demonstrated excellent agreement in most cases between the entropy extracted from measurements and from simulations, yet we noted a slight deficit of the measured entropy for large subsystems in the volume-law regime. We attribute this deficit to be a consequence of sampling statistics during a finite number of measurements. In this section, we explain this effect, reproduce the effect in simulations that include measurement sampling statistics, and discuss mitigation strategies.

\begin{figure*}[ht!]
\includegraphics{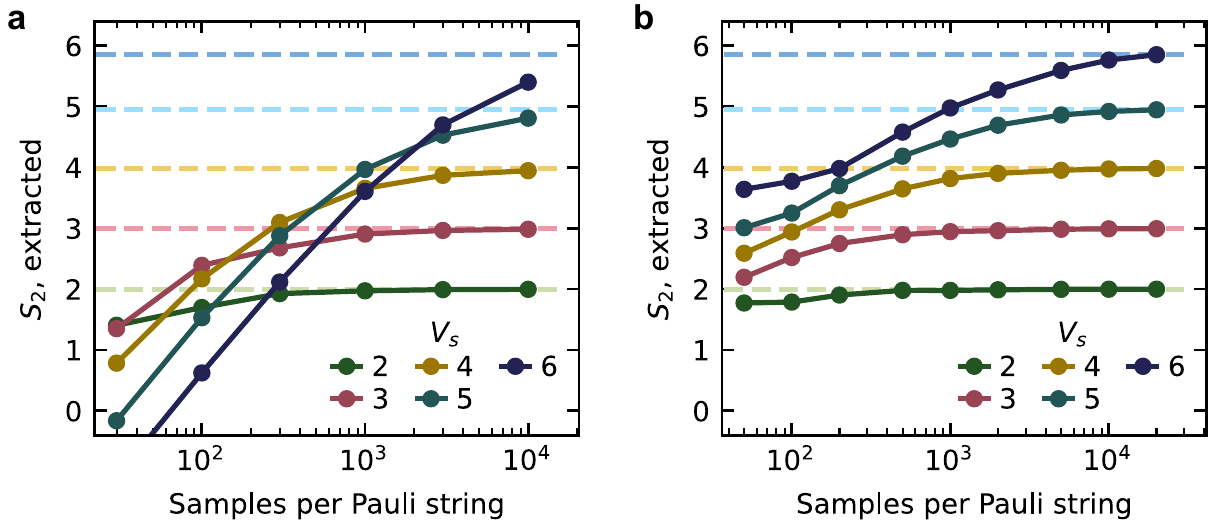}
\caption{\textbf{Simulation of the measurement sampling statistics problem.} The dark markers present the second R\'enyi entropy extracted from simulated tomography of subsystems as a function of the number of measurement samples of each Pauli string used for density matrix reconstruction. For each subsystem volume, results were averaged over the same subsystems used in the experiment. The probability distribution of measurement outcomes was determined from the simulated state prepared at $\Omega=J/2$ and $\delta = 0$. The light dashed lines represent the exact entropy of the simulated state at each volume. \textbf{(a)} Density matrices are reconstructed from the Stokes parameters obtained from Monte Carlo sampling of the probability distribution of eigenvalues of Pauli operator strings. \textbf{(b)} Density matrices are reconstructed using maximum-likelihood estimation on the bitstrings obtained from Monte Carlo sampling of the probability distribution of measurement outcomes of each Pauli operator string. Maximum-likelihood estimation is not used in \textbf{(a)}, whereas the simulation in \textbf{(b)} uses the same reconstruction procedure that was used for experimental data.}
\label{fig:sampling_statistics}
\end{figure*}

Full-state tomography of a subsystem $X$ containing $V_X$ sites involves measurement of Pauli strings $\prod_{i\in X} \sigma_i^{\alpha_i}$ for all combinations of Pauli operators $\alpha_i\in \{x,y,z\}$. For each Pauli string, we aim to accurately determine the distribution of measurement outcomes, of which there are $2^{V_X}$. For larger subsystems, the number of possible measurement outcomes is large; and as the state being measured approaches infinite temperature (an ideal volume law), the distribution of measurement outcomes approaches a uniform distribution (since at infinite temperature all states are equally likely). In these limits, the number of measurements required to accurately sample the outcome distribution becomes large.

The area law states generated when $|\delta|/J$ is larger have lower absolute effective temperatures meaning they feature far-from-uniform distributions of measurement outcomes. Reconstruction of these states is therefore less sensitive to finite sampling statistics. This observation is commensurate with the results of Ref.~\cite{song2017}, where only $3\times 10^3$ samples per Paul string were sufficient to accurately reconstruct 10~qubit GHZ states (which have area-law entanglement scaling).

\begin{figure*}[ht!]
\includegraphics{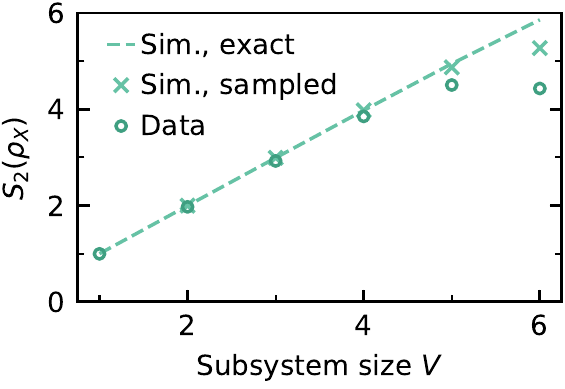}
\caption{\textbf{Comparing measurement sampling simulations to data.} (Crosses) the entropy extracted from Monte Carlo sampling of the simulated state at $\Omega=J/2$ and $\delta = 0$ is compared to (circles) the entropy extracted from experimental data (Fig.~3c of the main text). The values show the mean entropy of all subsystems at each volume. (Dashed line) the exact entropy of the simulated state is shown for comparison.}
\label{fig:sampling_statistics_comparison}
\end{figure*}

To quantify the impact of the number of samples $n_s$ on the extracted entropy, we take a Monte Carlo approach. Here, we consider the coherent-like state prepared at $\delta=0$ and $\Omega=J/2$ (a volume-law state), and begin by obtaining the final state via a decoherence-free simulation on a classical computer. For each subsystem and each Pauli string, we then sample from the distribution of bitstring measurement outcomes $n_s$ times. We reconstruct the subsystem density matrices from these samples and compute their entropy $S_2$. Density matrix reconstruction used the same maximum-likelihood estimation routine as was used to reconstruct density matrices from experimental data.

Results are shown in Fig.~\ref{fig:sampling_statistics} for $n_s$ ranging from 50 up to $2\times 10^4$. Density matrix reconstruction without maximum-likelihood estimation is shown for comparison. For low $n_s$, sampling bias causes a biased reconstruction of the distribution of measurement outcomes, resulting in a deficit of the extracted entropy. The extracted entropy increases and eventually saturates at the correct values as $n_s$ increases. The value of $n_s$ needed to accurately extract the subsystem entropy grows exponentially in subsystem volume. While $n_s=2\times 10^3$ was used for volume~6 subsystems in the present experiment, these simulations show that $n_s\gtrsim 10^4$ is needed to accurately extract the entropy of volume-law states for subsystems of volume~6.

Results from Monte Carlo simulation of measurement sampling effects are compared to experimental data in Fig.~\ref{fig:sampling_statistics_comparison}. We reiterate that, due to the simultaneous tomography of all subsystems, our data yields 2000$\times 3^{6-V}$ measurement samples for a volume~$V$ subsystem.

While the measurement sampling problem increases the time complexity of reconstructing volume-law states above $\mathcal{O}(3^V)$, our simulations confirm that the area-to-volume law transition remains clear in reasonable measurement time---a property that scales favorably to larger quantum systems, as tomography of larger subsystems is unneeded (see the following section).

\section{Extracting the entanglement scaling behavior in larger lattices}

The protocol used in this work to prepare coherent-like states is easily extensible to larger quantum lattices, as is the measurement of correlations. To extract entanglement entropy and Schmidt coefficients, we used full-state tomography of subsystems. Full-state tomography requires exponentially many measurements in system size, so the feasibility of such measurements in larger lattices may not be obvious. In this section, we show that our methods do extend favorably to larger quantum lattices because the size of subsystems that we must consider does not grow with the overall system size.

\begin{figure*}[ht!]
\subfloat{\label{fig:ratio_scaling_1d}}
\subfloat{\label{fig:ratio_scaling_V_max_1d}}
\subfloat{\label{fig:ratio_scaling_2d}}
\subfloat{\label{fig:ratio_scaling_V_max_2d}}
\subfloat{\label{fig:ratio_scaling_system_size}}
\includegraphics{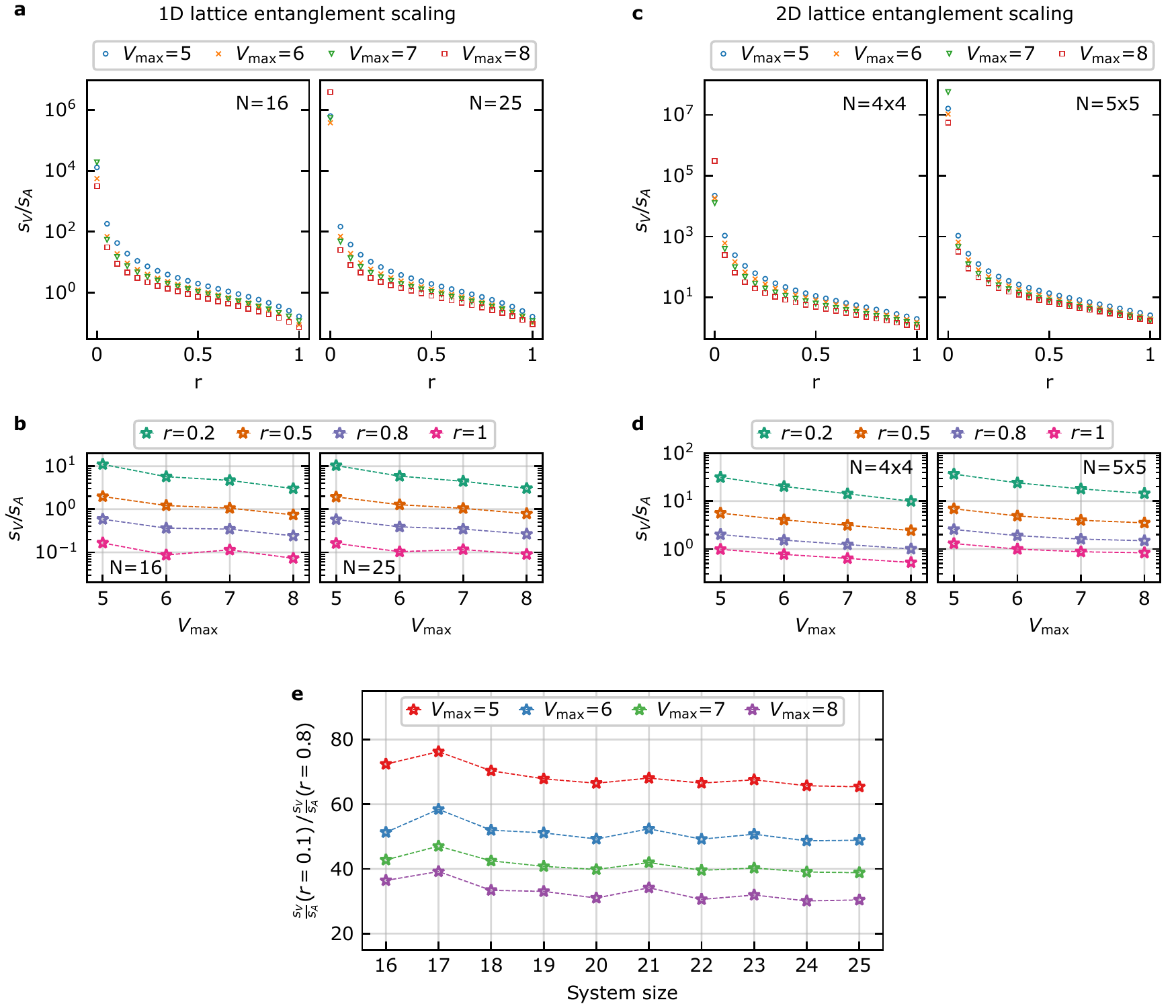}
\caption{\textbf{Scaling of the experimental approach in extracting the volume-to-area entropy per site ratio $s_V/s_A$.} \textbf{(a)} The $s_V/s_A$ ratio extracted for various states for 1d HCBH lattices with 16 and 25 qubits. 
\textbf{(c)} The $s_V/s_A$ ratio variation for different superposition states as a function of the maximum measured subsystem size $V_{\mathrm{max}}$ in 1d lattices.
\textbf{(c)} The $s_V/s_A$ ratio extracted for various states for 2d HCBH $4 \times 4$ and $5 \times 5$ lattices.
\textbf{(d)} The $s_V/s_A$ ratio variation for different superposition states as a function of the maximum measured subsystem size $V_{\mathrm{max}}$ in 2d lattices.
\textbf{(e) The contrast of the $s_V/s_A$ ratio extracted using maximum subsystem size $V_{\mathrm{max}}$ between superposition states with $r=0.1$ and $r=0.8$. }
}
\label{fig:ratio_scaling}
\end{figure*}

In order to study the geometric scaling of entanglement we do not need to perform full tomography of the entire system. Rather, we can learn the scaling behavior of entanglement within a quantum system by measuring the purity of its subsystems of finite size. In this section, we discuss the scalability of our approach in extracting the volume-to-area entropy per site ratio $s_V/s_A$ to larger lattices.

We begin by considering the quantum state $\ket{\psi}$ in a superposition of a volume-law state $\ket{\psi_{\mathrm{random}}}$ and an area-law state $\ket{\psi_{\mathrm{gs}}}$:
\begin{equation}
    \ket{\psi(r)} = \sqrt{1-r} \ket{\psi_{\mathrm{random}}} + \sqrt{r} \ket{\psi_{\mathrm{gs}}}\,.
\end{equation}
Here, $\ket{\psi_{\mathrm{random}}}$ is a randomly generated quantum state, which follows volume-law entanglement scaling with high probability~\cite{pavarini_2020}, and $\ket{\psi_{\mathrm{gs}}}$ is the ground state of the lattice Hamiltonian. By varying the parameter $r$ we can change the participation of the volume-law and area-law states in the superposition, and hence change the geometric entanglement scaling within the state. While fully diagonalizing the Hamiltonian of larger lattices is computationally inefficient, we are able to compute the ground state of HCBH with up to 25 qubits using sparse matrix methods on a commercially available desktop computer.

In Fig.~\ref{fig:ratio_scaling_1d} and ~\ref{fig:ratio_scaling_2d} we numerically study the scalability of our experimental approach in extracting the geometric entanglement scaling in 1D and 2D lattices. We extract the $s_V/s_A$ ratio using the method described in Section~\ref{sec:sV_sA_extraction} for lattices consisting of 16 and 25 qubits by considering subsystems with at most $V_{\mathrm{max}}=5,6,7,8$ qubits. 
We observe that the extracted $s_V/s_A$ ratio using different $V_{\mathrm{max}}$ values is consistent for the superposition states generated with various values of $r$. Next, we plot the scaling of the $s_V/s_A$ ratio extracted for superposition states with the four different $r$ factors in Fig.~\ref{fig:ratio_scaling_V_max_1d} and Fig.~\ref{fig:ratio_scaling_V_max_2d} for 1d and 2d lattices respectively. We notice there are minor variations in the extracted value as the maximum subsystem size increases, however, the $s_V/s_A$ ratios follow the correct trend (i.e. superposition states with a larger participation of the volume-law state have a higher $s_V/s_A$). 

Finally, to further investigate the capability of our approach in quantitatively distinguishing between area- and volume-law states in large systems, we consider the contrast in the extracted $s_V/s_A$ ratio. In Fig.~\ref{fig:ratio_scaling_system_size} we show the ratio between the extracted $s_V/s_A$ value for $\ket{\psi(r=0.1)}$ and $\ket{\psi(r=0.8)}$ for 1d lattices with different numbers of qubits. We observe that as the system size becomes larger, the contrast in $s_V/s_A$ values extracted using different maximum subsystem sizes remains consistent. It is worth noting that as $V_{\mathrm{max}}$ becomes larger the $\frac{s_V/s_A(r=0.1)}{s_V/s_A(r=0.8)}$ ratio decreases, however, for a given $V_{\mathrm{max}}$ the value is consistent for various lattice sizes. This observation suggests that our technique involving measuring the entanglement entropy of subsystems of finite size in a lattice can faithfully distinguish between area- and volume-law states regardless of the overall size of the system.

The numerical simulations discussed in this section confirm that we can study the scaling of entanglement within larger lattices by performing tomography on small subsystems. Furthermore, we note that the size of the subsystems required for studying the entanglement scaling behavior does not depend on the overall lattice size. Using our simultaneous tomographic readout capability we are able to reconstruct the density matrix for various subsystems throughout our lattice by measuring the same number of Pauli strings necessary to reconstruct a single density matrix describing the largest desired subsystem. This feature makes superconducting qubit arrays desirable for studying entanglement within many-body systems.

Lastly, we provide an estimation of the measurement time needed to perform full tomography of a $V=8$ subsystem. The measurement time is constrained by the time needed to upload measurement sequences from the measurement computer to the measurement hardware, plus the time needed to download measurement outcomes back to the measurement computer. In light of the rapid commercial development of improved measurement hardware, we expect these constraints to be assuaged in the near future. Based on the scaling shown in Fig.~\ref{fig:sampling_statistics}, a $V=8$ subsystem requires $3\times 10^5$ measurements for each of the $3^8$ Pauli strings, or $2\times 10^9$ total measurements. At the $\SI{100}{\micro s}$ measurement period ($10^4$ samples per second) used in this work, this corresponds to 55~hours of measurement time. Latency associated with measurement sequence upload can be reduced to constant time by programming the field programmable gate arrays onboard measurement hardware to synthesize combinations of pre-compiled pulse waveforms corresponding to each Pauli operator. Assuming state discrimination is performed onboard the measurement hardware, each measurement requires the transfer of 8~bits of data. Data transfer times will then be negligible, assuming $\SI{1}{Gbit/s}$ communication should become available. Adding extra time for occasional recalibration routines (to correct, for example, for slow drift in $\vec \Phi_{\rm offset}$), we, therefore, expect that near-term hardware developments will enable full tomography of a $V=8$ subsystem in approximately 3~days.

\section{Schmidt coefficients scaling}

\begin{figure*}[ht!]
\includegraphics{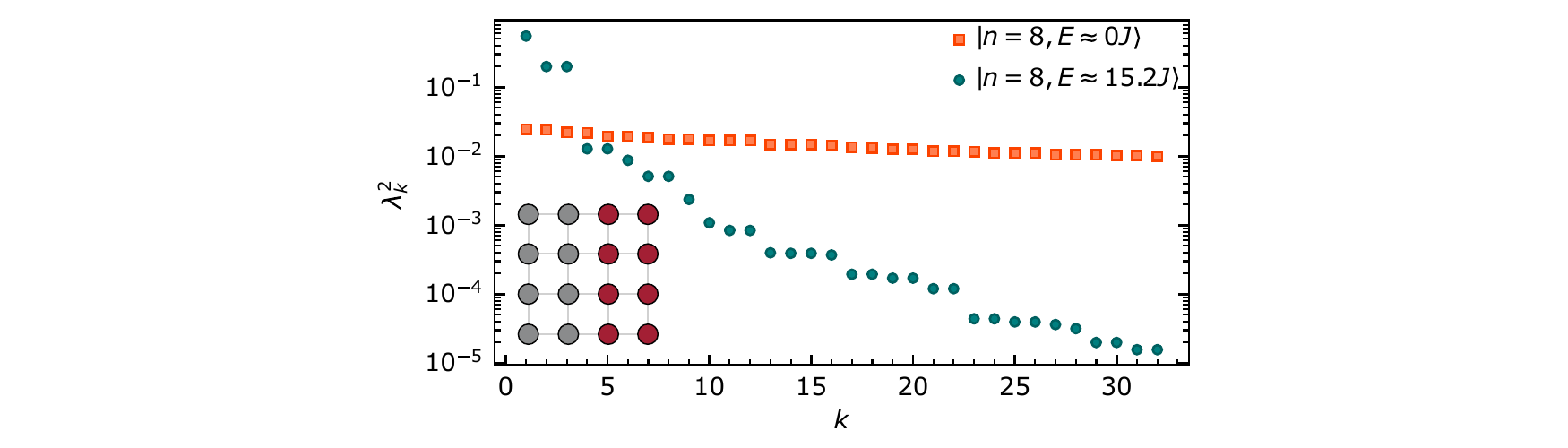}
\caption{\textbf{Schmidt decomposition} \textbf{(a)} The 32 largest Schmidt coefficients for a bipartition of a volume-law eigenstate at the center (energy $E \approx 0$) of the 16-qubit HCBH energy spectrum with an area-law eigenstate at the edge (energy $E \approx 15.2J$) of the spectrum.}
\label{fig:eigenstates_schmidt_numbers}
\end{figure*}

The Schmidt decomposition is a useful tool for representing entangled states.
For a system $AB$ in a pure state $\ket{\psi}$, the theorem states that there exist orthonormal states $\ket{k_A}$ for subsystem $A$, and $\ket{k_B}$ for subsystem $B$, referred to as the Schmidt bases, such that
\begin{equation}
    \ket{\psi} = \sum_k \lambda_k \ket{k_A} \ket{k_B}\,,
\end{equation}
where $\lambda_k$ coefficients are non-negative real numbers satisfying $\sum_k \lambda_k^2=1$, and are known as Schmidt coefficients~\cite{nielsen_2011}.
The Schmidt coefficients provide insight into the ``amount'' of entanglement between the two subsystems.
For two maximally entangled subsystems, each with $n$ qubits, $\lambda_k = 1/\sqrt{2^n}$, which results in an entanglement entropy $S_2(\rho_A) = n$.
The pairwise entanglement for any two qubits in such a system is, therefore, exponentially small in the system size.
In slightly entangled states, only a relatively small number of the Schmidt coefficients have a significant contribution to the weight of the state.
Therefore, area-law states can be compressed and represented by a truncated Schmidt decomposition~\cite{pavarini_2020} 
\begin{equation}
    \ket{\psi_{\mathrm{trunc}}} = \sum_k^\chi \lambda_k \ket{k_A} \ket{k_B}
\end{equation}
using the $\chi$ largest Schmidt coefficients.
Independent of the system size, we can truncate the Schmidt decomposition of an area-law state with a fine number $\chi$ and up to an arbitrary error $\epsilon > 0 $ such that
\begin{equation}
    \big|\big| \ket{\psi} - \ket{\psi_{\mathrm{trunc}}} \big|\big|^2 < \epsilon.
\end{equation}

Using the Schmidt decomposition, we can represent the reduced density matrix $\rho_A$ in the Schmidt basis
\begin{equation}
    \rho_A = \tr_{B}\left(\rho_{AB} \right) = \sum_i \lambda_i^2  \ket{i_A}  \bra{i_A}.
\end{equation}
Hence, by diagonalizing the measured density matrices of our lattice subsystems we can find the Schmidt coefficients of the decomposition between the different subsystems and their complement.
In Fig.~\ref{fig:eigenstates_schmidt_numbers}, we compare the largest 32 Schmidt coefficients for an equal bipartition of a volume-law eigenstate at the center of the 16-qubit HCBH energy spectrum with an area-law eigenstate at the edge of the spectrum.
We observe that a small number of Schmidt states contain nearly all the weight of the decomposition for the eigenstate at the edges of the spectrum, whereas, for the eigenstate at the center of the energy spectrum, the Schmidt coefficients are roughly equal to one another.
This observation suggests that the area-law states at the edges of the energy band can be efficiently numerically simulated using tensor network methods~\cite{eisert_2010}, whereas volume-law states are harder to approximate classically. 

\begin{figure*}[ht!]
\includegraphics{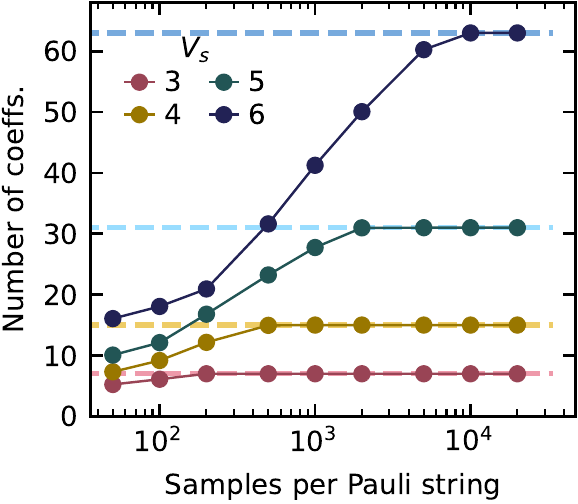}
\caption{\textbf{Simulation of Schmidt decomposition versus measurement samples.} The dark markers present the number of Schmidt coefficients needed to represent with 0.999~accuracy the density matrices extracted from simulated tomography of subsystems as a function of the number of measurement samples of each Pauli string used for density matrix reconstruction. The light dashed lines represent the number of Schmidt coefficients needed to decompose the exact simulated state with the same fidelity. Maximum-likelihood estimation is used for density matrix reconstruction. The coherent-like state generated at $\delta=0$ is used for this figure. For each subsystem volume $V_s$, results were averaged over the same subsystems used in the experiment.}
\label{fig:sampling_Schmidt}
\end{figure*}

Like the entropy, the Schmidt coefficients extracted from reconstructed density matrices are sensitive to the number of measurement samples recorded of each Pauli string in tomography. In Fig.~\ref{fig:sampling_Schmidt}, we show the number of Schmidt coefficients needed to decompose, with 0.999~accuracy, the density matrices reconstructed from simulations of sampled measurements discussed in Section~\ref{section:sampling}, as a function of number of samples.

\section{Estimating the impact of decoherence on measured system entropy}

In the duration of the experiment, our system is subject to decoherence in the form of qubit energy relaxation and dephasing. In our system the average relaxation time is $\bar{T_1}= \SI{22.8}{\micro s}$ and the average dephasing time is $\bar{T_\phi}= \SI{2.4}{\micro s}$ at $\omega_{\rm com}$. To prepare the highly-entangled, coherent-like states, we drive the lattice for $t=\SI{270}{ns}$, followed by the tomographic readout of each qubit. As all qubits undergo Rabi-like oscillations under the drive, which is near the qubit frequency, the decoherence process closely resembles relaxation during rotary echo~\cite{gustavsson2012, yan2013rotating}. In this case, the dominant decay process is exponential at the rate 
\begin{equation}
    \Gamma = \frac{3}{4}\Gamma_1 + \frac{1}{2}\Gamma_\nu,
\end{equation}
where $\Gamma_1=1/T_1$ and $\Gamma_\nu = \frac{1}{2}S_{z}(\Omega)$. Here, the spectral noise power $S_z$ is evaluated at the drive amplitude chosen in experiment $\Omega=J/2$. Assuming $1/f$-type dephasing noise, $S_z(\Omega)=\frac{A_\Phi}{2\pi\Omega}\left| \frac{\partial\omega}{\partial\Phi} \right|^2$, where $A_\Phi$ is the flux noise amplitude. Rotary echo and spin locking measurements were not explicitly performed; we instead extract $A_\Phi$ from a measurement of the spin-echo dephasing rate $\Gamma_\mathrm{echo} = \sqrt{A_\Phi \ln 2}\left|\frac{\partial\omega}{\partial\Phi}\right|$. We use measurements of QB5 at the common lattice frequency $\omega_{\rm com}$, where $T_1=\SI{17.6}{\micro s}$ and $1/\Gamma_\mathrm{echo}=\SI{19.7}{\micro s}$, from which we extract an inconsequential value $1/\Gamma_\nu>\SI{1}{ms}$. The dominant loss channel is then longitudinal decay in the lab frame at timescale $1/\Gamma\approx\frac{4}{3}T_1=\SI{23.5}{\micro s}$.

\begin{figure*}[ht!]
\includegraphics{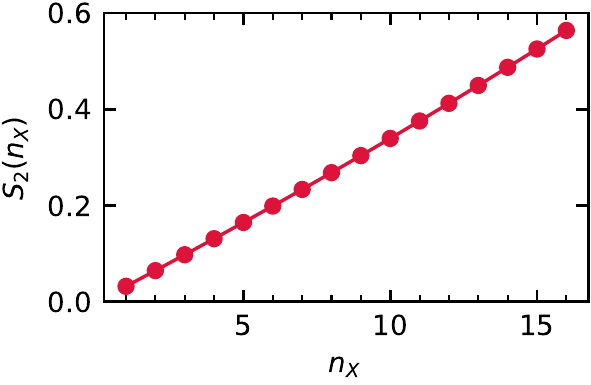}
\caption{\textbf{Entropy contribution from decoherence.} An estimate of the entanglement entropy contribution to the measurement from spin-locking dephasing (with timescale $1/\Gamma$).}
\label{fig:dephasing_entropy}
\end{figure*}

Consider a subsystem $X$ and its complement $\bar X$ within the system $S$. 
Defining the error $\gamma=1-e^{-\Gamma \tau}$, where $\tau=\SI{280}{ns}$ is the experiment duration, and approximating the system dephasing as a quantum channel to the first order in $\gamma$, the system density matrix after undergoing dephasing is
\begin{equation}
    \rho' = (1-n\gamma)\rho + \gamma \sum_i^n \sigma_i^z\rho \sigma_i^z,
\end{equation}
where $\rho = \ket{\psi}\bra{\psi}$ is the density matrix of the pure state of the system with $n=16$ qubits. We define the reduced density matrix without dephasing $\rho_X = \tr_{\bar X}\rho$.
Because $\tr_{\bar X} \sigma_z^i \rho \sigma_z^i = \rho_X$ for $i\in \bar X$, the reduced density matrix of the subsystem is:
\begin{equation}
    \rho'_X = \tr_{\bar X} \rho' = (1-n_X\gamma) \rho_X + \gamma\sum_{i\in X} \sigma_i^z\rho_X \sigma_i^z.
\end{equation}
The subsystem purity is then:
\begin{equation}
    \tr({\rho'}_X^2) = (1-n_X\gamma)^2 \tr[\rho_X^2] + \gamma^2\sum_{i,j\in X} \tr[\sigma_i^z\rho_X \sigma_i^z\sigma_j^z\rho_X \sigma_j^z] + 2\gamma(1-n_X\gamma)\sum_{i\in X} \tr[\rho_X \sigma_i^z\rho_X \sigma_i^z] ,
\end{equation}
or:
\begin{equation}
    \tr({\rho'}_X^2) = \left(1-n_X\gamma(2-n_X\gamma -\gamma)\right) \tr[\rho_X^2] + \gamma^2\sum_{i\neq j\in X} \tr[\sigma_i^z\rho_X \sigma_i^z\sigma_j^z\rho_X \sigma_j^z] + 2\gamma(1-n_X\gamma)\sum_{i\in X} \tr[(\rho_X \sigma_i^z)^2] .
\end{equation}
Notice that the latter two terms are density-density correlators and the squared density expectation values, respectively.
In the volume-law limit, both terms vanish~\cite{srednicki_1994} (intuitively, this is because volume-law states are similar to infinite temperature states). 
Approximating these terms as zero, the subsystem entropy with dephasing is then:
\begin{equation}
    S_2(\rho'_X) \approx S_2(\rho_X) - \log(1+n_X^2\gamma^2-2n_X\gamma + n_X\gamma^2).
\end{equation}
Therefore, the entropy contribution due to dephasing is approximately
\begin{equation}
    S_{2, \mathrm{dephasing}}=- \log(1+n_X^2\gamma^2-2n_X\gamma + n_X\gamma^2).
\end{equation}
This estimate for the additional entropy due to decoherence is used in Fig.~3c of the main text, and is reproduced here for all subsystem sizes in Fig.~\ref{fig:dephasing_entropy}.

Away from the volume-law limit $\delta/J\lesssim 1$, we can estimate the leading correction to the above approximation $2\gamma(1-n_X\gamma)\sum_{i\in X} \tr[(\rho_X \sigma_i^z)^2\approx 0$, by the mean-field expression $2\gamma n_X \left(\frac{N/2 - \langle n \rangle}{N/2}\right)^2$ where $N=16$ is the system size, and measurements of $\langle n \rangle$ at various $\delta$ are shown in Fig.~\ref{fig:detuned_drive_population}.

While the intended initial state in this experiment is a vaccum state $\langle n\rangle=0$, in reality there may be finite initial population due to non-zero qubit temperature. As indicated in Table~\ref{tab:sample_parameters}, the average thermal population is roughly $\varepsilon_0 = 0.04$. Subsystem initial states therefore have entropy $S_2^{\rm thermal} \approx -n_X\log( 1 - 2\varepsilon )$. However, because the qubits are driven on or near resonance during the experiment, we do not expect $S_2^{\rm thermal}$ to contribute directly to additional entropy of the final state. Instead, an initial thermal excitation can be thought of as initializing the experiment in a superposition of the single-particle states. The experiment will then populate a swath of the many-body spectrum slightly more broad than the ideal coherent-like state, reducing the observed contrast between area-law-like and volume-law states.

\section{Time dynamics simulations}

\begin{figure*}[ht!]
\subfloat{\label{fig:3x3_vs_4x4}}
\subfloat{\label{fig:purity_time_dynamics}}
\subfloat{\label{fig:sV_sA_time_dynamics}}
\subfloat{\label{fig:correlation_length_time_dynamics}}
\includegraphics[width=\textwidth]{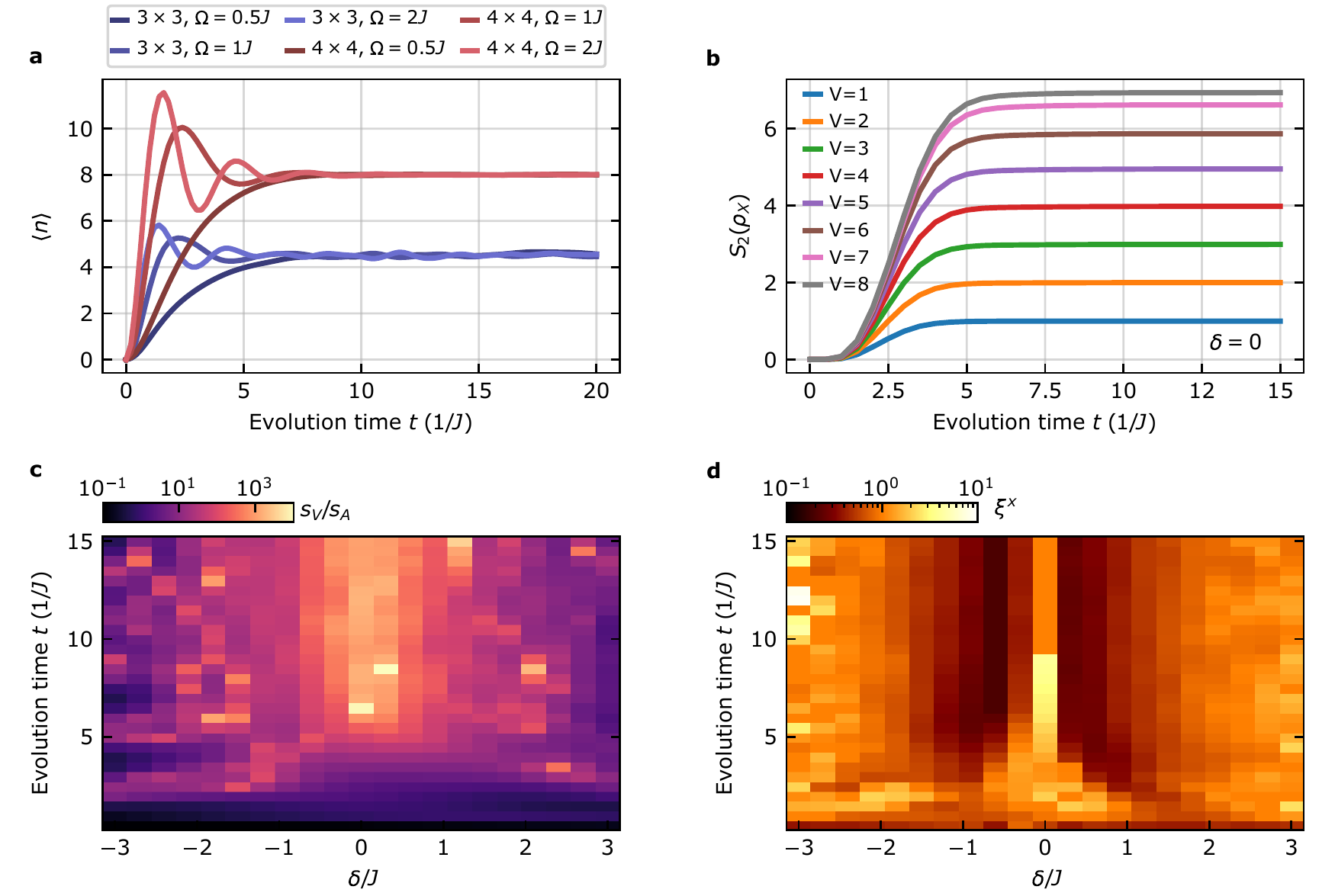}
\caption{\textbf{Time dynamics simulations.} \textbf{(a)} Excitation population for driven $3 \times 3$ and $4 \times 4$ lattices. \textbf{(b)} The time evolution of the entropy of subsystems of different sizes in lattice driven with drive strength $\Omega=J/2$ and detuning $\delta=0$. \textbf{(c)} The time dynamics of $s_V/s_A$ ratios  for various drive detunings $\delta$. \textbf{(d)} The time dynamics of correlation lengths $\xi^x$ for various drive detunings $\delta$.}
\label{fig:time_dynamics simulations}
\end{figure*}

In this section, we report time dynamics simulation results to get a better understanding of the steady-state behavior of a driven lattice. First, we compare the excitation population for a $3 \times 3$ and a $4 \times 4$ lattice in Fig.~\ref{fig:3x3_vs_4x4}. We observe that both systems reach a steady-state population of half-filling by $t=10/J$ for different drive strengths, suggesting that the steady-state time scale does not strongly depend on the lattice size. However, the  $3 \times 3$ lattice exhibits small fluctuations in population due to the finite lattice size and deviation from the thermal limit.

Next, we consider the evolution of the subsystem entropy in a driven $4 \times 4$ lattice with a drive strength $\Omega=J/2$, and detuning $\delta=0$ (Fig.~\ref{fig:purity_time_dynamics}). We observe the formation of entanglement within the system as indicated by the increase in entropy. The entanglement entropy of the subsystems of various sizes approaches the value predicted by the Page curve a volume-law state by $t=10/J$. 

In Fig.~\ref{fig:sV_sA_time_dynamics} and \ref{fig:correlation_length_time_dynamics} we report the simulated time dynamics of $s_V/s_A$ ratios and correlation lengths $\xi^x$ for various drive detunings $\delta$. We notice minor variations in these metrics for drive times exceeding $t=10/J$. The time evolution simulations in this section suggest that the states prepared with a drive time of $t=10/J$ in our experiments are a very close approximation to the desired steady-state coherent-like states.

\section{Coherent-like states in 1D}

\begin{figure*}[ht!]
\subfloat{\label{fig:1d_population}}
\subfloat{\label{fig:1d_population_avg}}
\subfloat{\label{fig:1d_purity}}
\subfloat{\label{fig:1d_correlations}}
\subfloat{\label{fig:1d_correlations_avg}}
\includegraphics[width=\textwidth]{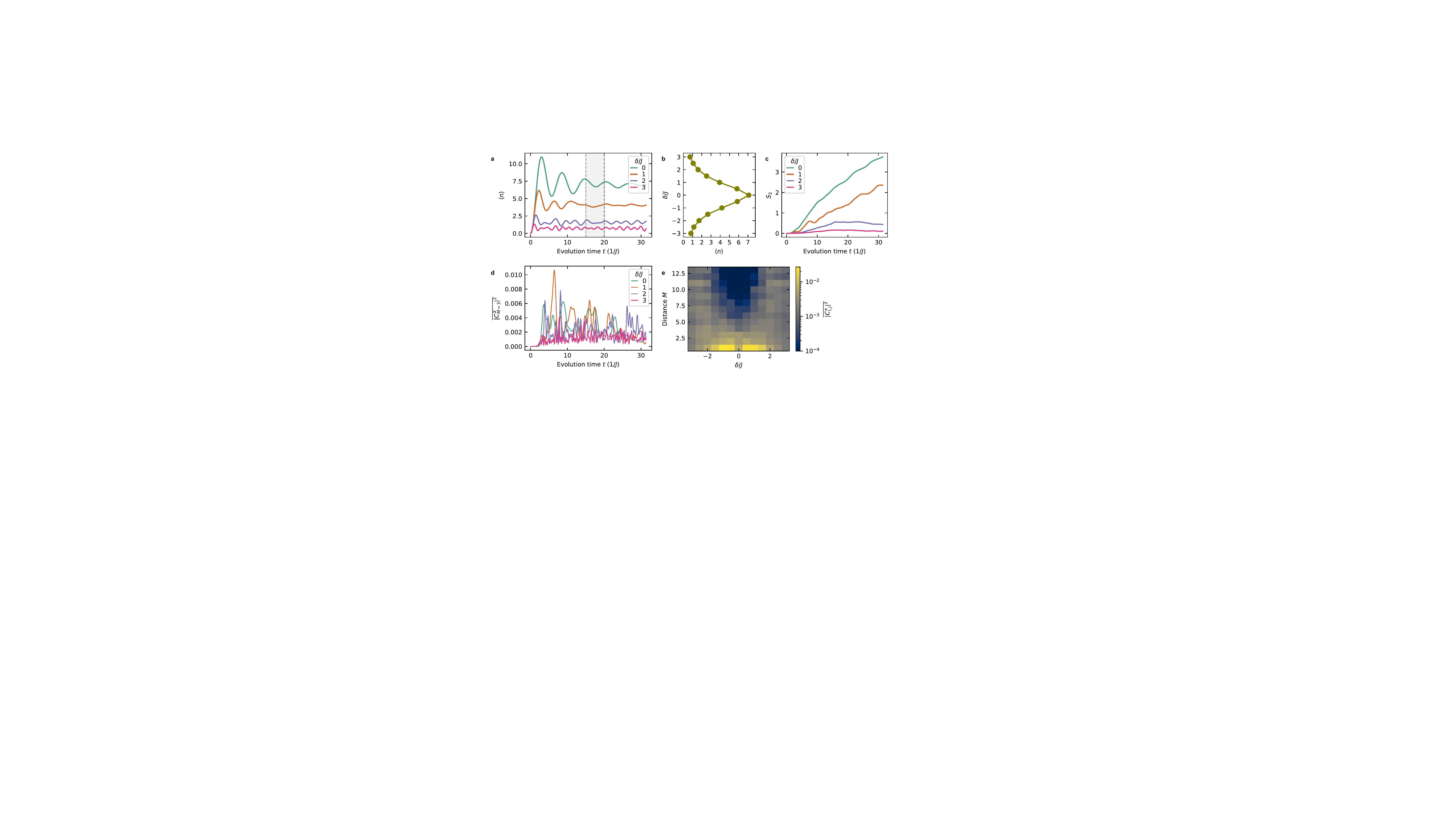}
\caption{\textbf{Simulations of coherent-like states in a one-dimensional 14 qubit system.} \textbf{(a)} Measured total number of excitations $\langle n\rangle$ while driving the system with drive strength $\Omega = J/2$ and various detunings $\delta$ for time $t$. \textbf{(b)} Total number of excitations versus $\delta$, averaged between $t/J=15$ and $t/J=20$. \textbf{(c)} The second R\'enyi entropy of the subsystem consisting of the first seven qubits in the chain, shown as a function of time for various drive detunings. \textbf{(d)} The 2-point distance-3 correlator $|C^X_{M=3}|^2$, averaged over all distance-3 pairs in the lattice, as a function of time for various drive detunings. \textbf{(e)} The time-averaged 2-point correlator shown as a function of $\delta$ and pair distance $M$. The shaded region in \textbf{a} indicates the arbitrarily-chosen time window throughout which the results shown in subfigures \textbf{b} and \textbf{e} were averaged.}
\label{fig:1d_simulations}
\end{figure*}

In two dimensions, the hard-core Bose-Hubbard model is non-integrable, while in one dimension the model is integrable.
While a general statement is beyond the scope of the present work, eigenstate thermalization is known to fail in many integrable systems.
It is therefore interesting to compare the behavior of the two-dimensional (2D) system presented in this work with that of an analogous one-dimensional (1D) system.
In this section, we present numerical simulations of coherent-like states prepared in a 14-qubit one-dimensional hard-core Bose-Hubbard chain, both with and without next-nearest neighbor exchange interactions.
These simulations do not include decoherence.

\begin{figure*}[ht!]
\subfloat{\label{fig:1d_population_NNN}}
\subfloat{\label{fig:1d_population_avg_NNN}}
\subfloat{\label{fig:1d_purity_NNN}}
\includegraphics[width=\textwidth]{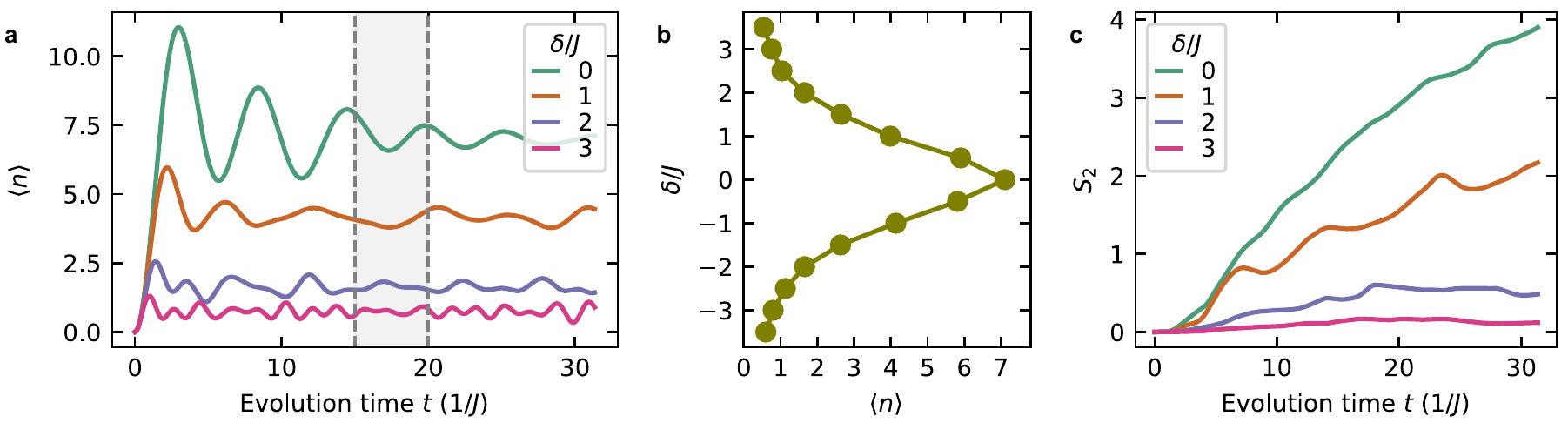}
\caption{\textbf{Simulations of coherent-like states in a one-dimensional 14 qubit system with a next-nearest neighbor coupling of strength $J_\text{NNN}=J/6$.} \textbf{(a)} Measured total number of excitations $\langle n\rangle$ at drive strength $\Omega = J/2$. \textbf{(b)} Total number of excitations versus $\delta$, averaged between $t/J=15$ and $t/J=20$ (indicated by the shaded region in \textbf{a}). \textbf{(c)} The second R\'enyi entropy of the subsystem versus time, for the subsystem consisting of the first seven qubits in the chain.}
\label{fig:1d_simulations_NNN}
\end{figure*}

Unlike in two dimensions, in the one-dimensional system, the population does not reach a steady state in time of order $1/J$.
The average number of excitation continues to oscillate at for tens of characteristic periods $1/J$, with the specifics of the oscillation amplitude and ringdown depending on $\delta$ (see Fig.~\ref{fig:1d_population} and \ref{fig:1d_population_avg}). The analogous dynamics in the presence of exaggerated next-nearest neighbor coupling $J_\text{NNN}=J/6$ are shown in Fig.~\ref{fig:1d_population_NNN} and \ref{fig:1d_population_avg_NNN}.

In Fig.~\ref{fig:1d_correlations} we present the two-point correlations averaged across all distance-3 qubit pairs in the 1D lattice as a function of time.
Unlike in the 2D case, here the correlations fluctuate significantly in time, even after time $10/J$.
We cannot extract meaningful correlation lengths because the correlations do not reach a steady state. Correlations at different distances, presented as time-averaged values between $t/J=15$ and $t/J=20$, are shown in Fig.~\ref{fig:1d_correlations_avg}.

As in the 2D case, we find the largest entanglement entropy for $\delta=0$.
In 2D, the entropy saturates after a time of order $1/J$ (Fig.~\ref{fig:qubit_purity_evolution}).
In 1D, the entropy continues to grow after many multiples of the characteristic time $1/J$.
As shown in Fig.~\ref{fig:1d_purity}, the continued entropy growth at $t\gg 1/J$ is particularly pronounced when $\delta/J\lesssim1$ and persists in the presence of weak next-nearest neighbor exchange interactions (Fig.~\ref{fig:1d_purity_NNN}).

\clearpage

\section{Extended data}

\begin{figure*}[ht!]

\includegraphics{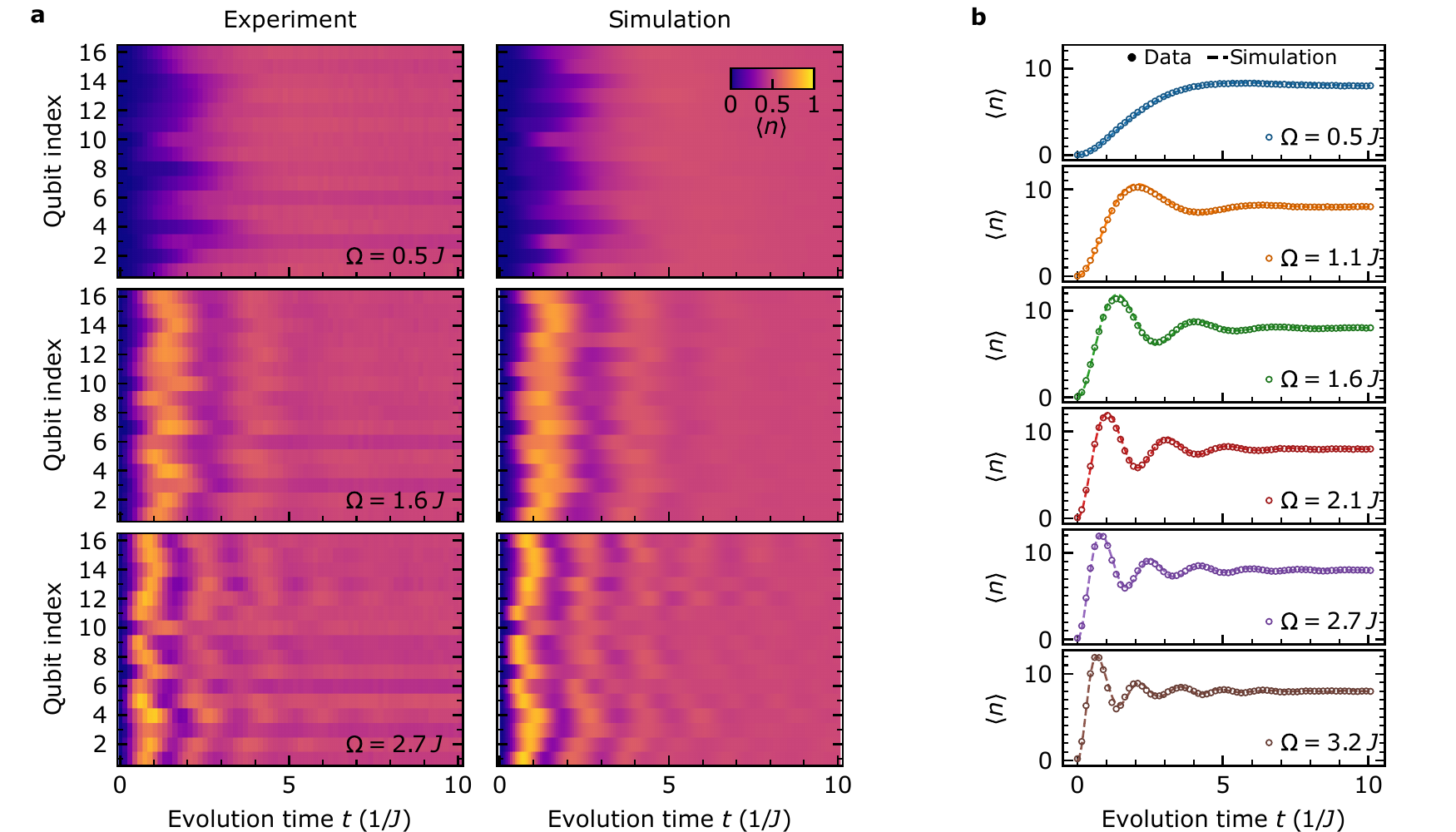}
\caption{\textbf{Extended data on generating coherent-like states}. \textbf{(a)} Measured population on each qubit in the uniform lattice while driving the system on resonance for time $t$ with different values of drive strength $\Omega$. \textbf{(b)} The total number of excitation $\langle n \rangle$ in the uniform lattice while driving the system on resonance for time $t$. We observe excellent agreement between experimental data and numerical simulations using the characterized Hamiltonian parameters.}
\label{fig:Coherent-like_states}
\end{figure*}

\begin{figure*}[ht!]

\includegraphics{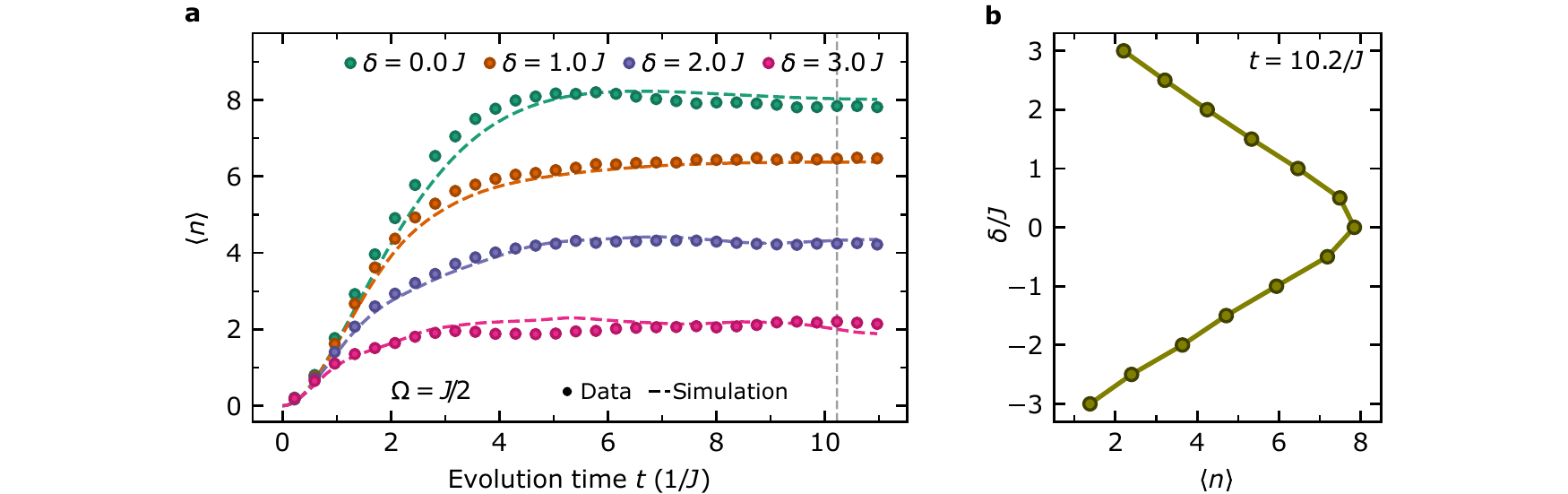}
\caption{\textbf{Extended data on coherent-like states population with detuned drive}. \textbf{(a)} Measured total number of excitation $\langle n \rangle$ in the uniform lattice while driving the system with drive strength $\Omega=J/2$ and detuning $\delta$ for time $t$. \textbf{(b)} The lattice particle number $\langle n \rangle$ measured in experiments for states at approximately $t\approx10/J$. The number of excitations at equilibrium will vary depending on $\delta$. When $\delta=0$, the lattice reaches half-filling, and as the magnitude of the detuning increases, the value of $\langle n \rangle$ at equilibrium decreases.}
\label{fig:detuned_drive_population}
\end{figure*}

\begin{figure*}[ht!]

\includegraphics{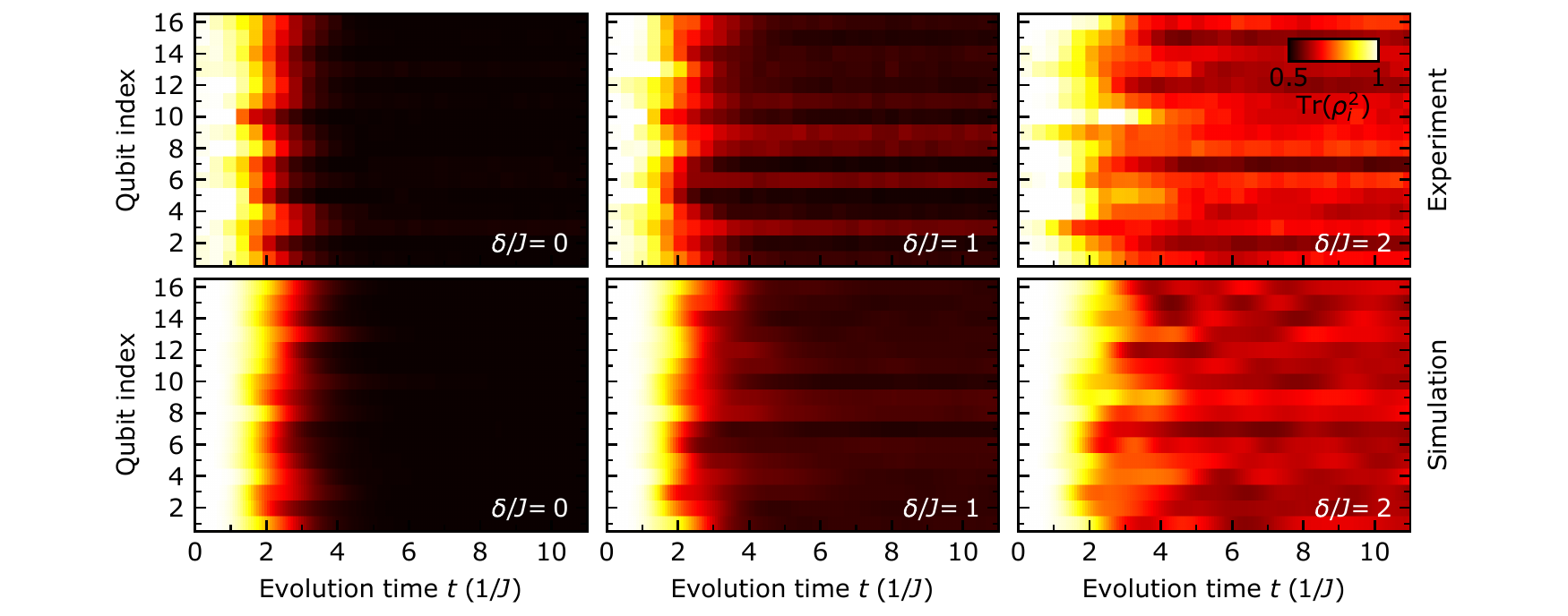}
\caption{\textbf{Extended data on single-qubit purity evolution}. }
\label{fig:qubit_purity_evolution}
\end{figure*}

\begin{figure*}[ht!]

\includegraphics{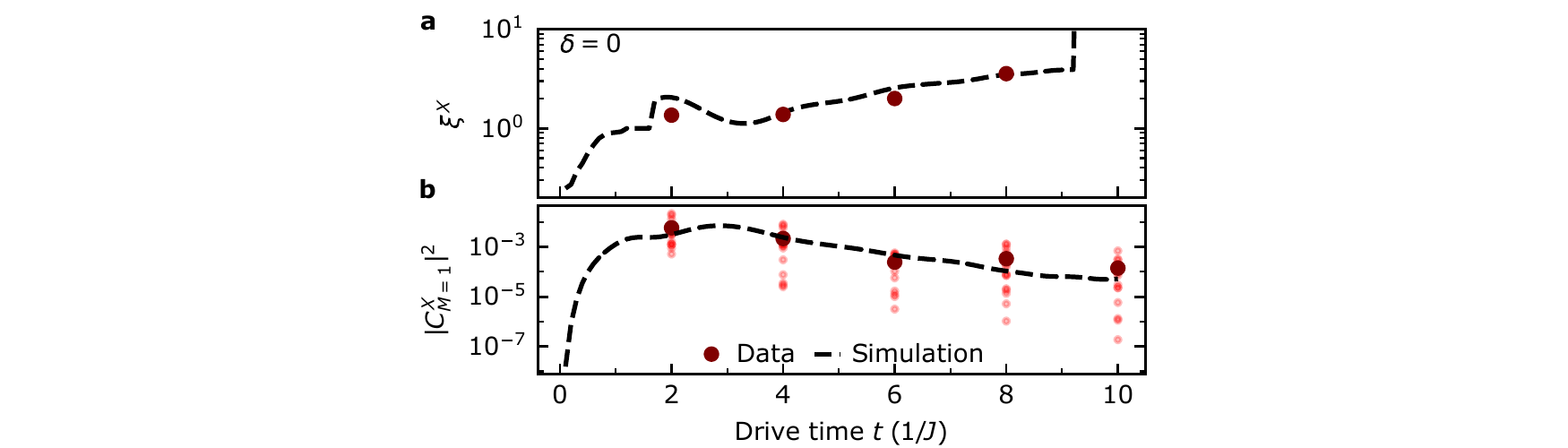}
\caption{\textbf{Evolution of 2-point correlations} at $\delta=0$. The evolution of the correlation lengths \textbf{(a)} and the 2-point correlator values for neighboring qubits $|C^{X}_{M=1}|^2$ \textbf{(b)} as a function of the duration of the common drive at $\delta=0$. After $t=1/J$, the correlation lengths become longer, and the $|C^{X}_{M=1}|^2$ decreases.}
\label{fig:2_point_correlations_evolutio}
\end{figure*}

\begin{figure*}[ht!]

\includegraphics{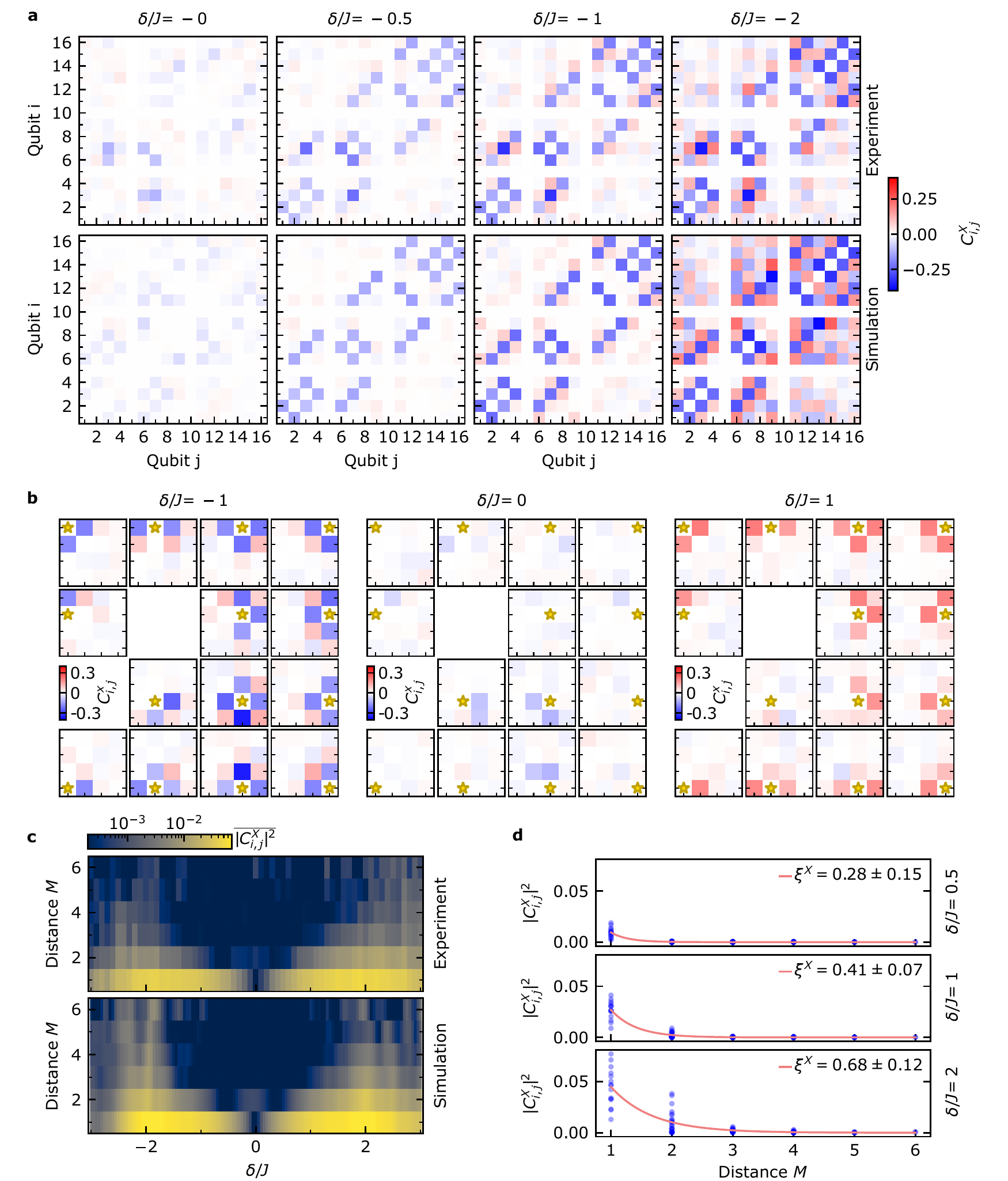}
\caption{\textbf{Extended data on 2-point correlations}. \textbf{(a)} Correlation matrix constructed using the pairwise correlators $C^{X}_{i,j} \equiv \langle \hat{\sigma}^x_i \hat{\sigma}^x_j \rangle - \langle \hat{\sigma}^x_i \rangle \langle \hat{\sigma}^x_j \rangle$ between the different qubit pairs in our system for different values of the drive detuning $\delta$.
\textbf{(b)} Spatial structure of the measured pairwise correlations at $\delta=-J,0,J$. We notice that the sign of the drive detuning determines the sign of the correlations between the nearest neighbors. For $\delta < 0$ ($>0$), corresponding to population in lower (higher) energy half of the many-body band, we observe antiferromagnetic (ferromagnetic) correlations between nearest neighbors. This observations reflects that $J>0$ in our device. Correlations among next-nearest neighbors are weaker, reflecting that the coherent-like states lack long-range order.
\textbf{(c)} Experimental data and numerical simulations of the average 2-point correlators squared along the $x$-basis, $\overline{|C^{X}_{i,j}|^2}$, between qubit pairs at distance $M$ for drive duration $t=10/J$, strength $\Omega=J/2$ and detuning $\delta$ from the lattice frequency. \textbf{(d)} The correlation length $\xi^{X}$ are extracted with an exponential fit following $|C^{X}_{i,j}|^2 \propto \exp(-M/\xi^{X})$ using all measured 2-point correlators as a function of the qubit pair distance. The reported correlation length fit value is accompanied by the standard fit error.}
\label{fig:2_point_correlations}
\end{figure*}

\clearpage

\bibliography{supplement}